\documentclass[twocolumn,usenames,dvipsnames]{aastex62}
\usepackage{amsmath}
\usepackage{apjfonts}
\usepackage{amssymb}
\usepackage{xspace}
\usepackage{hyperref}
\usepackage{natbib}
\usepackage{multirow}
\usepackage{graphicx}
\usepackage{epstopdf}
\usepackage{epsf}
\usepackage{subfigure}
\usepackage{xcolor}
\usepackage{rotating}
\usepackage{mathrsfs}
\usepackage{slashbox}
\usepackage{multirow}
\usepackage{lipsum}
\usepackage[normalem]{ulem}

\def\Msun{M_\odot}
\def\dim#1{\mbox{\,#1}}
\def\xHI{{x_{\rm HI}}}
\def\xHII{x_{\rm HII}}
\def\xHeI{x_{\rm HeI}}
\def\xHeII{x_{\rm HeII}}
\def\xHeIII{x_{\rm HeIII}}

\begin{document}

\title{The Distribution and Evolution of Quasar Proximity Zone Sizes}

\author{Huanqing Chen}
\affiliation{Department of Astronomy \& Astrophysics; 
The University of Chicago; 
Chicago, IL 60637, USA}

\author{Nickolay Y.\ Gnedin}
\affiliation{Theoretical Physics Department; 
Fermi National Accelerator Laboratory;
Batavia, IL 60510, USA}
\affiliation{Kavli Institute for Cosmological Physics;
The University of Chicago;
Chicago, IL 60637, USA}
\affiliation{Department of Astronomy \& Astrophysics; 
The University of Chicago; 
Chicago, IL 60637, USA}

\correspondingauthor{Huanqing Chen}
\email{hqchen@uchicago.edu}

\begin{abstract}
In this paper, we study the sizes of quasar proximity zones with synthetic quasar absorption spectra obtained by post-processing a Cosmic Reionization On Computers (CROC) simulation. CROC simulations have both relatively large box sizes and high spacial resolution, allowing us to resolve Lyman limit systems, which are crucial for modeling the quasar absorption spectra. We find that before reionization most quasar proximity zone sizes grow steadily for $\sim 10$ Myr, while after reionization they grow rapidly but only for $\sim 0.1$ Myr. We also find a slow growth of $R_{\rm obs}$ with decreasing turn-on redshift. In addition, we find that $\sim 1-2\%$ of old quasars ($30$ Myr old) display extremely small proximity zone sizes ($<1$ proper Mpc), of which the vast majority are due to the occurrence of a damped Ly$\alpha$ absorber (DLA) or a Lyman limit system (LLS) along the line of sight. These DLAs and LLSs are contaminated with metal, which offers a way to distinguish them from the normal proximity zones of young quasars.

\end{abstract}

\section{Introduction}

Understanding when and how the universe underwent reionization is a frontier in both cosmology and astrophysics. Direct constraints on the process and timing of reionization come from probes of neutral hydrogen in the intergalactic medium (IGM). Since at present the direct emission in the (redshifted) 21 cm line has not yet been detected, the only available direct probe of intergalactic gas is Ly$\alpha$ absorption in the spectra of distant quasars. The Ly$\alpha$ line is a resonant line with extremely large cross section; at lower redshifts ($z=2\sim4$) the residual neutral hydrogen in the IGM creates dense absorption features called the Ly$\alpha$ forest. At higher redshifts both the neutral hydrogen fraction and the density of the universe increase, and the absorption features blend together to form ``dark gaps'' between isolated ``transmitted spikes'', which eventually disappear in the complete ``Gunn-Peterson trough'' above $z\gtrsim6.5$ {\citep{becker2001,fan2006b,becker2015,mcgreer2015,mazzucchelli2017,  bosman2018,eilers2018,lu2020, yang2020}}. When that happens and the transmitted flux drops to zero, probing the IGM  with Ly$\alpha$ absorption becomes impractical, except for special environments in the vicinity of bright quasars, the so-called "quasar proximity zones". Inside these proximity zones, quasar radiation ionizes the surrounding gas much in excess of the cosmic mean and, hence, makes it much more transparent to Ly$\alpha$ radiation.

\begin{figure*}
    \centering
    \includegraphics[width=0.33\textwidth]{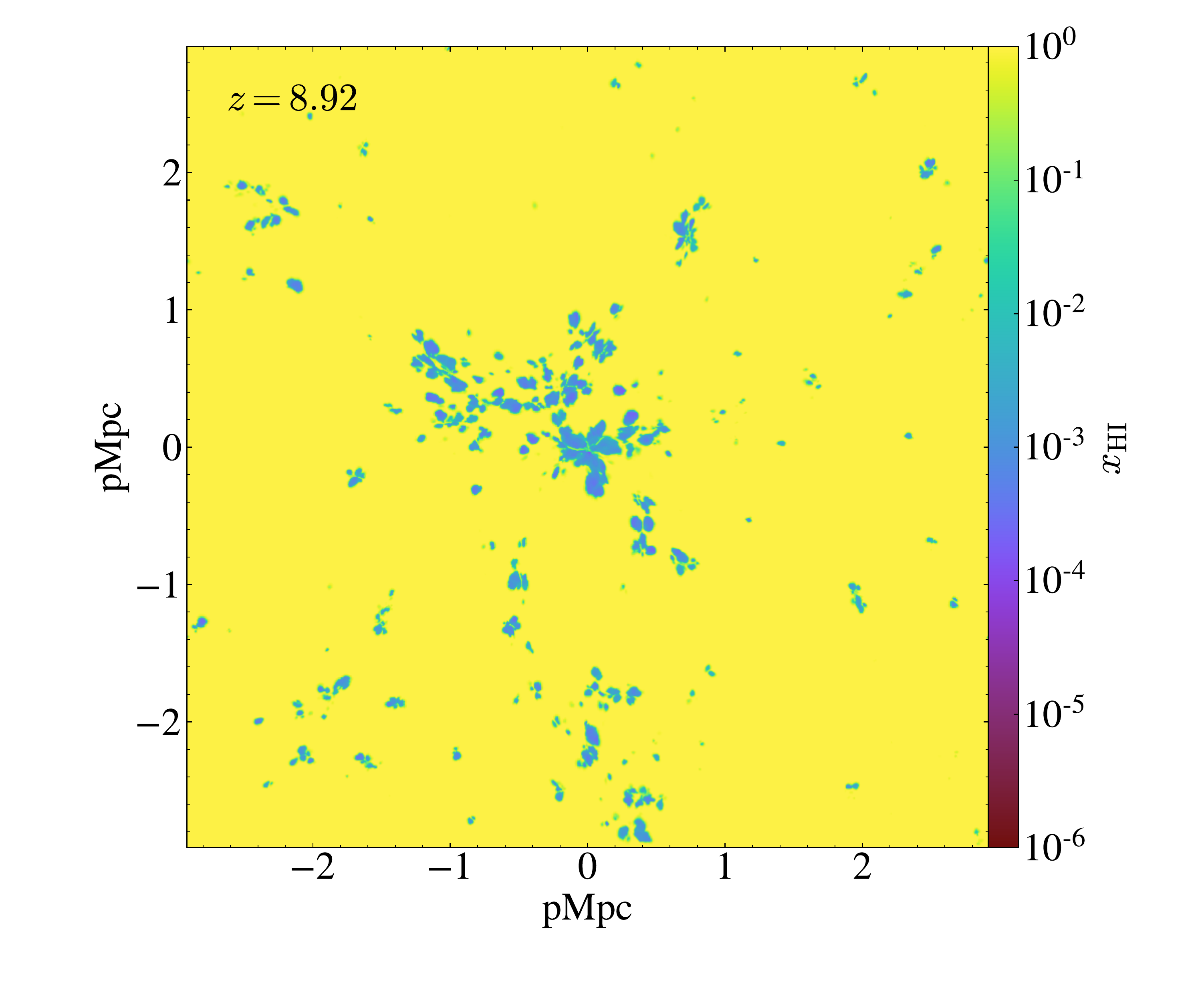}\hspace{-1.147cm}
    \includegraphics[width=0.33\textwidth]{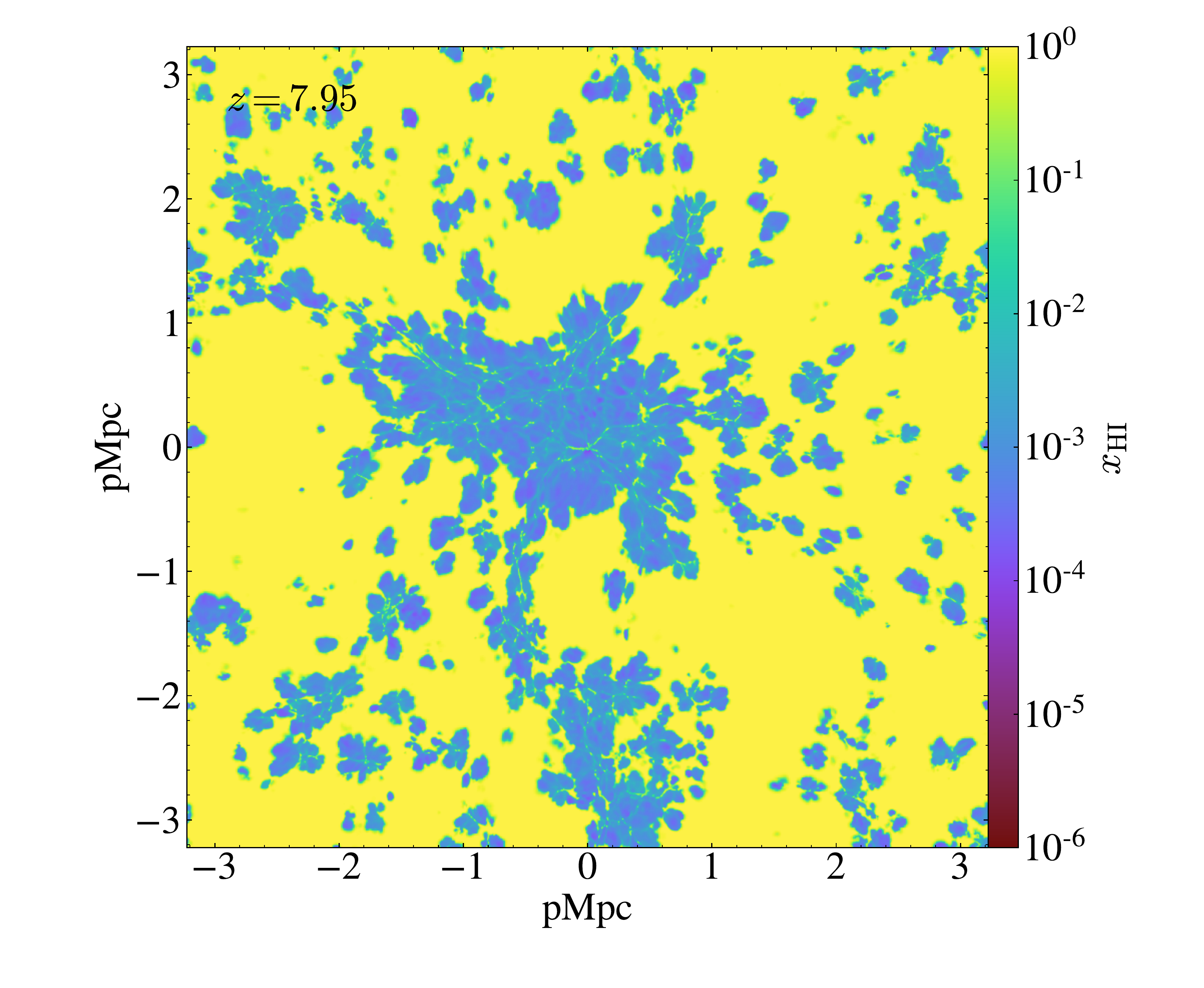}\hspace{-1.147cm}
    \includegraphics[width=0.33\textwidth]{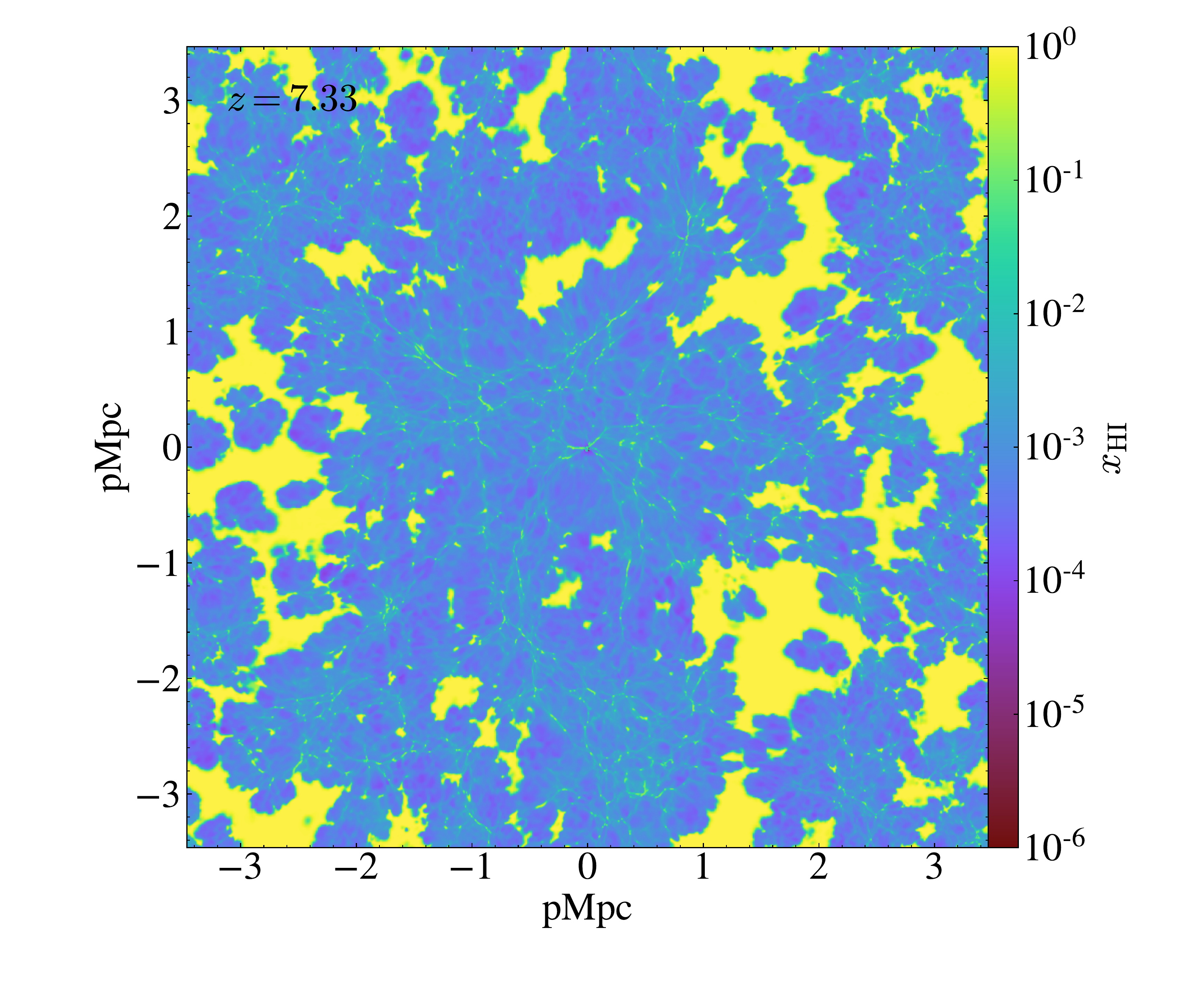}
    \includegraphics[width=0.33\textwidth]{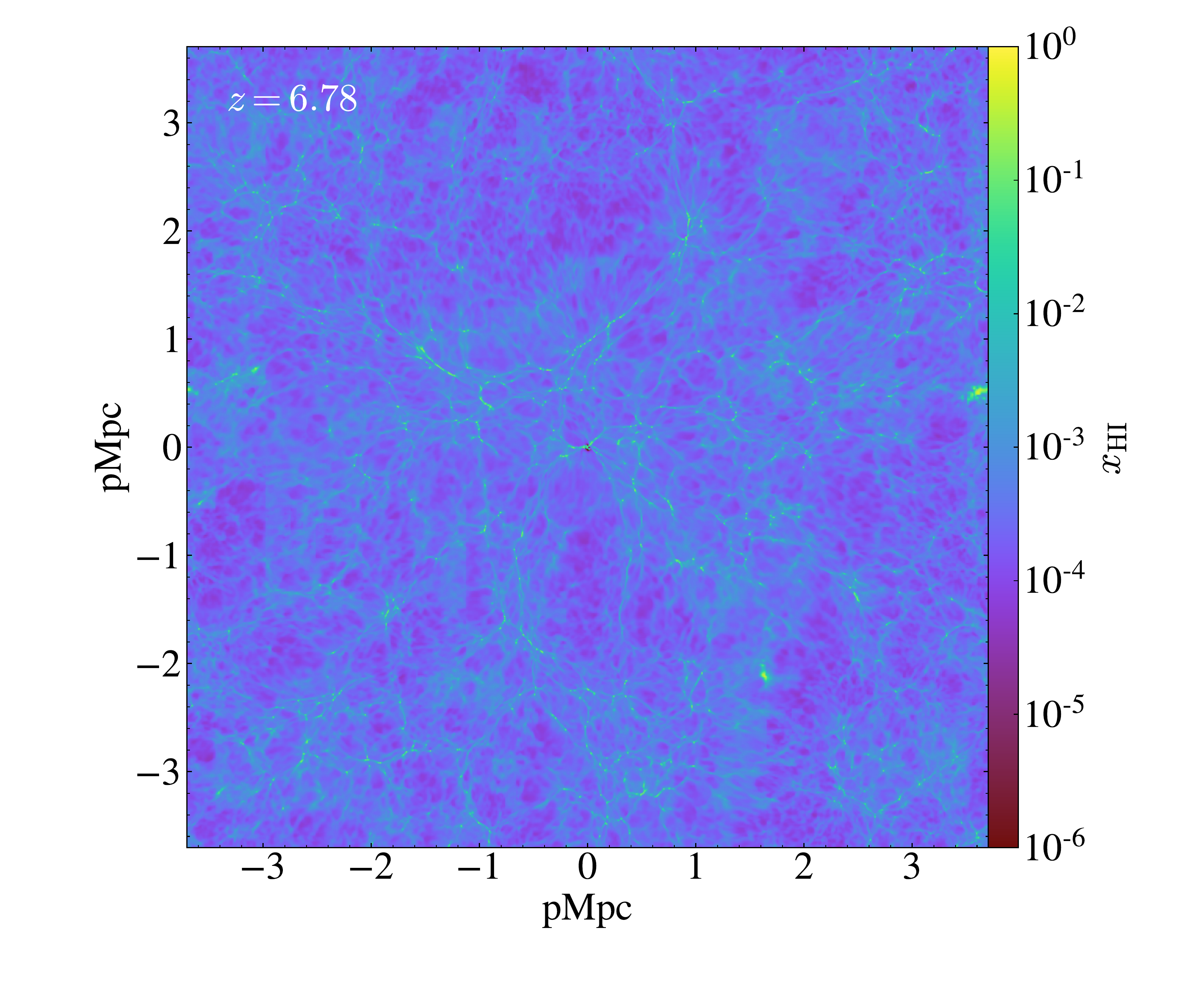}\hspace{-1.147cm}
    \includegraphics[width=0.33\textwidth]{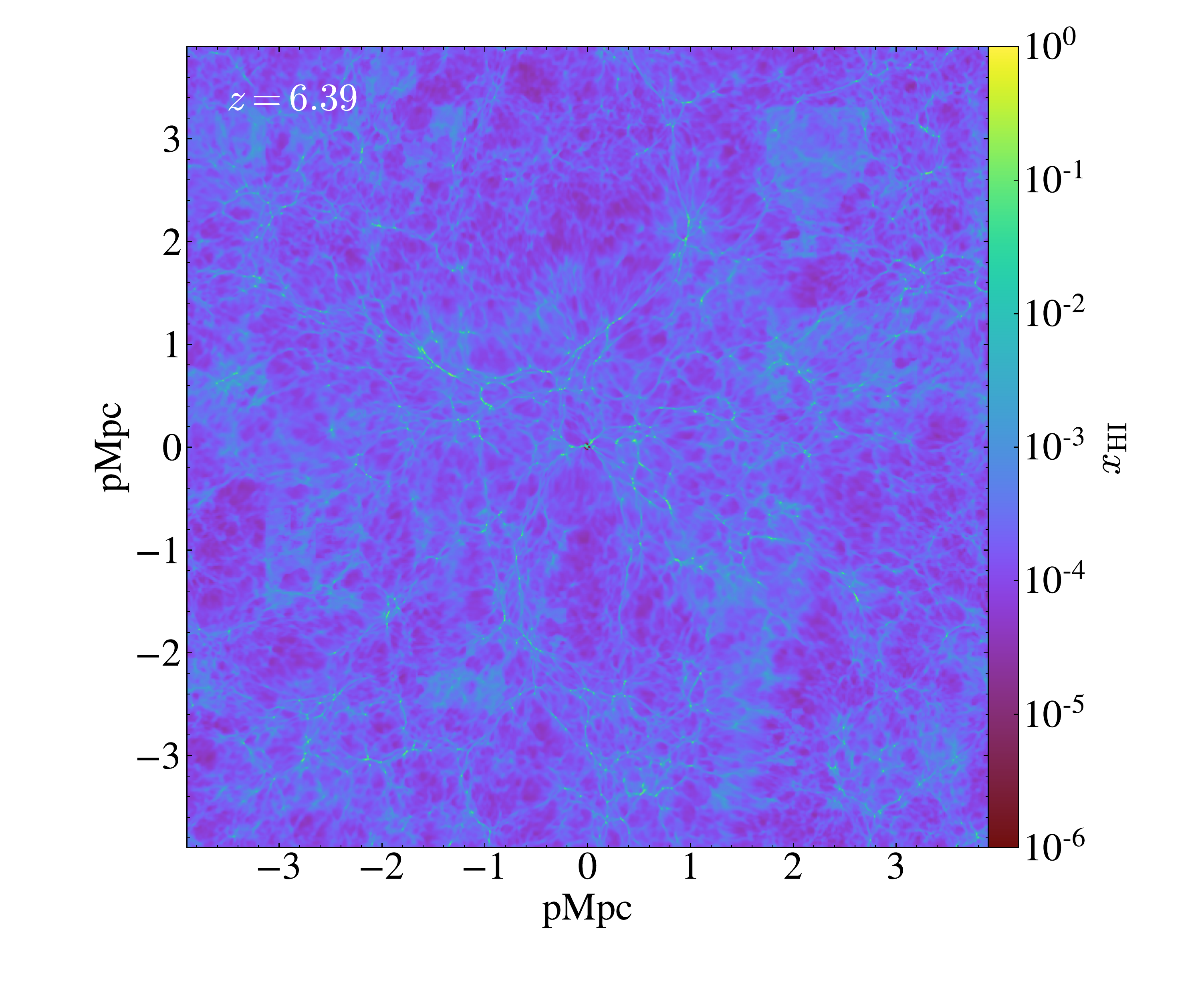}\hspace{-1.147cm}
    \includegraphics[width=0.33\textwidth]{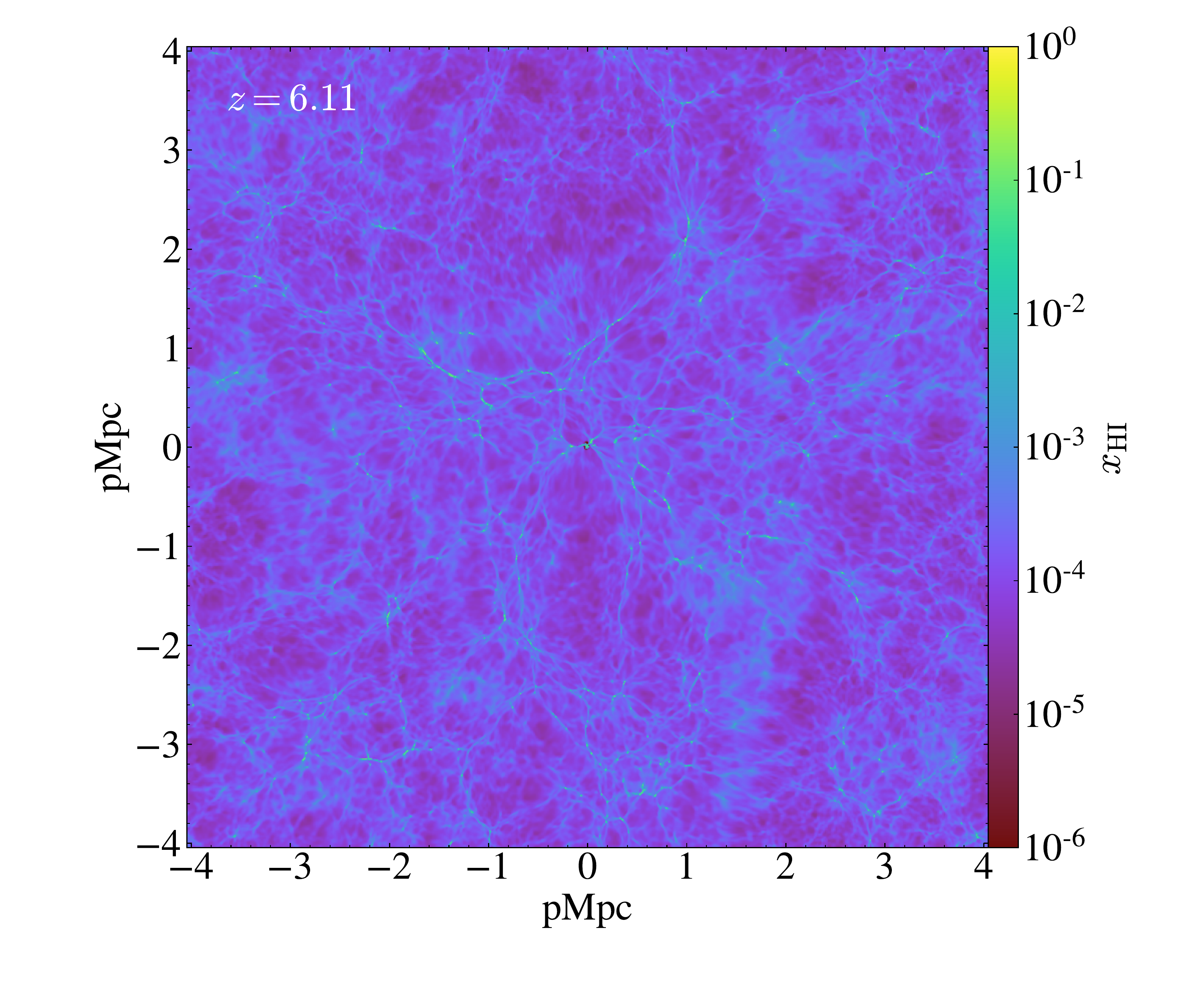}
    \caption{Neutral hydrogen fraction at six different redshifts from one of the CROC simulations used as the initial condition. { All the figures have a projection depth of $10$ pkpc and slice through the same massive halo in the center.} All six panels share the same colormap.}
    \label{fig:snapshots}
\end{figure*}

In spectra of high redshift quasars proximity zones appear as regions blueward of the quasar Ly$\alpha$ line with transmitted flux in excess of the average transmission at this redshift. Proximity zones have been studied both in observation and theory in the past two decades. However, with limited spectral resolution, most studies have been focused only on measuring or modeling the ``sizes'' of proximity zones. 
In observational work, the size of the quasar proximity zone is often defined following the pioneering work of \citep{fan2006b} as the distance from the quasar to the first point along the line of sight where the transmitted flux drops below 10\% in a spectrum smoothed by a $20$ {\AA} boxcar. \footnote{Another definition of the last transmitted spike is also discussed by \citet{lidz2007}.} In a mostly neutral universe, the thus defined proximity zone in the beginning is limited by the position of the quasar ionization front (I-front), which further depends on the quasar age and the neutral fraction of the ambient IGM \citep{cen2000, haiman2001,madau2000}. Therefore, these two important physical quantities can be constrained by measuring the proximity zone size distribution and evolution. 
For example, \citet{fan2006b} and \citet{carilli2010} measured the sizes of quasar proximity zones at $z=5.7 \sim 6.4$ and found the rapid growth of proximity zone sizes during that redshift interval. They argued that this reflected the rapid progress of cosmic reionization.
On the other hand, using a slightly larger sample of 34 quasars at $z=5.77\sim6.54$, \citet{eilers2017} found the slower growth of proximity zone sizes during the same redshift range, with most proximity zone sizes being $\sim 5$ physical Mpc (pMpc) with intrinsic scatter of $\sim 2$ pMpc after re-scaling all proximity zone sizes to a fixed quasar magnitude of $M_{1450}=-27$. In addition, they found several quasars with exceptionally small proximity zones. Such exceptionally small sizes of proximity zones, they argued, are most likely because the quasars are extremely young, with quasar ages $t_Q<1\times 10^5$ yr.

The apparent inconsistency among different studies can come from many factors. For example, to measure the transmitted flux, one needs to determine the unabsorbed, intrinsic quasar spectrum (i.e.\ the quasar ``continuum''). Different models give slightly different continua, which introduces significant uncertainty in the proximity zone size measurement. Also, transmitted spikes and noise make it hard to determine the point $z_{\rm GP}$ where transmitted flux drops below 10\%. Quasar host redshift $z_Q$ may also have large uncertainties due to the gas motion around the SMBH \citep{fan2006b}. All these uncertainties can significantly bias the measured proximity zone sizes. 
Furthermore, the number of quasars at $z>6$ with high quality data is still small. They have different luminosities,  and may have large intrinsic scatter from sightline to sightline, weakening further constraints on quantities like quasar age and redshift evolution of neutral hydrogen fraction in the IGM. The good news is that the number of discovered $z>6$ quasars is increasing rapidly, { and more and more high resolution spectra are being obtained to study the proximity zones \citep{eilers2020,ishimoto2020}. Moreover,} with 30-meter class optical telescopes coming online in the next decade, it is expected that a large increase in high quality spectra will greatly improve the observational statistics.

In order to better interpret the upcoming data, we need to correspondingly improve our theoretical models. In a simple idealized scenario where the quasar I-front expands into the uniform IGM with high neutral fraction $\xHI$, the I-front position $R_{ion}\propto ({\dot{N} t_Q}/{x_{\rm HI}})^{\frac{1}{3}}$. However, in the real Universe, cosmic structures are complex and reionization process is patchy, and the sizes of quasar proximity zones show significant scatter \citep{lidz2007}. Also, $R_{\rm ion}$ is not always directly measurable. This is because inside $R_{\rm ion}$, the radiation intensity from quasar usually drops as $1/r^2$. For large enough $R_{\rm ion}$ at sufficiently high redshift, the quasar is unable to ionize all hydrogen inside $R_{\rm ion}$ enough to keep the transmitted flux above the observable threshold.
Therefore, the observed proximity zone size will eventually reach a maximum value. Assuming that inside the proximity zone the IGM is always optically thin, \citet{bolton2007a} showed that the maximum size of proximity zone is then $R_{\rm obs}^{\rm max} \propto \dot{N}^{\frac{1}{2}}$, which is independent of the quasar lifetime and the neutral fraction of the general IGM. They also showed that for mostly ionized IGM, the proximity zone sizes reach $R_{\rm obs}^{\rm max}$ and do not change significantly after $\sim 1$ Myr. This was confirmed in several subsequent works \citep{keating2015,eilers2017}.

However, most previous theoretical studies (with a few exceptions like \citet{keating2015} and \citep{kakiichi2018}) suffered from limited spatial resolution of their simulations, with numerical resolution being insufficient to resolve small cosmic structures like the Lyman limit systems (LLSs). The LLSs may significantly lower the transmitted flux from the quasar, thus biasing predictions for the proximity zone size distribution in simulations that do not resolve them. 
Also, some studies assume a uniform UV background, which is not realistic, especially when reionization is not complete. If the quasar sightline hits the neutral patch in the IGM, proximity zone can terminate suddenly. Therefore, modeling the ionized bubble caused by galaxies is important when studying proximity zones during reionization. 

In this study, we use simulations from the Cosmic Reionization On Computers (CROC) project \citep{ng14} as the background model for cosmic reionization. CROC is a suite of radiative transfer cosmological hydrodynamic simulations with different comoving box sizes of $20 {\rm~ cMpc}/h$, $40  {\rm~ cMpc}/h$ and $80 {\rm~ cMpc}/h$ and the peak spatial resolution of $100$ pc in proper units. Therefore, it can model both the global reionization process and internal properties of galaxies. We draw lines of sight from CROC simulation snapshots and post-process them with a new 1D RT code using adaptive time steps. We study quasar proximity zones during the entire reionization period, from the beginning of reionization, through ionized bubble growing and overlapping, all the way to the end of reionization at $z\sim6$. We use a statistically significant sample to study the distribution of proximity zone sizes and their evolution, and discuss the observational applications.

This paper is organized as follows. In Section 2, we describe how we model the quasar absorption spectra. We describe CROC simulation we use as the initial condition, as well as the new 1D RT code for post-processing the lines of sight. In Section 3, we show our results for proximity zone sizes, including individual examples at different stages of reionization and their distribution and evolution with quasar age and redshift. In Section 4, we focus on the extremely small proximity zones and compare our results with previous studies. In Section 5, we discuss some observational applications and note some caveats. A summary is provided in Section 6.

\begin{figure*}
    \centering
    \includegraphics[width=\textwidth]{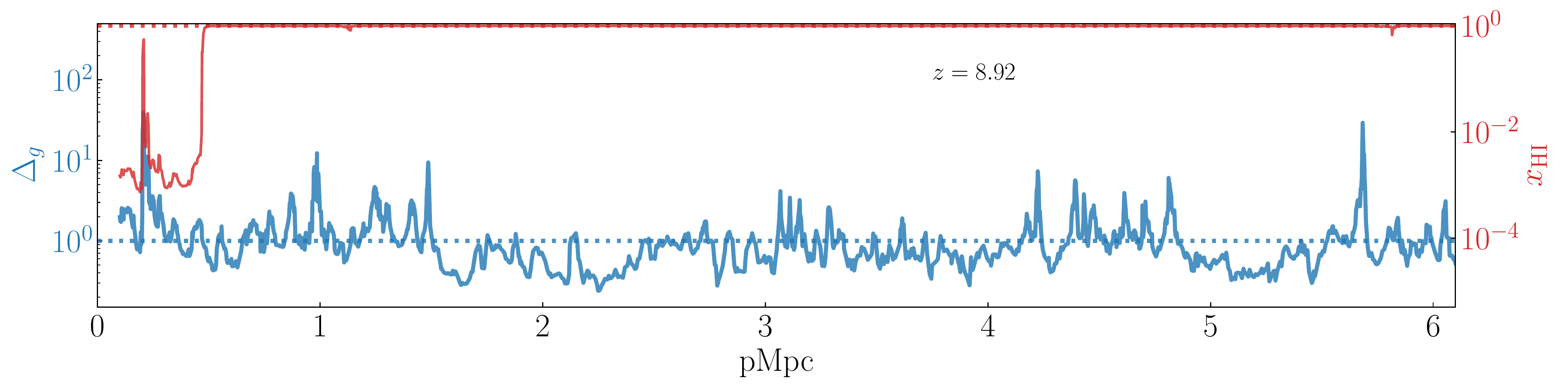}\vspace{-1cm}
    \includegraphics[width=\textwidth]{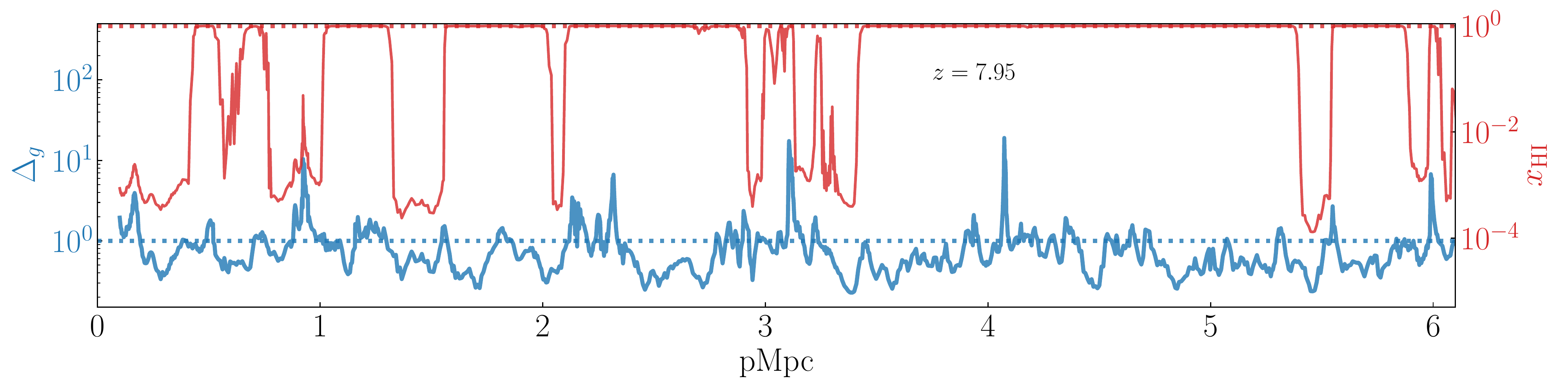}\vspace{-1cm}
    \includegraphics[width=\textwidth]{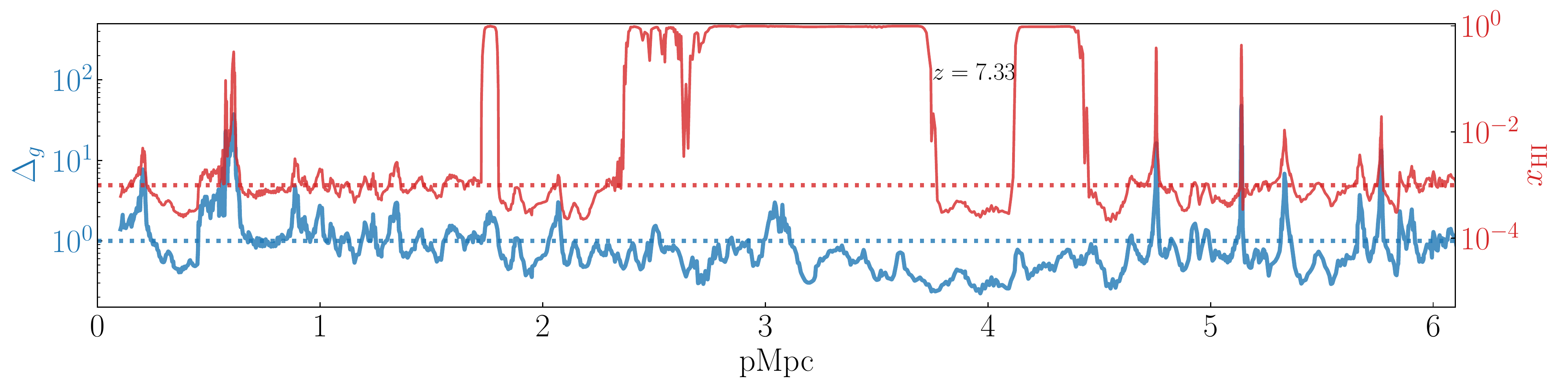}\vspace{-1cm}
    \includegraphics[width=\textwidth]{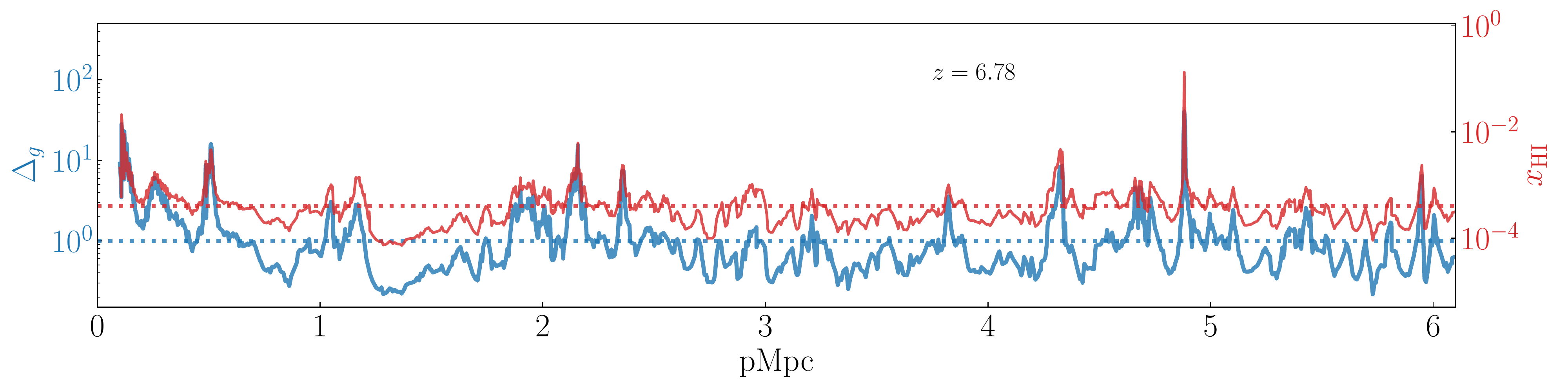}\vspace{-1cm}
    \includegraphics[width=\textwidth]{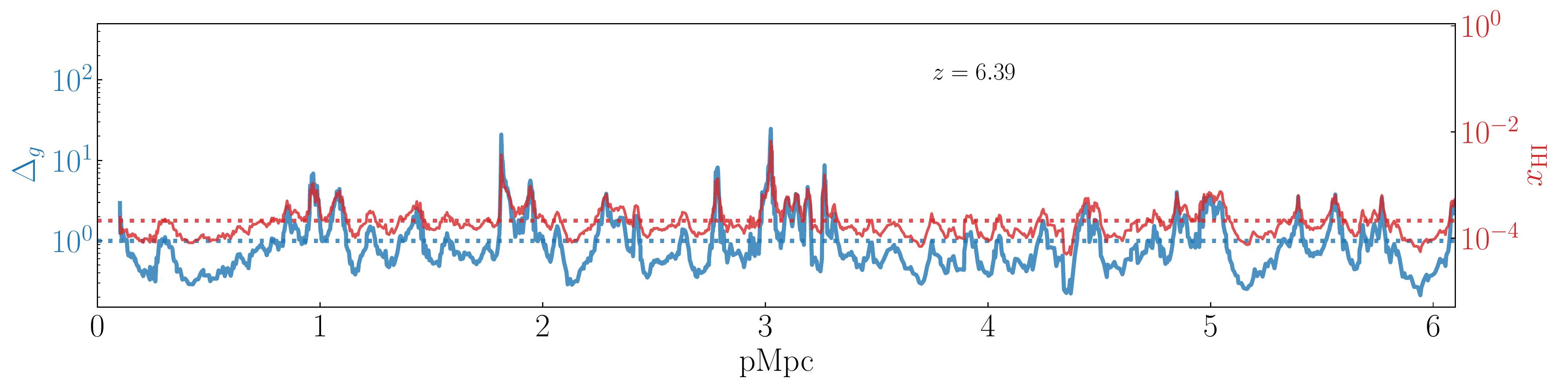}\vspace{-1cm}
    \includegraphics[width=\textwidth]{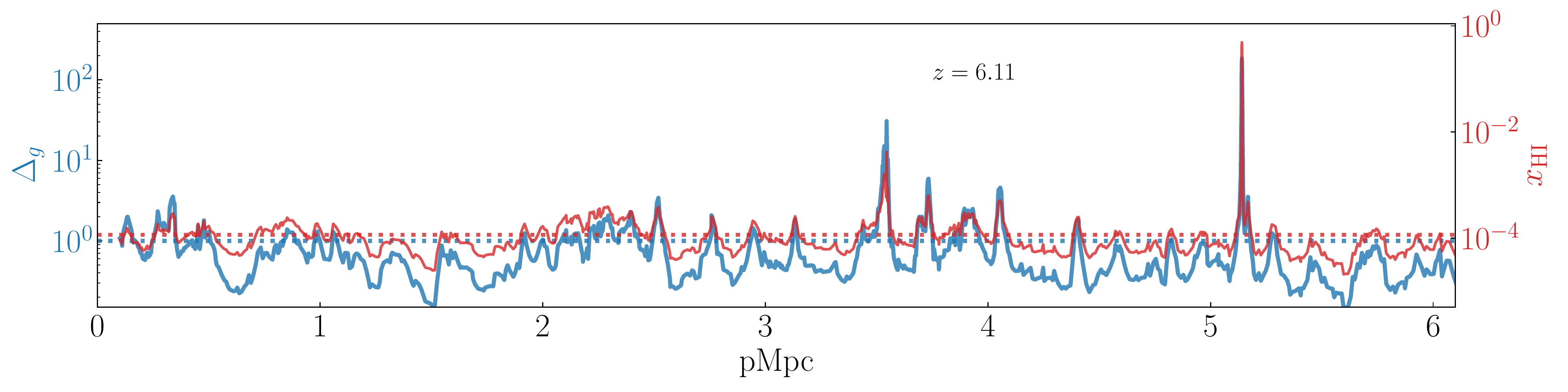}
    \caption{Initial condition for some sightlines at six different redshifts. Blue lines with left $y$-axes show the gas density contrast $\Delta_g\equiv \rho/\bar\rho$ and red lines with right $y$-axes show hydrogen neutral fraction $\xHI$. Dotted lines of both colors show the corresponding mean values for both quantities at each redshift. Before $z=7.3$ when large ionized bubbles are overlapping, regions around high density peaks are ionized to $\xHI\sim 10^{-3}$, while other regions remain neutral. The density peaks with $\Delta_g >10$ are usually partially ionized ($\xHI>10\%$) because of high density and, hence, more shielding and recombination. After the ionized bubble overlap, $\xHI$ traces gas density.}
    \label{fig:init_los}
\end{figure*}

\section{Methodology}

We study the sizes of the proximity zones using synthetic absorption spectra. In this section we describe our procedure for post-processing lines of sight drawn from a Cosmic Reionization on Computer (CROC) simulation.

\subsection{Initial Condition}

\begin{table}[]
    \centering
    \begin{tabular}{|c|c|c|c|c|c|c|}
    \hline
 z      &8.92 &7.95 &7.33 &6.78 &6.39 &6.11 \\
  \hline
$\langle\xHI\rangle_V$ &0.95 & 0.60& 0.13& $6.7\times10^{-4}$& $3.3\times10^{-4}$& $2.1\times10^{-4}$\\
\hline
$\langle\xHI\rangle_M$   & 0.93&0.56 &0.15 &0.037 &0.038 & 0.039\\
\hline
    \end{tabular}
    \caption{{ Volume and mass weighted neutral fraction of the simulation.}}
    \label{tab:B40ExHI}
\end{table}

\subsubsection{Simulation}

We use snapshots from one of the CROC simulations in the $40\dim{Mpc}/h$ box as the initial condition. { The CROC project uses the Adaptive Refinement Tree (ART) code  \citep{kravtsov1999, kravtsov2002, rudd2008} to reach high spatial resolution using adaptive mesh refinement approach. The base grid is $40\  h^{-1}~\rm~cMpc$ in size, and the peak resolution is $\sim 100$ pc (in physical units). CROC simulations include relevant physics such as gas cooling, heating, star formation and stellar feedback. After each star particle is formed, it becomes an individual radiation source. Quasar radiation, on the other hand, is only treated as the background. The radiative transfer is done using the Optically Thin Variable Eddington Tensor (OTVET) method \citep{gnedin2001}, which is fully coupled temporally (i.e.\ being updated with the same time-step) and spatially (modeled at the same spatial resolution) to gas dynamics and other simulated physics. For more details on the CROC project we refer the readers to \citet{gnedin2014}}.

We choose snapshots at six different redshifts $z=8.9$, $8.0$, $7.3$, $6.8$, $6.4$, $6.1$. In Figure \ref{fig:snapshots} we show the simulated neutral hydrogen map at these six redshifts, slicing through a massive halo at the center of each panel. In this realization the volume-weighted hydrogen ionized fraction reaches $0.1$ at $z\approx8.5$ and $0.9$ at $z\approx7.2$ {(a table of neutral fraction of this simulation can be found in Table \ref{tab:B40ExHI})}. However, as we can see from Figure \ref{fig:snapshots}, this process is highly inhomogeneous. Early at $z=9$, only regions around the most massive halos are ionized, due to the collective ionizing photons from both the most massive galaxies and less massive galaxies around them. The ionized bubbles grow quickly and leave only small patches neutral at $z\approx 7.3$. After $z\approx 7$, almost all the IGM is ionized in this simulation box. 
Note that voids reionize later, but reach a lower neutral fraction after reionization is complete, because of the lower recombination rate in lower density regions.

\subsubsection{Lines of Sight}

We have run ROCKSTAR halo finder \citep{Behroozi2013} to identify halos in the simulation. We choose halos with dark matter mass  $M_h>1.5 \times 10^{11}\Msun$ at each redshift as potential quasar host halos (except $z=8.9$ in which we choose $M_h>8\times 10^{10}\Msun$ because massive halos are extremely rare at that high redshift). The numbers of halos at each redshift are listed in Table \ref{tab:halos}.  Then we use the LightRay.make\_light\_ray function in the analysis and visualization package yt\footnote{https://yt-project.org/}\citep{turk2011}  to draw lines of sight, centered on these halos and with random distribution in all directions. We achieve this by generating three random number $x$, $y$, and $z$ from Gaussian distribution with the mean value $0$ and the standard deviation $1$, and the direction $(x, y, z)$ can be proven to be uniformly distributed over the sphere. For each halo, we draw hundreds of sightlines (see Table \ref{tab:halos}), so that at each redshifts we have thousands of lines of sight in total. Each line of sight is $15$ pMpc long, larger than any observed quasar proximity zones currently reported. Because CROC simulations use adaptive mesh refinement, the sampled resolution elements are smaller when the gas density is higher. In a typical sightline, most cell sizes are between 1 and 10 pkpc, and about 10\% of cells have sizes below 1 pkpc.

In Figure \ref{fig:init_los} we show six examples of sightlines drawn from snapshots at $z=8.9-6.1$ from top to bottom. The blue line shows the gas density contrast $\Delta_g\equiv \rho/\bar\rho$, where $\bar\rho$ is the mean gas density of the universe at each redshift, and the red line shows the neutral fraction of the gas. Again, as we can see at redshifts before $z=7.5$ (the first two panels), reionization starts around high density peaks. The high density gas hardly fall below $\xHI<0.01$ because of its short recombination time, but diffuse gas around it easily reaches $\xHI<10^{-3}$. Note that in the second and third panels, there are ionized regions which are not adjacent to apparent high density peaks. This is  because the corresponding high density peaks do not lie exactly along the sightline but close to it. After $z=7$ (the last three panels), the ionized bubble overlap and the whole IGM becomes ionized. In this case the neutral fraction $\xHI$ correlates with the gas density.

One important point we have learned from the simulation is that at any given time, the bulk of the IGM is usually either very neutral $\xHI\approx 1$ or very ionized $\xHI \lesssim 10^{-3}$. In other words, the ionization process is patchy, and every patch of the ionized gas has $\xHI \lesssim 10^{-3}$. There are no such times that the bulk of the IGM is uniformly ionized to a modest degree of, say, $\xHI \sim 0.1$.

\begin{table}
\centering
\caption{Quasar host halo parameters at different redshifts
}
\label{tab:halos}
\begin{tabular}{|c|c|c|c|c|}\hline
redshift &   $M_{h} [\Msun]$ & \# halos & \# l.o.s. each & total \# l.o.s. \\ \hline
  8.9      &  $> 8.0 \times10^{10}$ & 5  & 210  & 1150 \\ \hline
  8.0      &  $> 1.5 \times10^{11}$ & 6  & 210  & 1260 \\ \hline
  7.3    &    $> 1.5 \times10^{11}$ & 13 & 110  & 1430 \\ \hline
  6.8    &    $> 1.5 \times10^{11}$ & 23  & 110  & 2530 \\ \hline
  6.4    &    $> 1.5 \times10^{11}$ & 40   & 110  & 4400 \\ \hline
  6.1    &    $> 1.5 \times10^{11}$ & 63  & 110  & 6930 \\ \hline
\end{tabular}

\end{table}

\subsection{1D Radiative Transfer Code}

To model the quasar proximity zone spectra, we post-process a statistically significant sample of sightlines with 1D RT code. We discuss the pros and cons of the post-processing approach in \S \ref{sec:pp}.

\begin{figure*}
    \centering
    \includegraphics[width=1\textwidth]{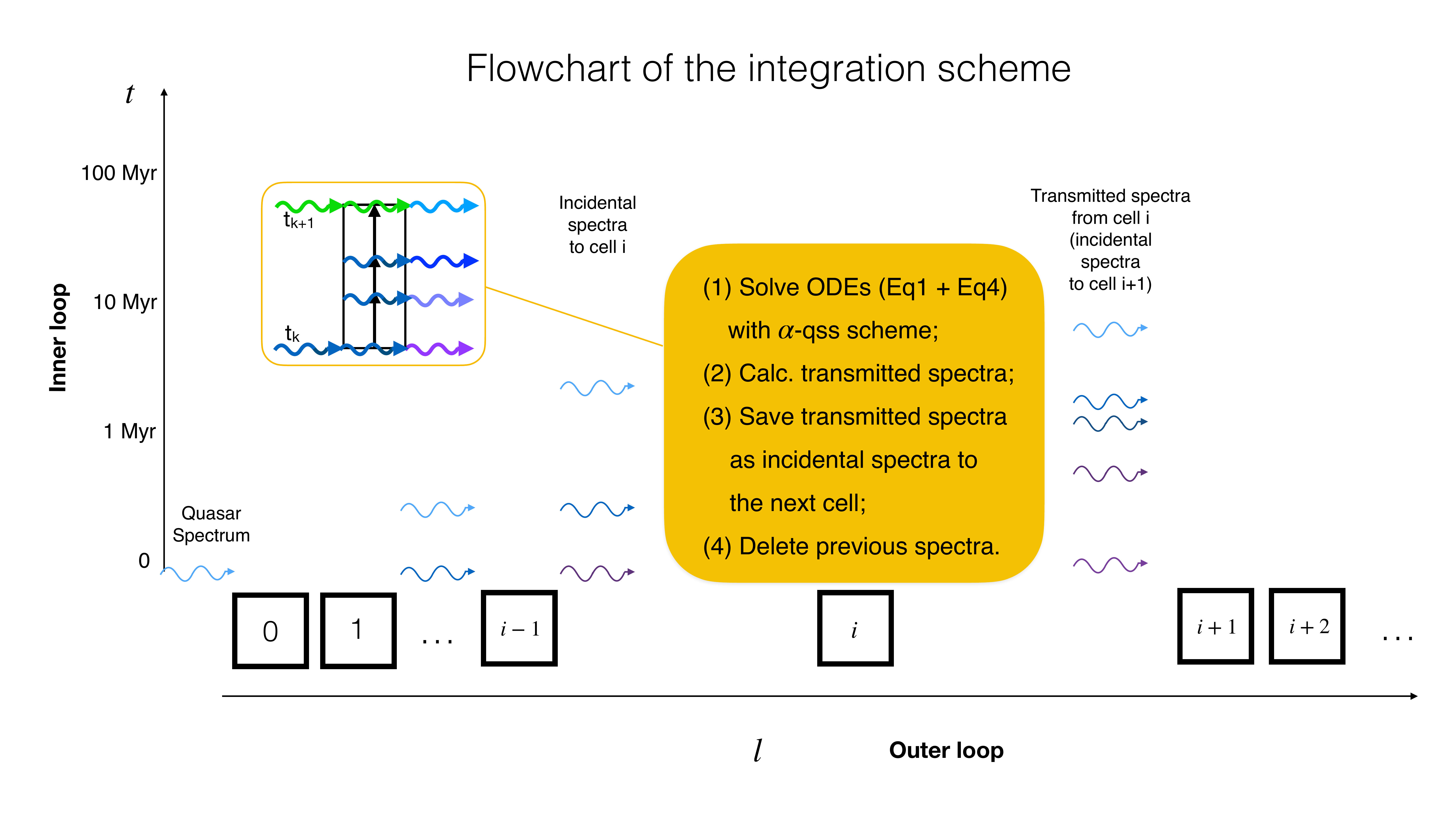}
    \caption{The algorithm of the 1D RT post-processing code. It calculates the cells one by one, from the closest to the farthest from the quasar. For each cell the code solves the ODEs and calculates the transmitted spectra at all times and then passes the spectra to the next cell. The spectra are represented by wavy lines with arrows. Since the cell are usually more optically thick at the beginning, the transmitted spectra are harder, which is represented by lines having more violet colors. { The subfigure framed in yellow illustrates how one step (from $t_{\rm k}$ to $t_{\rm k+1}$) is calculated. The incidental spectrum at $t_{\rm k}$ is represented by the blue wavy line and the one at $t_{\rm k+1}$ is represented by the green wavy line. The incidental spectrum at $t_{\rm k}$ is used to calculate the photo-ionization/heating rates for the ODEs. The $\alpha$-QSS scheme is used to solve the ODEs and to output the ionization fractions and temperature at $t_{\rm k+1}$ as well as at several intermediate times if the time-steps chosen by the $\alpha$-QSS scheme require that (in this plot we show two, represented by the black arrowheads in the center). Finally we calculate the transmitted spectra with updated ionization fractions and temperature at all substeps. Note that the transmitted spectrum at $t_{\rm k+1}$ is calculated using the incidental spectrum at $t_{\rm k+1}$.}}
    \label{fig:integration}
\end{figure*}

Following \citet{bolton2007a} and \citet{davies2016}, we solve the ordinary differential equation system for each cell with width $dr$ at distance $r$ from the quasar. The equations for the three ionization fractions are:

\begin{equation} \label{eq1}
\begin{split}
\frac{d\xHI}{dt} & = - (\Gamma_{\rm QSO}^{\rm HI}+\Gamma_{\rm bkg}^{\rm HI}+ n_e \Gamma_e^{\rm HI}) \xHI + \alpha^{\rm HI} n_e \xHII, \\
\frac{d \xHeI}{dt} & = -(\Gamma_{\rm QSO}^{\rm HeI}+\Gamma_{\rm bkg}^{\rm HeI}+n_e \Gamma_e^{\rm HeI})\xHeI + \alpha^{\rm HeI} n_e \xHeII, \\
\frac{d \xHeII}{dt} &= -(\Gamma_{\rm QSO}^{\rm HeII}+\Gamma_{\rm bkg}^{\rm HeII}+n_e \Gamma_e^{\rm HeII})\xHeII + \alpha^{\rm HeII}  n_e \xHeIII\\
& \ \ 
\ + (\Gamma_{\rm QSO}^{\rm HeI}+\Gamma_{\rm bkg}^{\rm HeI}+ n_e \Gamma_e^{\rm HeI}) \xHeI - \alpha^{\rm HeI} n_e \xHeII. 
\end{split}
\end{equation}

The quantity $\Gamma_e^{i}$ is the collisional ionization rate and $\alpha^{i}$ is the recombination rate of species $i$ (HI, HeI, or HeII), which are both functions of gas temperature $T$. The term $\Gamma_{\rm QSO}^{i} $ is the photo-ionization rate from quasar, which can be expressed as
$$
\Gamma_{\rm QSO}^{i}=\frac{1}{n_iV_{\rm shell}} \int^{\infty}_{\nu_{i}}\dot{N}_\nu^{\rm abs} P_i d\nu,
$$
where $V_{\rm shell}$ is the volume of the spherical shell with width $dr$ and radius $r$,  $\nu_i$ is the ionization threshold for species $i$, and $\dot{N}_\nu^{\rm abs}$ is the incidental photon production rate at distance $r$ from the quasar, which is equal to the intrinsic photon production rate attenuated by the absorption between the cell and the quasar:
$$ 
\dot{N}_\nu^{\rm abs} = \dot{N}_\nu e^{-\tau_\nu}.
$$
$P_i$ is the probability that an ionizing photon is absorbed by species $i$:
\begin{equation}
\begin{split}
P_{\rm HI}=p_{\rm HI}q_{\rm HeI}q_{\rm HeII} (1-e^{-\tau_\nu^{\rm tot}})/D \\
P_{\rm HeI}=q_{\rm HI}p_{\rm HeI}q_{\rm HeII}(1-e^{-\tau_\nu^{\rm tot}})/D \\
P_{\rm HeII}=q_{\rm HI}q_{\rm HeI}p_{\rm HeII}(1-e^{-\tau_\nu^{\rm tot}})/D \\
\end{split}
\end{equation}
where $p^i=1-e^{-\tau_\nu^i}$, $q_i=e^{-\tau_\nu^i}$, $\tau_\nu^{\rm tot}$ is the total optical depth of the cell and $D=p_{\rm HI}q_{\rm HeI}q_{\rm HeII}+q_{\rm HI}p_{\rm HeI}q_{\rm HeII}+q_{\rm HI}q_{\rm HeI}p_{\rm HeII}$ \citep{bolton2007a}.

The term $\Gamma_{\rm bkg}^{i}$ is the background photo-ionization rate of species $i$ (HI, HeI, or HeII) of the cell. The full radiation field has not been stored in the CROC output files due to the limited disk space available. Therefore we calculate $\Gamma_{\rm bkg}^{i}$ assuming the gas is in ionization equilibrium before the quasar turns on:

\begin{equation} \label{eq3}
\begin{split}
\Gamma^{\rm HI}_{\rm bkg} & = \frac{\alpha^{\rm HI}(T_0) n_{e,0} {\xHII}_{,0}  }{{\xHI}_{,0}} - n_{e, 0} \Gamma_e^{\rm HI}(T_0)\\
\Gamma^{\rm HeI}_{\rm bkg} & = \frac{\alpha^{\rm HeI}(T_0) n_{e,0} {\xHeII}_{,0}  }{{\xHeI}_{,0}}  - n_{e, 0} \Gamma_e^{\rm HeI}(T_0)\\
\Gamma^{\rm HeII}_{\rm bkg} & = \frac{\alpha^{\rm HeII}(T_0) n_{e,0} {\xHeIII}_{,0}}{{\xHeII}_{,0}} - n_{e, 0} \Gamma_e^{\rm HeII}(T_0)\\
\end{split}
\end{equation}

Along with ionization fraction equations we also solve for the temperature evolution:
\begin{equation}
\frac{dT}{dt}=\frac{2}{3k_Bn_{\rm tot}}(\mathcal{H}-\Lambda)-2HT-\frac{T}{n_{\rm tot}}\frac{dn_{\rm tot}}{dt}
\end{equation}

where $\mathcal{H}$ is the photo-heating rate from the quasar, $\Lambda$ is the cooling rate, $H$ is the Hubble parameter, and  $n_{\rm tot}$ is the total density of particles $n_{\rm tot}=n_{\rm H}+n_{\rm He}+n_e$. The cooling rate here includes recombination cooling, collisional ionization cooling, collisional excitation cooling, Bremstrahlung cooling, and inverse Compton cooling. { We adopt the same rates as those in \citet{bolton2007a}: photo-ionization cross-sections from \citet{osterbrock1989}, recombination rates from \citet{abel1997}, collisional ionization rates from \citet{theuns1998}, collisional excitation and bremsstrahlung rates from \citet{cen1992}, inverse Compton cooling rates from \citet{peebles1971}. Note that in the current version of the code, secondary ionizations \citep{shull1985, furlanetto2010} are not included. Secondary ionizations can reduce the gas temperature and increase the ionization rate, which affects the transmitted flux \citep{davies2016}. However, their effect on the size of the proximity zone is expected to be very small, and hence not change the main results of this paper.}

Our code differs from previous codes used in \citet{bolton2007a} and \citet{davies2016} mainly in how we advance the radiation field in time and space. Figure \ref{fig:integration} shows the flowchart of our integration scheme. Previous codes choose many global time steps and evolved all the cells for each global time step. Our code solves the evolution of each cell for the entire time of interest ($\sim 30 \rm Myr$) using an adaptive prediction-correction scheme.  At each adaptive output time, we calculate the transmitted spectra to pass to the next cell. This algorithm is motivated by the very different temporal behavior of cells very close to the quasar and very far away from the quasar, for which the timescales to reach ionization equilibrium can differ by several orders of magnitude. With such large range of physical time scales in different spatial locations using a fixed global time step is inefficient. To solve the ODE for each cell, we use the $\alpha-$QSS scheme \citep{mott01}, which is designed to solve stiff ordinary differential equations of the form
$$ \frac{d y_i}{dt}=q_i -p_i y_i,$$
where $q_i$ and $p_i$ are functions of time. We refer readers to the original paper for details. 

{ 
The sub-figure framed in yellow shows how one step (from $t_{\rm k}$ to $t_{\rm k+1}$) is calculated in detail. The incidental spectrum at $t_{\rm k}$ is used to calculate the photo-ionization/heating rates for the ODEs. The $\alpha$-QSS scheme is used to solve the ODEs and to output the ionization fractions and temperature at $t_{\rm k+1}$ as well as at several intermediate times if the time-steps chosen by the $\alpha$-QSS scheme require that (in this plot we show two, represented by the black arrowheads in the center). Finally, we calculate the transmitted spectrum with th updated ionization fractions and temperature at all sub-steps. Note that the transmitted spectrum at $t_{\rm k+1}$ is calculated using the incidental spectrum at $t_{\rm k+1}$ rather than that at $t_{\rm k}$. This is crucial to ensure the correct I-front speed.}

We set the integration tolerance to be $1\%$ for all four variables. After solving the ODEs for the cell for all time-steps, we calculate the transmitted spectra for that cell and store them to be used as incident spectra for the next cell. Storing the full temporal evolution of the incident spectra on each spatial cell is memory demanding. For some time-steps, the transmitted spectra can be extremely similar - for example, after the gas has reached a new ionization equilibrium, the change in the ionization fraction is mainly due to gas cooling. Therefore, we do not store the spectra that differ by less than $1\times 10^{-5}$ from the previous time-step. Note that this is a very conservative choice and introduces a negligible error. By trimming the number of transmitted spectra, we save memory and reduce the number of time-steps for the following cell. We show some tests of our code in the Appendix.

{ In this paper we use a simple power-law quasar spectrum with the spectral index $\alpha=-1.5$:   $L_\nu\propto \nu^{-1.5}$. The spectra are evenly divided into $80$ bins on the log scale, with the lowest energy of $13.6$ eV and highest energy of $1$ keV. This choice mimics the frequency sampling in the RT solver of the ART code, which was optimized after extensive testing.} The luminosity for the quasar is also fixed, with the production rate of ionizing photon being:
$$
\dot{N}_{\rm tot}=\int^{\infty}_{\rm 13.6 eV} \dot{N}_{\nu}d\nu=1\times10^{57} s^{-1}.
$$
This translates into the quasar magnitude of $M_{\rm 1450}=-26.66$  { , assuming the same spectral index $\alpha=-1.5$ from $1450$\AA\ to $912$\AA\ with no break. With the double power-law spectrum model of \citet{Lusso2015} this translates into the quasar UV absolute magnitude of -26.2.}

After post-processing, we
generate synthetic Ly$\alpha$ absorption spectra from these lines of sight with the analytical Voigt profile formula in \citep{tepper-garcia2006}. We account for the peculiar velocity of the gas, as well as the quasar host halo.

\section{Results}

\subsection{Example Lines of Sight}

\begin{figure*}
    \centering
    \includegraphics[width=0.9\textwidth]{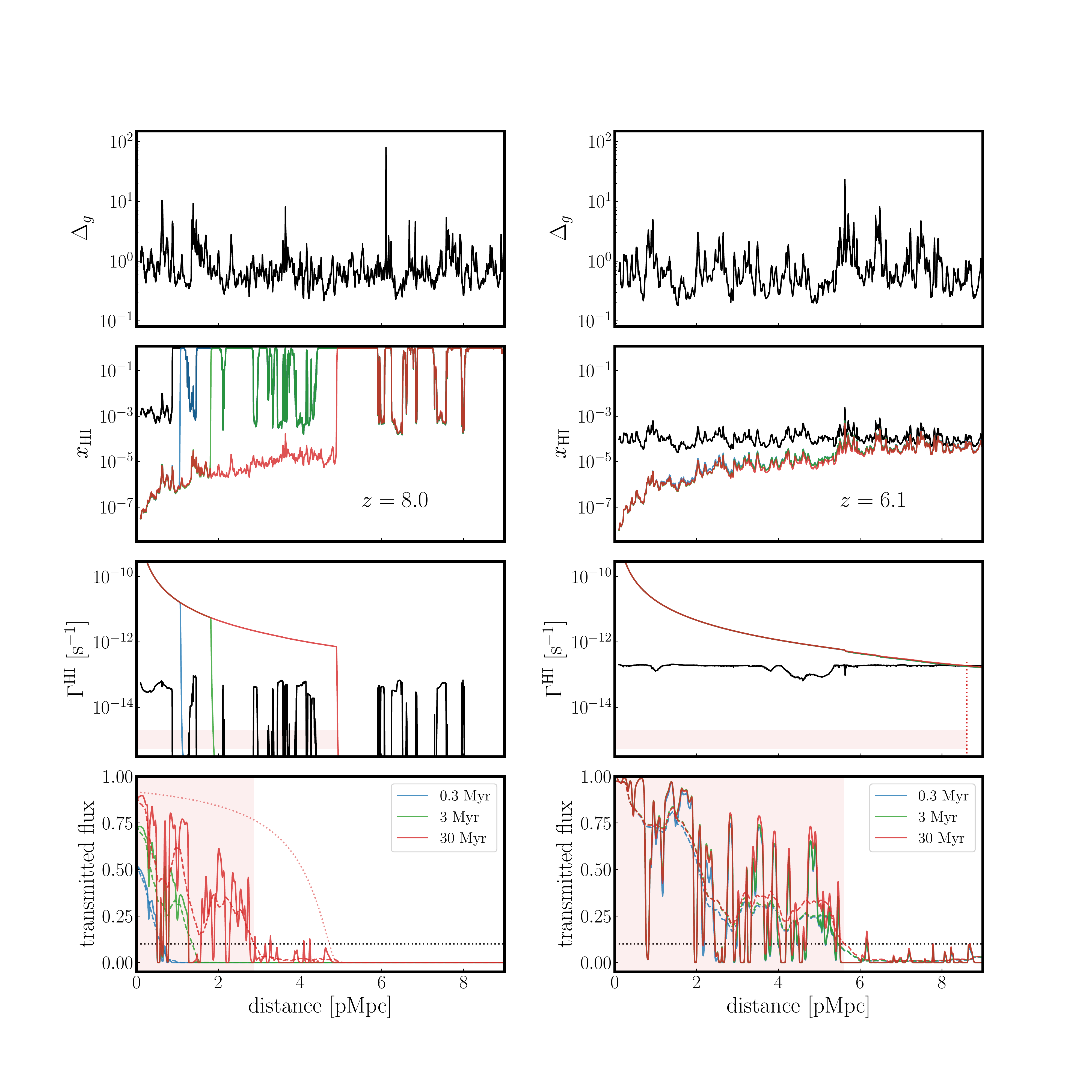}
    \caption{ Sightlines drawn from the snapshot at $z=8.0$ (left) and $z=6.1$ (right), respectively. Plotted from top to bottom are the gas density contrast $\Delta_g (=\rho/\bar{\rho}$), the neutral hydrogen fraction $x_{\rm HI}$, the HI ionization rate, and the transmitted flux. Black lines represent the initial values at $t_Q=0$ (no quasar radiation), while colored lines represent the values at different quasar ages of $t_Q=0.3$ Myr (blue), $3$ Myr (green), and $30$ Myr (red). In the third row, the colored lines represent $\Gamma^{\rm HI}$ due to quasar only, and the black line represents the cosmic background. { The  dotted red line in the third row on the right marks the position where $\Gamma^{\rm HI}_{\rm QSO}=\Gamma^{\rm HI}_{\rm bkg}$ at $t_Q=30$ Myr.} The transparent red bands show the span of the physical proximity zone $R_{\rm phy}$ at $t_Q=30$ Myr. In the bottom row, the dashed colored lines are the spectra smoothed by 20 {\AA}, and the horizontal dotted line marks the $0.1$ threshold. Traditionally, the first point where the smoothed spectra drop below this threshold is considered the edge of the proximity zone. The transparent red zones show the size of the observational proximity zone $R_{\rm obs}$ at $t_Q=30$ Myr.  {  The faint dotted red curve in the bottom left panel shows the transmitted flux solely due to the absorption of the neutral patch outside of $R_{\rm phy}$ at $t_Q=30$ Myr. }}\label{fig:example_los}
\end{figure*}

In Figure \ref{fig:example_los} we show two typical sightlines at two different redshifts $z=8.0$ (left) and $z=6.1$ (right).
For the sightline at $z=8.0$ (left panels) when the universe is predominantly neutral, the background ionization rate of HI has significant spatial fluctuations, as shown by the black line in the third panel. Inside HII bubbles created by clustered galaxies, the value is $\Gamma^{\rm HI}_{\rm bkg}\sim 3\times10^{-14} \rm~s^{-1}$.
After the quasar turns on, the ionization front (I-front) gradually moves outwards (second row), and so does the region dominated by the quasar radiation (third row). The neutral gas outside the quasar I-front creates the damping wing in the spectrum and this explains the overall spectral shape at $t_Q=0.3$ Myr and $3$ Myr. After $\sim 30$ Myr, the I-front propagates further away {, and the damping wing due to the neutral patch is shown as the faint dotted red line.} { At this time, the absorption at $\sim 1-2$ pMpc is mostly due to the large scale overdense structure around the quasar.}

At $z=6.1$, the IGM in the simulation box is highly ionized. The background radiation $\Gamma^{\rm HI}_{\rm bkg}$  rises to above $10^{-13} s^{-1}$ and becomes rather uniform. As is shown by the black line in the right column, the neutral fraction of the IGM is only $\sim10^{-4}$ before the quasar turns on. Therefore, after the quasar turns on, there is no traditional ``I-front'' of the quasar, and the timescale for the IGM to reach a new ionization equilibrium is no longer limited by the speed of the quasar I-front as at $z=8.0$. Instead, the neutral fraction drops on a timescale of $t_{\rm eq}\sim 1/\Gamma^{\rm HI}_{\rm QSO}$. This value is extremely short ($\approx 2\times 10^3$ yr at $1$ pMpc and $\approx 3\times 10^4$ yr at $4$ pMpc), therefore the gas near the quasar re-establishes the ionization equilibrium quickly within $\sim 0.3$ Myr and there is not much difference between the blue, green and red lines in the right panel. The same explanation applies to the transmitted flux. Notice that there is a small increase in the transmitted flux at $\sim 2$ pMpc after $t_Q=3$ Myr and $\sim 3-5$ pMpc after $t_Q=30$ Myr due to the additional photo-heating by the moving HeII I-front (not shown in the figure).

\subsection{Definitions of the Proximity Zone Size}

Traditionally, the edge of the proximity zone has been defined as the point where the transmitted flux drops below 10\% after the transmitted flux has been smoothed by a $20${\AA} boxcar \citep{fan2006b}. We label this observationally-motivated definition as $R_{\rm obs}$. In Figure \ref{fig:example_los}, we show the $R_{\rm obs}$ at $t_Q=30$ Myr as the red-shaded region in the bottom row. This quantity, albeit straightforward to measure, is hard to interpret, as it depends on the arbitrary values of the threshold and spectral smoothing and does not correspond to any physical scale. The precise value of $R_{\rm obs}$ is set by the complicated interplay between the damping wing outside of the I-front and the gas density distribution inside the proximity zone.
Therefore, here we define another, physically motivated proximity zone size $R_{\rm phy}$ and in the next subsection we will study the distribution and the evolution of both $R_{\rm obs}$ and $R_{\rm phy}$.

We use the radiation field to define the physical size of the proximity zone $R_{\rm phy}$ - a region where the ionization rate of HI due to the quasar is larger than that of the background. In practice, we calculate the  $\Gamma^{\rm HI}_{\rm QSO}$ and $\Gamma^{\rm HI}_{\rm bkg}$ for each cell and define the physical proximity zone as the region adjacent to the quasar where
$$
    \Gamma^{\rm HI}_{\rm QSO}>{\rm max}(\Gamma^{\rm HI}_{\rm bkg}, 10^{-15} s^{-1}).
$$
The threshold of $10^{-15} s^{-1}$ is included to account for the situation when a sufficiently neutral patch (a super-LLS or a DLA) blocks the sightline, reducing both $\Gamma^{\rm HI}_{\rm QSO}$ and $\Gamma^{\rm HI}_{\rm bkg}$ to almost zero (see the third row in the left column of Figure \ref{fig:example_los}). Note that changing this threshold by several orders of magnitude does not impact $R_{\rm phy}$, because whenever the neutral patch terminates the proximity zone, the drop in $\Gamma^{\rm HI}_{\rm QSO}$ is very sharp. In Figure \ref{fig:example_los} we show $R_{\rm phy}$ as the light red band in the third row.

\begin{figure*}
    \centering
    \includegraphics[width=0.52\textwidth]{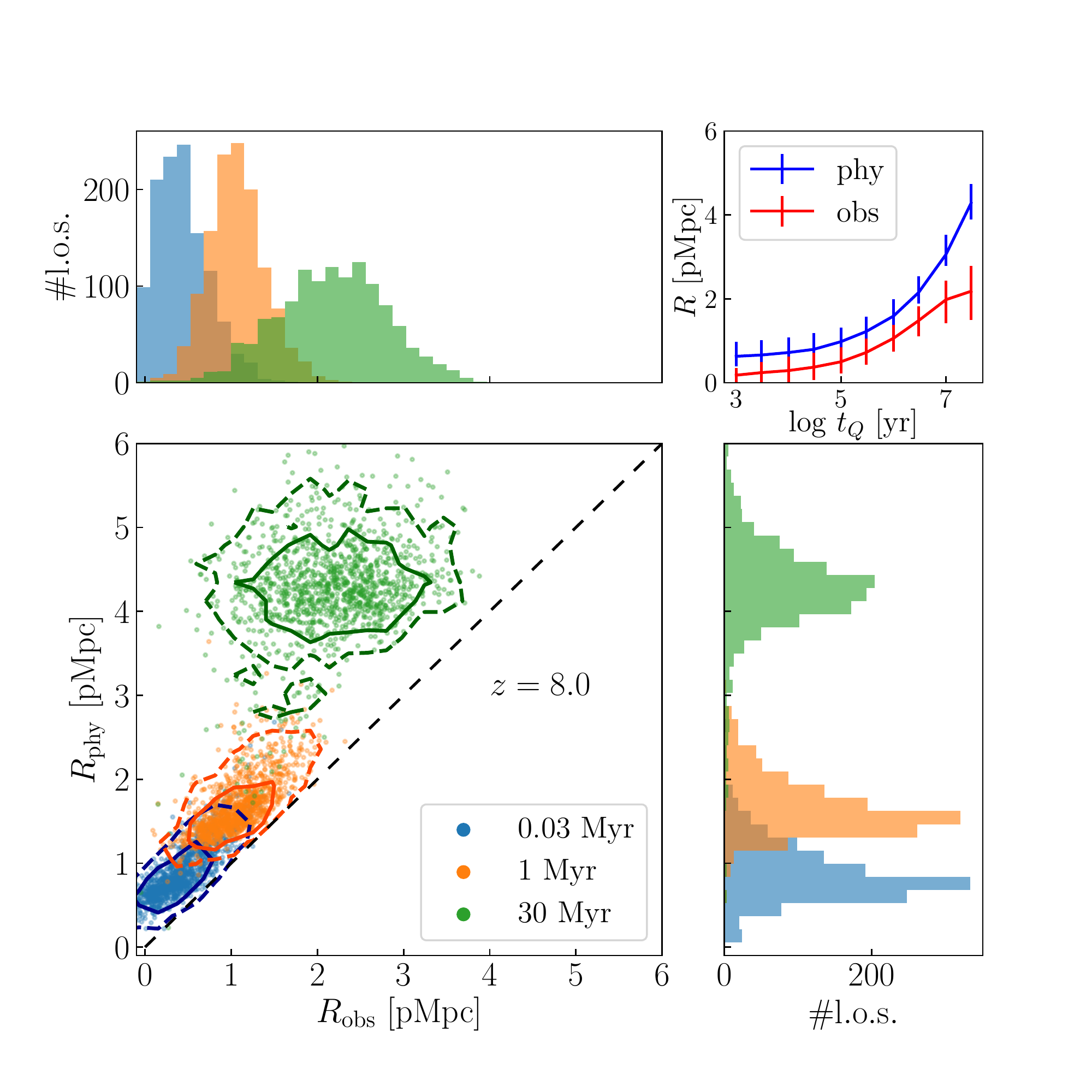} \hspace{-.9cm}
    \includegraphics[width=0.52\textwidth]{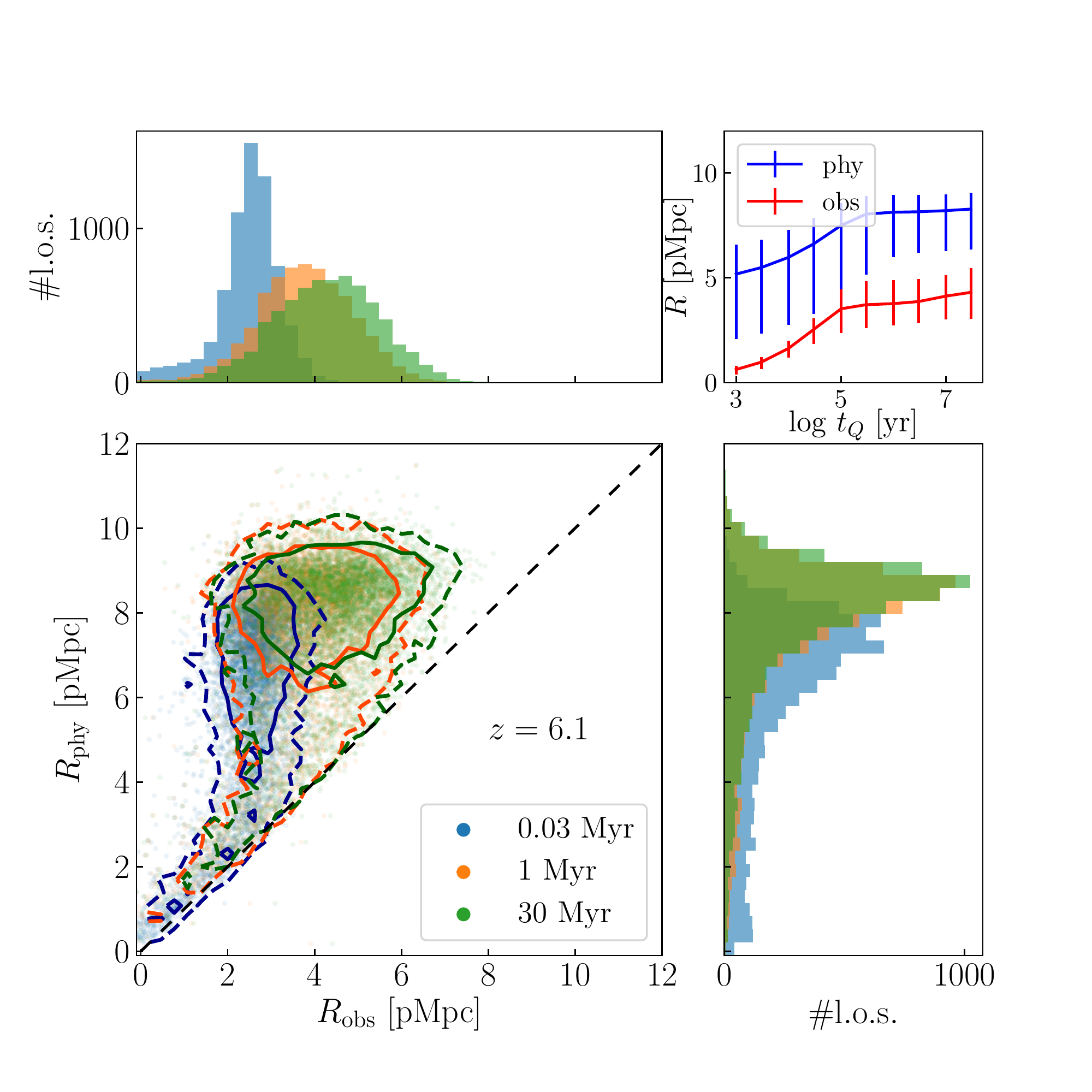}
    \caption{Proximity zone size distribution and and its dependence on $t_Q$ when the quasar turns on at $z=8.0$ (left) and $z=6.1$ (right). The lower left panel shows the distribution of the observational proximity zone size ($R_{\rm obs}$) vs the physical proximity zone size ($R_{\rm phy}$) at $t_Q=0.03$ Myr (blue), $1$ Myr (orange), and $30$ Myr (green). The solid and dashed lines are the $68\%$ and $95\%$ contours. Upper left and lower right are marginal distributions of $R_{\rm obs}$ and $R_{\rm phy}$ respectively. Plotted in the upper right panel is the evolution of $R_{\rm obs}$ (red) and $R_{\rm phy}$(blue) as a function of quasar age $t_Q$.}
    \label{fig:Robs_Rtheo_tQ}
\end{figure*}

\begin{figure}
    \centering
    \includegraphics[width=0.5\textwidth]{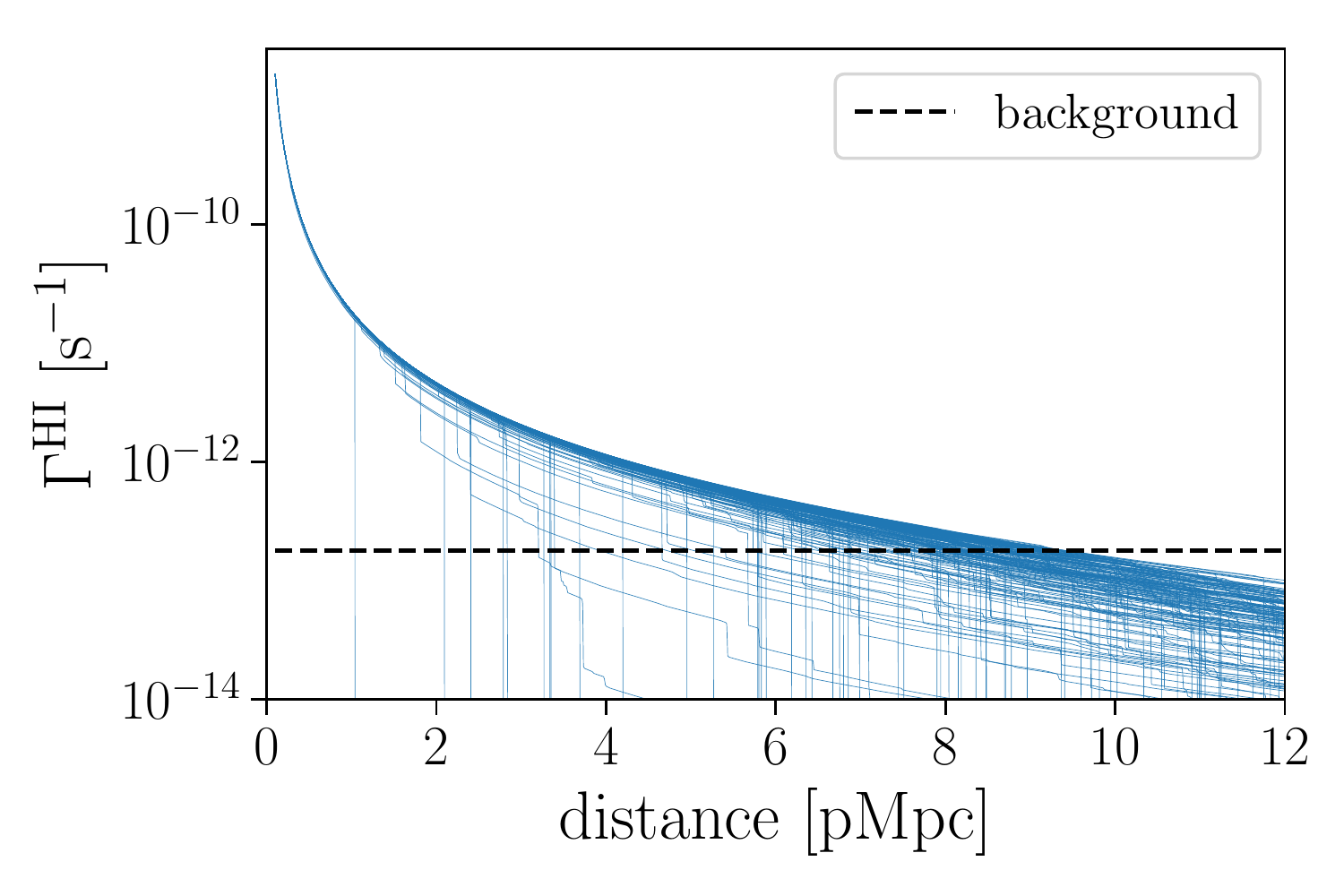}
    \caption{{ Blue lines are the ionization rate $\Gamma_{\rm QSO}^{\rm HI}$ profiles of $210$  randomly selected sightlines for $t_Q=30$ yr quasars at $z=6.1$. The black dashed line is the median background ionization rate $\Gamma_{\rm bkg}^{\rm HI}$}.}
    \label{fig:GammaProfile}
\end{figure}

\begin{figure}
    \centering
    \includegraphics[width=0.48\textwidth]{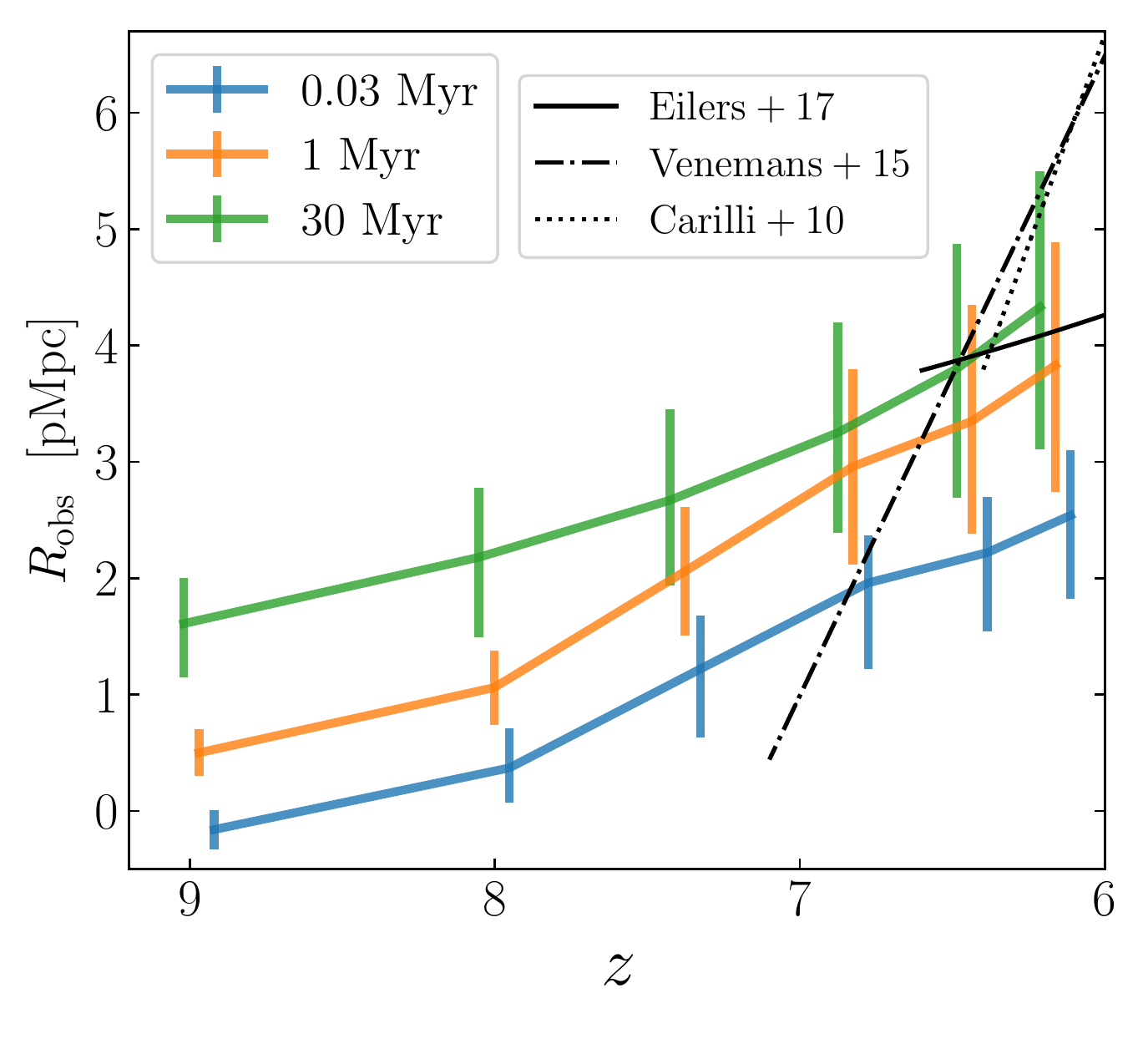}
    \caption{Evolution of the observational proximity zone size $R_{\rm obs}$ as a function of the quasar turn-on redshift for a fixed quasar age of $0.03$ Myr (blue), $1$ Myr (orange) and $30$ Myr (green), respectively. The different black lines are fits to observational data in three previous studies; in each the redshift range of the quasar sample spans from the start of the corresponding black line to $z\sim 5.8$. Our results are most consistent with \protect\citet{eilers2017}.}
    \label{fig:R_z}
\end{figure}

\subsection{Distribution and Evolution of Proximity Zone Sizes}

\subsubsection{As a Function of Quasar Age $t_Q$}

In the two four-panel graphs in Figure \ref{fig:Robs_Rtheo_tQ} we show the distribution and evolution of $R_{\rm obs}$ and $R_{\rm phy}$. At $z=8.0$ (before the global reionization) $R_{\rm obs}$ and $R_{\rm phy}$ track each other well with moderate scatter for the first 1 Myr, and they approximately follow the $R_{\rm phy}=R_{\rm obs}+0.5\dim{pMpc}$ relation. This is because at short quasar ages the observational proximity zone size is usually limited by the damping wing of the neutral gas just ahead of the I-front, which is within $\sim 2$ pMpc from the quasar. After $t_Q>1$0 Myr,  the I-front moves far enough away from the quasar that the damping wing does not suppress the transmitted flux significantly (see the example sightline in Figure \ref{fig:example_los}). Instead, the truncation of $R_{\rm obs}$ is mostly due to the  absorption features from the cosmic large-scale structure that correlates with the quasar host halo. Therefore, after  $t_Q>10$ Myr $R_{\rm phy}$ keeps growing while $R_{\rm obs}$ is saturated, and $R_{\rm obs}$ does not trace $R_{\rm phy}$ anymore.

The evolution of $R_{\rm phy}$ for $t_Q>0.1$ Myr can be well fitted by a power-law with a slope of $0.27$, slightly smaller than the value of $1/3$ for the uniform IGM. Also, because by $z=8$ quasar host halos are already embedded in ionized bubbles created by galaxies that cluster around them, $R_{\rm phys}$ starts at a non-zero value and its evolution with $t_Q$ is very flat in the first  $0.1$ Myr . 

At $z=6.1$ (right panel) the IGM is completely ionized, and the distribution and evolution of both $R_{\rm obs}$ and $R_{\rm phy}$ (right panel in Figure \ref{fig:Robs_Rtheo_tQ}) are very different from the $z=8.0$ case. The physical proximity zone size $R_{\rm phy}$ (the histogram on the right) has much larger scatter than at $z=8.0$ simply because the edge of $R_{\rm phy}$ is no longer a sharp I-front. Far away from the quasar, $\Gamma^{\rm HI}_{\rm QSO}$ drops very slowly and slight fluctuations in either $\Gamma^{\rm HI}_{\rm QSO}$ or  $\Gamma^{\rm HI}_{\rm bkg}$ can change $R_{\rm phy}$ significantly.
{ The value of $R_{\rm phy}$ is sensitive to the Lyman limit systems along the line of sight. When a quasar is very young $t_Q=0.03$ Myr, there is a large scatter of $R_{\rm phy}$ down to extremely small values. This is because sightlines often hit pre-existing  LLSs with overdensities $\Delta_g\sim 100$ and $\xHI>10^{-3}$, which initially attenuate the radiation from the quasar. However, after $\gtrsim 0.1$ Myr the gas in these LLSs becomes more ionized and transparent to the quasar radiation. This explains the change of the distribution of $R_{\rm phy}$ from $t_Q=0.03$ Myr to $1$ Myr.  After $1$ Myr the ionization equilibrium is re-established for most of the pre-existing LLSs and thus the distribution of $R_{\rm phy}$ does not evolve any further. At $z=6.1$, the tail of $R_{\rm phy}<6$ pMpc is mostly due to the remaining LLSs/DLAs that are too dense to be over-ionized by the quasar and block the its radiation. Examples of $\Gamma_{\rm HI}^{\rm QSO}$ profiles for such sightlines can be found in Figure \ref{fig:GammaProfile}. The sudden drop in $\Gamma_{\rm HI}$ corresponds to the positions of these LLSs/DLAs. As for $R_{\rm obs}$, its value is sensitive to any gas with $\xHI\gtrsim 10^{-4}$. When $t_Q=0.03$ Myr, even the mean density gas at $\gtrsim 4$ pMpc has not reestablished ionization equilibrium and still has $\xHI\sim 10^{-4}$, therefore at such young quasar age the majority of $R_{\rm obs}<4$ pMpc. After $\sim 0.1$ Myr the new ionization equilibrium is established in most places within $8$ pMpc (see Figure \ref{fig:example_los}) and $R_{\rm obs}$ does not change significantly afterwards. The slight change in the distribution of the $R_{\rm obs}$ peak (orange and green histograms on the upper panel) is due to the subsequent additional heating from the HeII I-front.}

\subsubsection{As a Function of Redshift $z$}

As is shown above, for a given quasar age $t_Q$, the proximity zone sizes at higher redshifts, when the IGM is mostly neutral, are smaller than the proximity zone sizes at lower redshifts, when the IGM is mostly ionized. Studying the evolution of proximity zone sizes as a function of redshift thus helps us to understand how the ionization state of the IGM evolves with redshift.

In Figure \ref{fig:R_z} we show the evolution of the median observed quasar proximity zone size $R_{\rm obs}$ as a function of redshift. The error bars capture the $68\%$ spread of all sightlines at each snapshot. For a fixed quasar age $t_Q=30$ Myr, we find the evolution is slow and smooth, because here the growth of $R_{\rm obs}$ is due to the decrease in mean density of the universe. For short quasar age $t_Q\sim 1$ Myr, we notice that the growth of $R_{\rm obs}$ is slightly faster after $z\sim 8$ than before $z\sim 8$. This is because the universe is mostly neutral at high redshift, and for $t_Q$ as short as $\sim 1$ Myr neutral patches limits the growth of $R_{\rm obs}$ (see the left panels in Figure \ref{fig:example_los} or \ref{fig:Robs_Rtheo_tQ}). After the universe becomes mostly ionized, only rarely neutral patches stand in the way of quasar sightlines, resulting in the slight up-tilt of the orange line in Figure \ref{fig:R_z}.  

Also plotted in Figure \ref{fig:R_z} in black are three best fits to observational data from the previous studies \citep{carilli2010,venemans2015,eilers2017}. In these studies the fits are normalized to the fixed quasar magnitude of $M_{\rm 1450}=-27$, slightly brighter than $M_{\rm 1450}=-26.66$ in this study. The black lines in Figure \ref{fig:R_z} are re-normalized using the correcting formula from each of the three observational studies respectively.  Note that the numbers of quasars used in these fits are very limited ($<40$), and these quasars are at redshifts between $z \sim 5.8$ and the end points of the black lines, with majority at $z\sim 6.1$. Our simulation is most consistent with the shallow slope measured by \citet{eilers2017}.

\begin{table}
\begin{tabular}{|l|c|c|c|c|}\hline
\backslashbox{$z$\kern-1em}{\kern-1em\ $t_Q$ [yr]}&\makebox[3em]{$1\times10^3$}&\makebox[3em]{$3\times10^4$}&\makebox[3em]{$1\times10^6$}&\makebox[3em]{$3\times10^7$}\\\hline
  8.9  &1050, 1050  &1050, 1050&510, 1050& 12, 96\\\hline
  8.0 &1227, 1260&830, 1202&50, 547&15, 43\\\hline
  7.3 &1166, 1430 &176, 505 &17, 55& 13, 35\\\hline
  6.8 &1492, 2511&190, 304&31, 58&22, 41\\\hline
  6.4  &1663, 4380&246, 401&55, 89&39, 55\\\hline
  6.1 &1818, 6782 &324, 497 &86, 118& 66, 93\\\hline
\end{tabular}
\caption{Number of slightlines that have observational proximity zone sizes $R_{\rm obs}$ smaller than $0.5$ pMpc (the first number) and $1$ pMpc (the second number) at different redshifts and quasar ages, respectively. The total number of sightlines at each redshift can be found in Table \ref{tab:halos}.}
\label{tab:stat}
\end{table}

\section{Extremely Small Proximity Zones}

\citet{eilers2017} has analyzed 34 medium-resolution quasar spectra at redshift $5.77\leq z\leq6.54$ and found several quasars with exceptionally small proximity zone sizes. Specifically, for the 11 quasars in the magnitude bin $-27.5\leq M_{\rm 1450} \leq -26.5$, which are similar to the one we simulate in this paper ($\sim M_{\rm 1450}=-26.66$), there is one quasar with the observed proximity zone size of $0.78$ pMpc. They argue that the most likely explanation for such an extremely small size is that the quasar is extremely young ($t_Q<10^5$ yr). { In a more recent study, \citet{eilers2020} estimate the fraction of such young quasars at $z\sim 6$ to be between 5\% and 10 \%.} If this is true, it makes it hard to explain how a supermassive black hole can form within the one billion year age of the universe \citep{martini2004,smith2019}. In this section, we search for such small proximity zones in our sample and analyze their properties as well as the statistics of finding such small proximity zones.

\subsection{Probability of Finding an Extremely Small Zone}

\begin{figure*}
 \centering \includegraphics[width=0.45\textwidth]{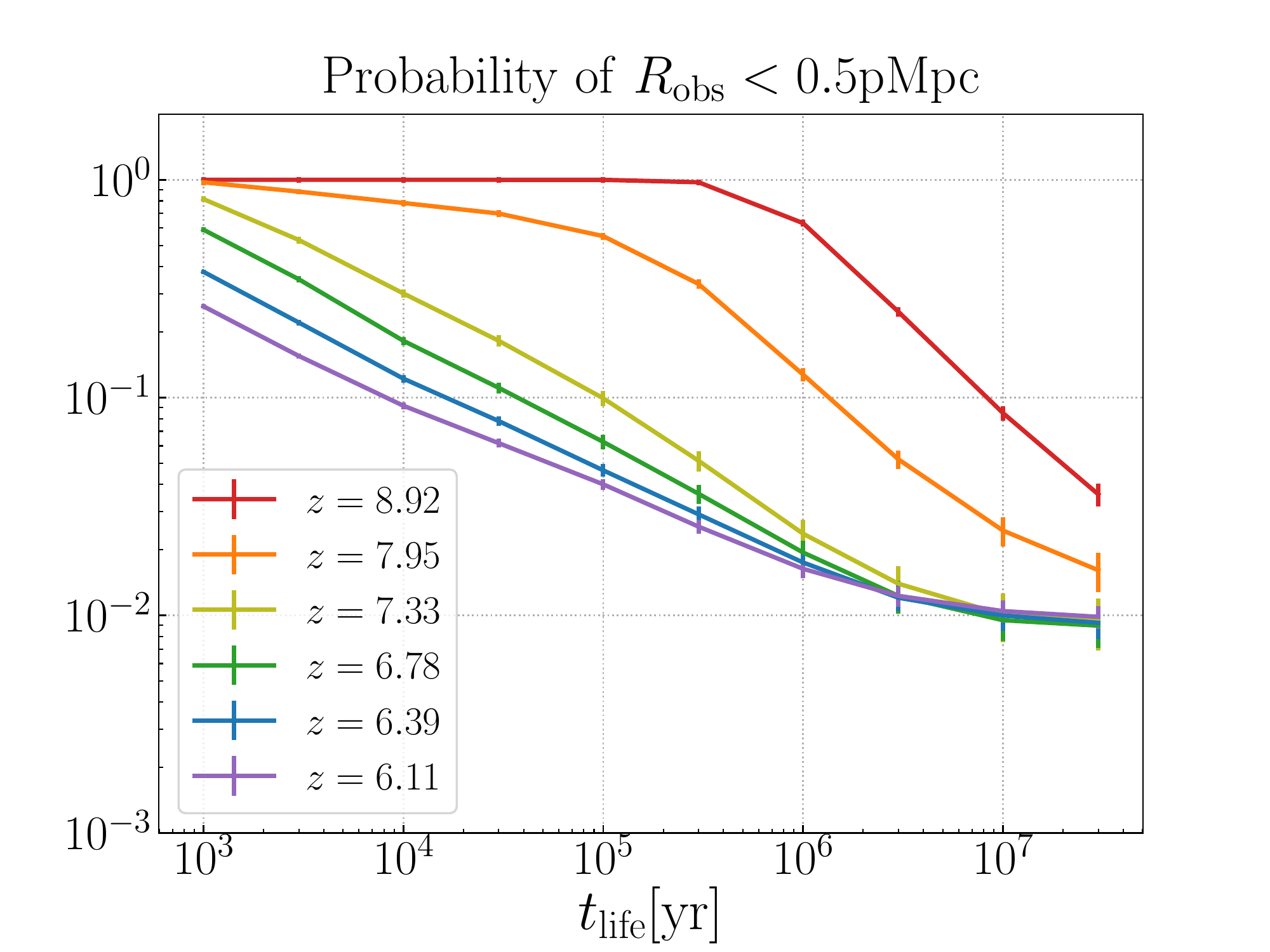}
\includegraphics[width=0.45\textwidth]{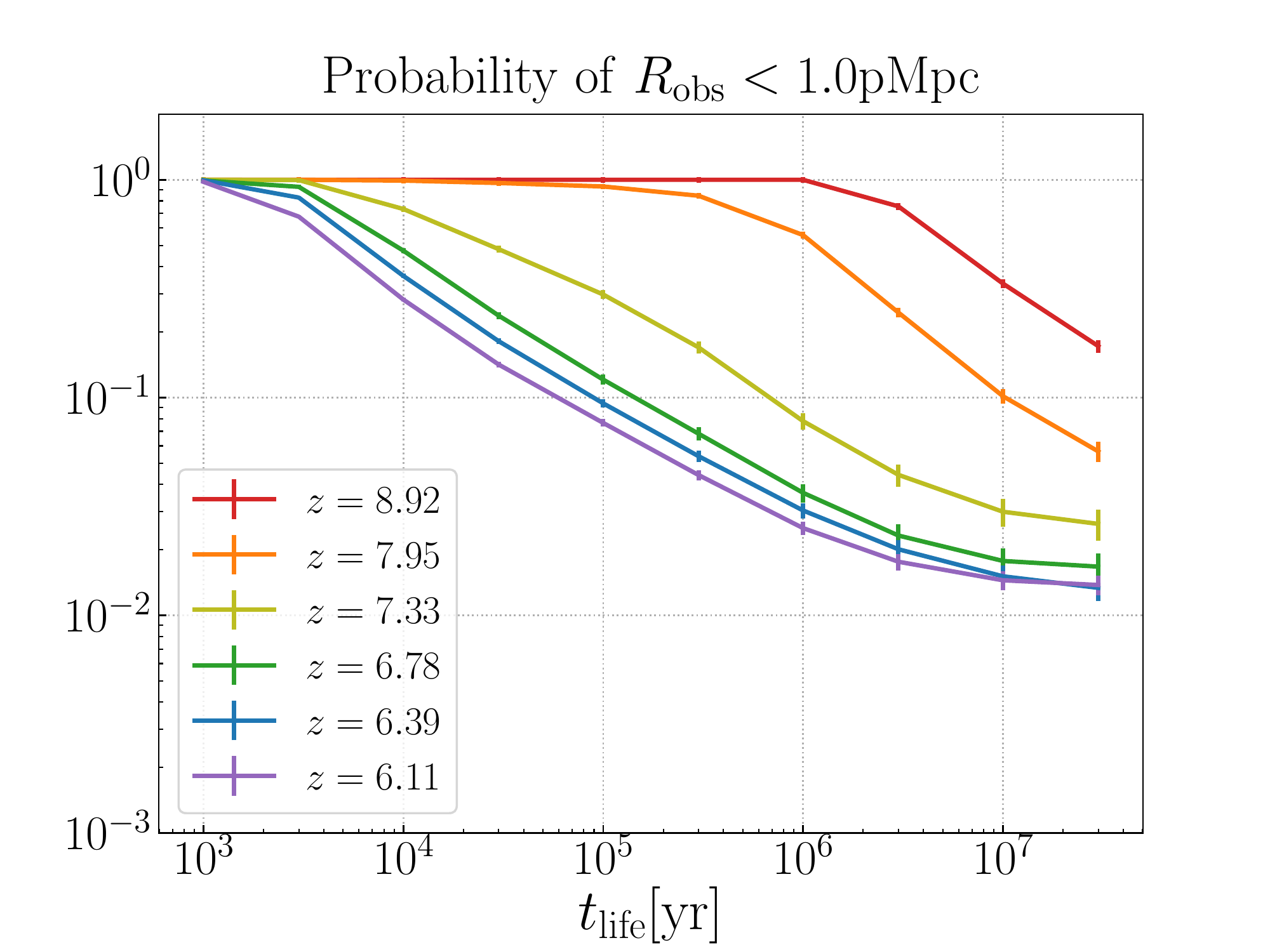}
\caption{Probability of finding a small proximity zone with $R_{\rm obs}<0.5$ pMpc (left) and $R_{\rm obs}<1$ pMpc (right) respectively as a function of the quasar lifetime at different redshifts.}\label{fig:prob}
\end{figure*}

In Table \ref{tab:stat} we show the number of sightlines that are smaller than $0.5$ pMpc (left) and $1$ pMpc (right) at a given redshift and quasar age. 
Using the simulation we can estimate the probability of observing a small proximity zone size $R_{\rm obs}$ for quasars with lifetime $t_{\rm life}$. At the moment we happen to observe a quasar, its age $t_Q$ can be significantly smaller than its total lifetime\footnote{We only consider that quasars shine for one epoch.}. Therefore, the probability is calculated as:
\begin{eqnarray}
P(R_{\rm obs}<R|t_{\rm life}) & = & \int_0^{t_{\rm life}} \frac{dt_{Q}}{t_{\rm life}} P(R_{\rm obs}<R|t_{Q})\nonumber\\
  & \approx & \sum\limits_{i} \frac{t_{Q_i}-t_{Q_{i-1}}}{t_{\rm life}} P(R_{\rm obs}<R|t_{Q_i}).
\end{eqnarray}

We plot the probability of observing small $R_{\rm obs}$ in Figure \ref{fig:prob}. Well before the global reionization ($z> 7$) the neutral patches of the IGM limit the growth of $R_{\rm obs}$, and most sightlines have $R_{\rm obs}<1$ Mpc for $t_{\rm life}\sim 1$ Myr. Therefore, for a short quasar lifetime the probability of finding a $R_{\rm obs}<1$ pMpc is very high ($>0.5$). After the global reionization ($z<7$) there are few neutral patches and the IGM reaches new ionization equilibrium very fast. As a result, probability curves start with a low value and flatten after $\sim 1$ Myr.

At $z=6.1$ the fraction of small proximity zone saturates for $t_{Q}=30$ Myr is $93/6930 \approx 1.3\%$. This number is larger than the simulation result of \citet{eilers2017}, who found that only one in $1100$ of their modeled sightlines has $R_{\rm obs}<1$ pMpc. One possible reason for this discrepancy is that the spatial resolution of \citet{eilers2017} simulations ($\sim 5$ pkpc at $z=6$ is not enough to resolve LLSs -  a dense enough LLS of sufficiently large size (or, equivalently, column density) along the simulated quasar sightline would stop the quasar ionization front and hence limit the proximity zone size. With adaptive mesh refinement the CROC simulations are able to reach $100$ physical pc peak resolution to resolve the LLSs - this peak resolution is, of course, only reached in the highest density regions, and so may not reflect the effective resolution the CROC simulations actually achieves in LLSs.

\subsection{Extremely Small Proximity Zones in Old Quasars}

Studying the features and properties of the extremely small proximity zones helps us to constrain the quasar age with more confidence and decide if the ``quasar age tension'' is real \citep{eilers2017,eilers2018b,khrykin2019,davies2019}. In particular, we want to know what can make the proximity zone extremely small for even an old quasar.
To this end, we inspect all 93 proximity zones with $R_{\rm obs}<1$ pMpc at $t_Q=30$ Myr in the $z=6.1$ snapshot.

\begin{figure*}
 \centering \includegraphics[width=0.48\textwidth]{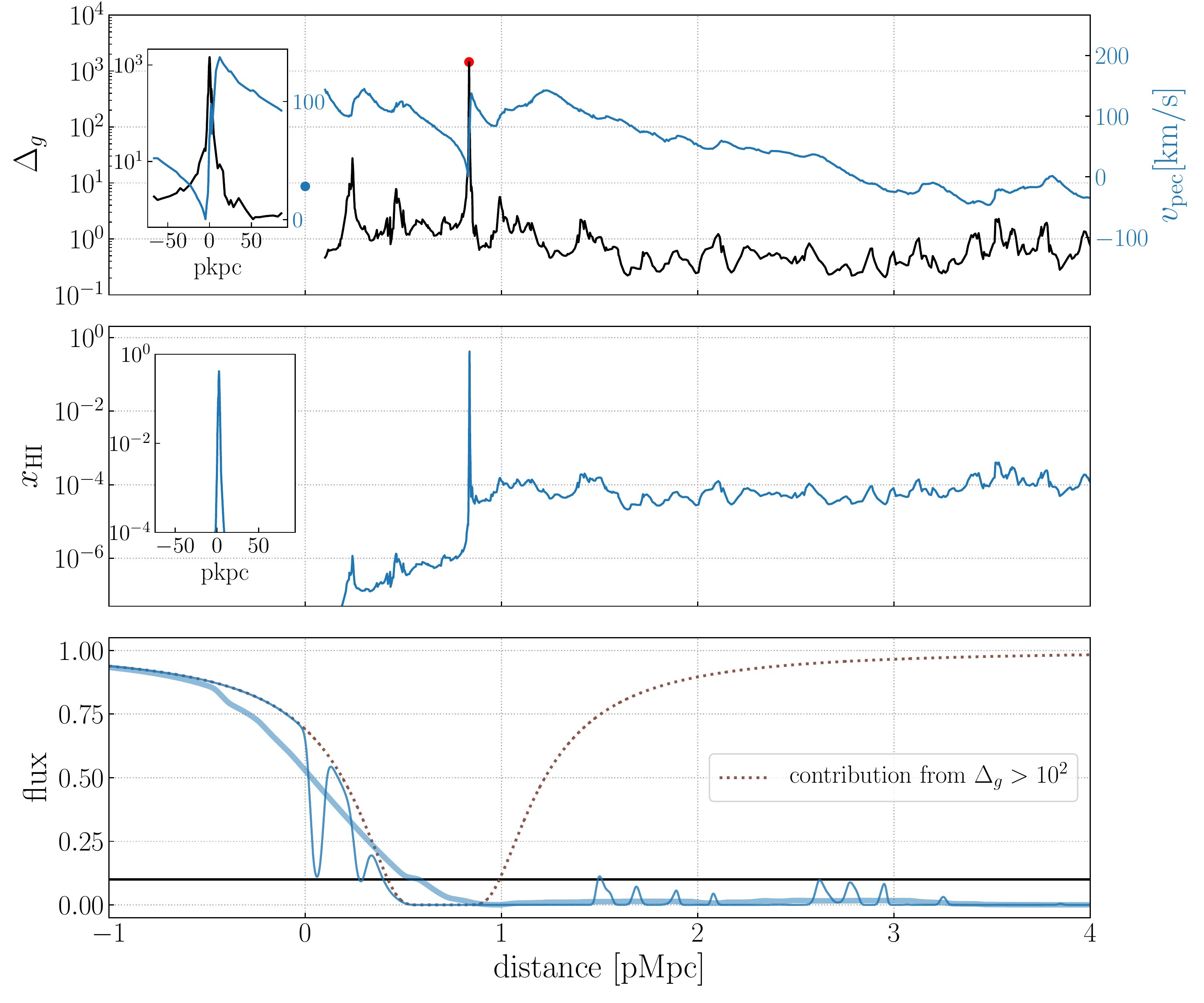}
 \includegraphics[width=0.48\textwidth]{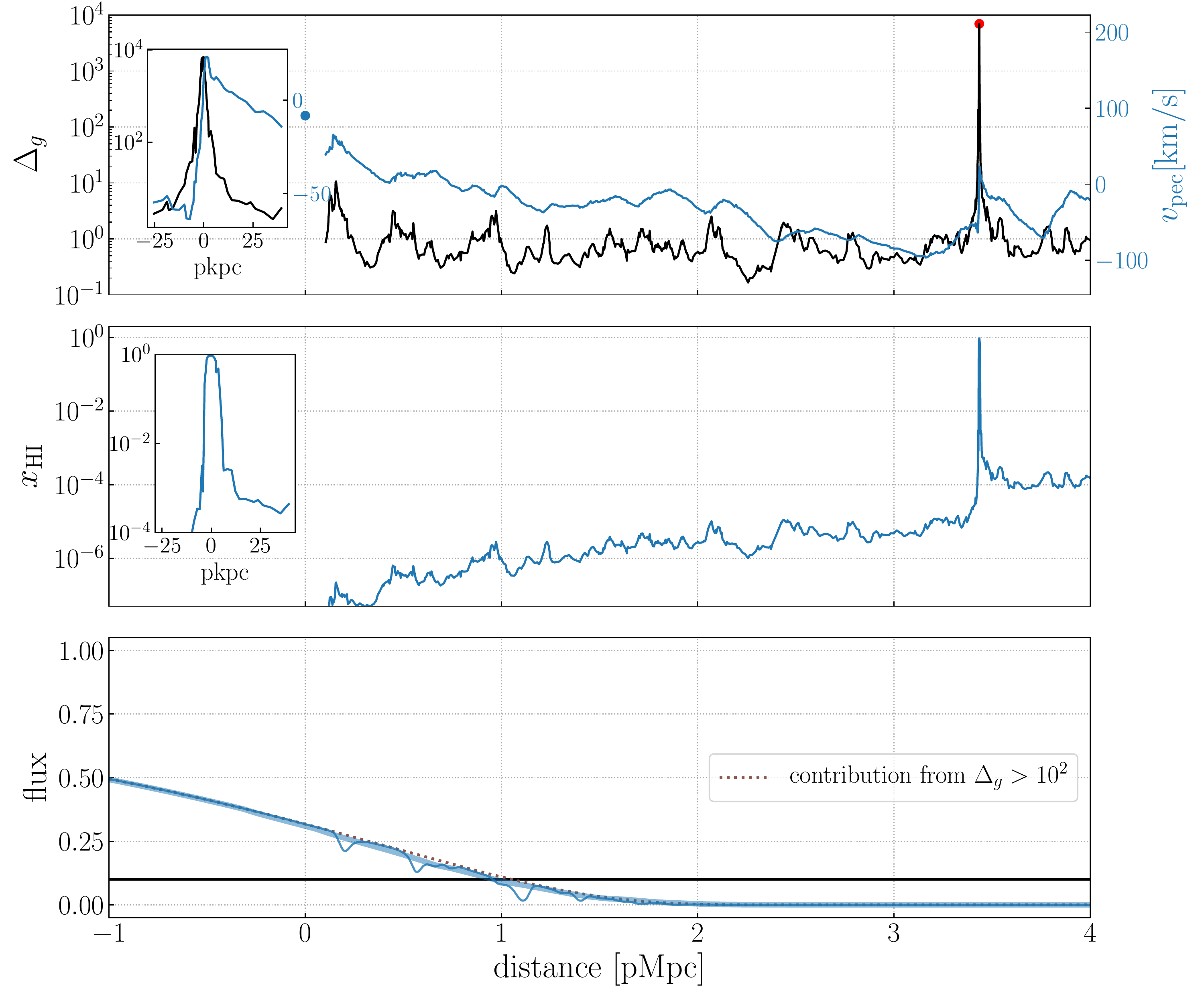}
\caption{Two example sightlines with extremely small observational proximity zone ($R_{\rm obs} < 1$ pMpc) caused by DLAs. There are $85$ such DLA-terminated proximity zooms in the whole $93$ extremely small proximity zone sample at $z=6.1$.
The black lines in the first row show the gas density in units of the mean density $\Delta_g\equiv \rho_g/\bar{\rho_g}$
and the red dot marks the local density peak with $\Delta_g>10^2$. The embedded panel zooms into this region. The blue line in the first row shows the peculiar velocity along the line of sight, with positive velocity pointing away from the observer. Thus the regions with sharp positive velocity jump are the regions with 
strong inflow, usually around a halo. The blue dot at $0$ pMpc shows the peculiar velocity of the quasar host halo along the line of sight. The second row shows the 
neutral hydrogen fraction of the gas at $t_Q=30$ Myr, and the third row shows the transmitted flux at the corresponding time. The thin blue line is the flux without 
smoothing, while the thick blue line is smoothed by a $20$ \AA\ boxcar. The dotted 
brown line shows the contribution from the cells with 
$\Delta_g > 10^2$.
}
\label{fig:DLAexample}
\end{figure*}

 We find that 85 out of 93 extremely small proximity zones show visible damping wings which are caused by very dense gas clumps with $\Delta_g \gtrsim 10^3$ within the $6$ pMpc from the quasar. 
 In the histogram of the neutral hydrogen column density  (the left panel of Figure \ref{fig:DLAprop}),  these $85$ sightlines with visible damping wings correspond to the main peak at $N_{\rm HI}\gtrsim 10^{20} \rm cm^{-2}$. 
 Such systems are commonly called DLAs or sub-DLAs. If we adopt the $N_{\rm HI}>2\times10^{20} \rm cm^{-2}$ threshold for DLA as defined in \citet{wolfe2005}, the number of sightlines that encounter DLAs is $70$, or $1\%$ of all sightlines. 
 We show two typical examples in Figure \ref{fig:DLAexample}. Both of them encounter a very dense clump of gas. The clump in the left panel has an overdensity of $\sim 10^3$, lying within $1$ pMpc from the quasar. Based on the high density and the peculiar velocity structure, we know that this clump is a part of a galaxy. Because this clump is so close to the quasar, this clump is no longer ``neutral'' ($\xHI<0.5$ after $30$ Myr), as is shown in the second row. Still, this structure has neutral hydrogen column density $N_{\rm HI}=3.6\times10^{19} \rm cm^{-2}$ and displays a damping wing in the spectra. In the right panel, the sightline hits an even denser clump of gas $\sim 3.4$ pMpc away from the quasar. Because it is both denser and further away than the example on the left, this clump is mostly neutral even after the quasar has been shining for $30$ Myr. The column density is extremely high, with $N_{\rm HI}=3.3\times 10^{21} \rm cm^{-2}$. Thus the suppression of the transmitted flux is even more significant, with the damping wing extending redward of the quasar Ly$\alpha$ line.

\begin{figure*}
 \centering \includegraphics[width=0.48\textwidth]{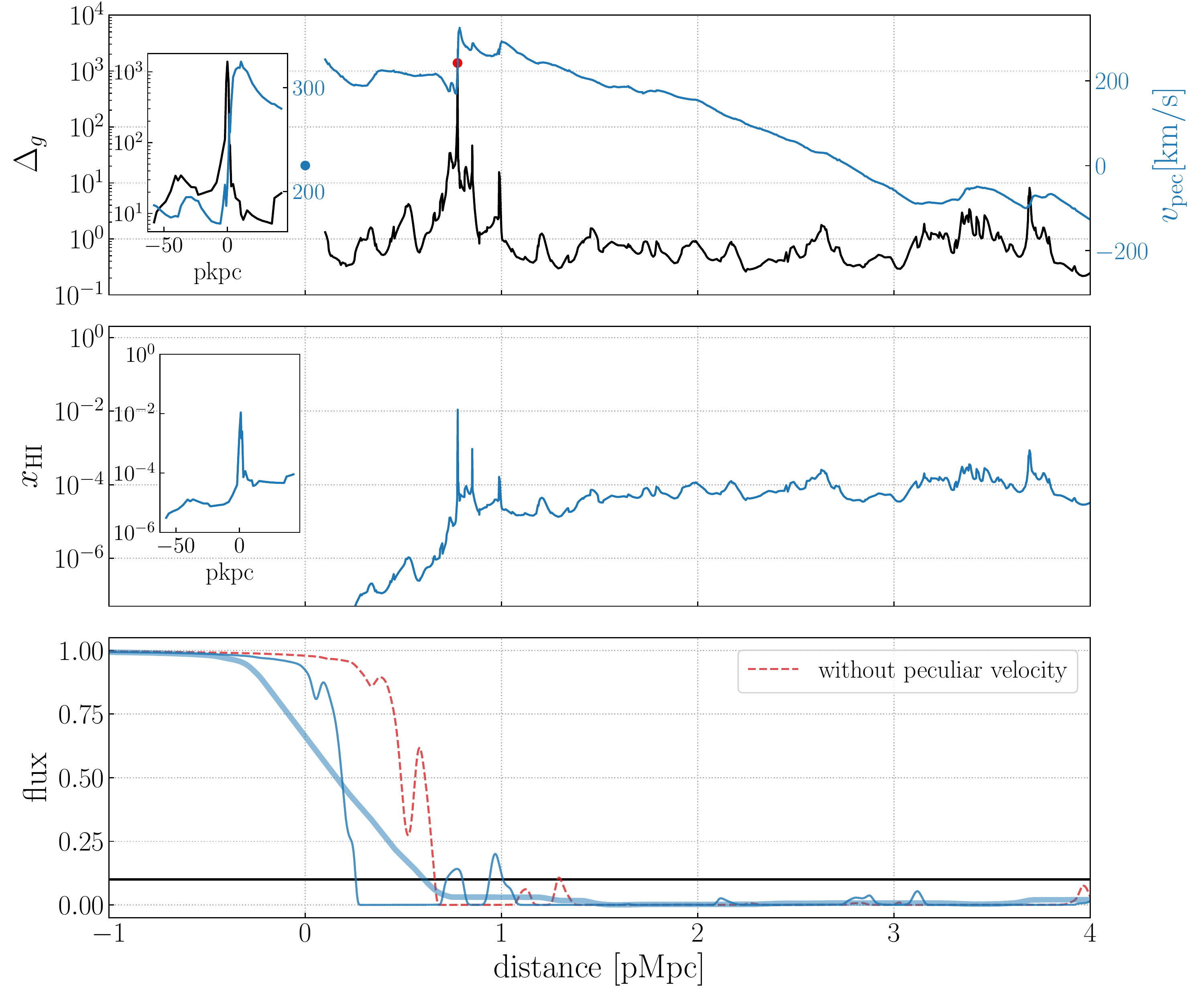}
 \includegraphics[width=0.48\textwidth]{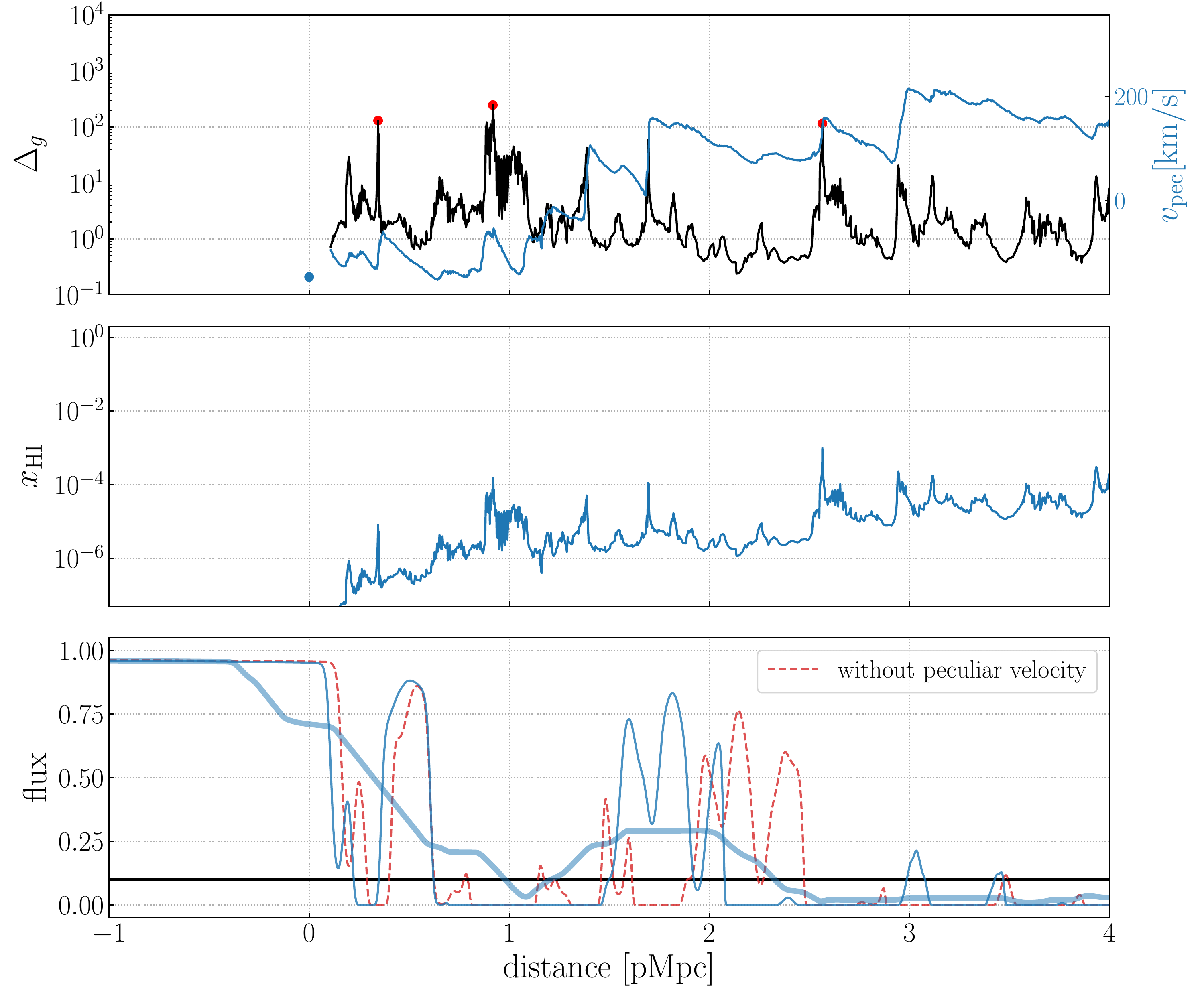}
\caption{Same as Figure \ref{fig:DLAexample}, but for a LLS-terminated proximity zone (left) and a proximity zone cut short by an extended overdensity region. The red dotted line in the bottom row shows the transmitted flux if the gas has no peculiar velocity.}\label{fig:nonDLAexample}
\end{figure*}

\begin{figure*}
 \centering 
 \includegraphics[width=0.3\textwidth]{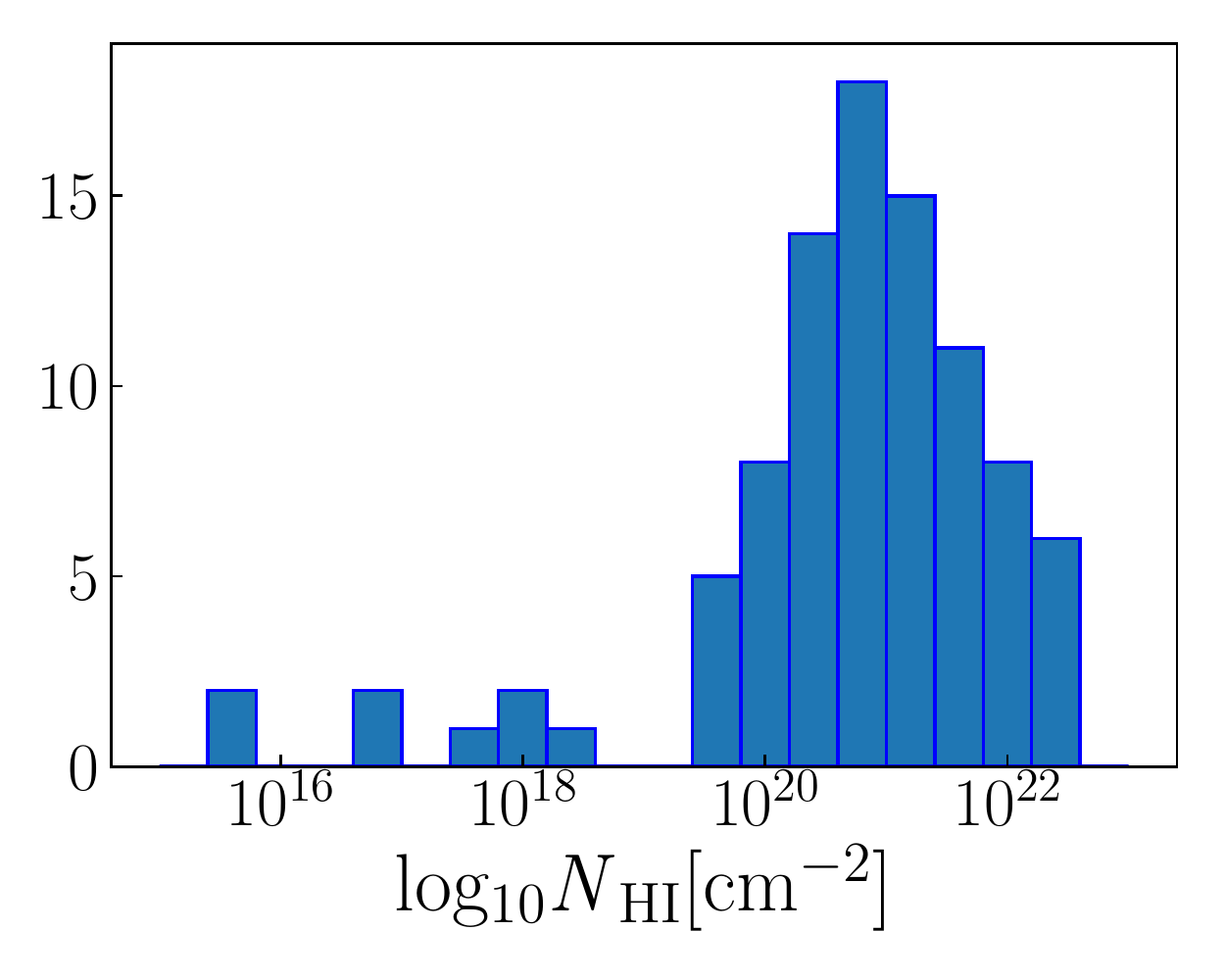}
\includegraphics[width=0.3\textwidth]{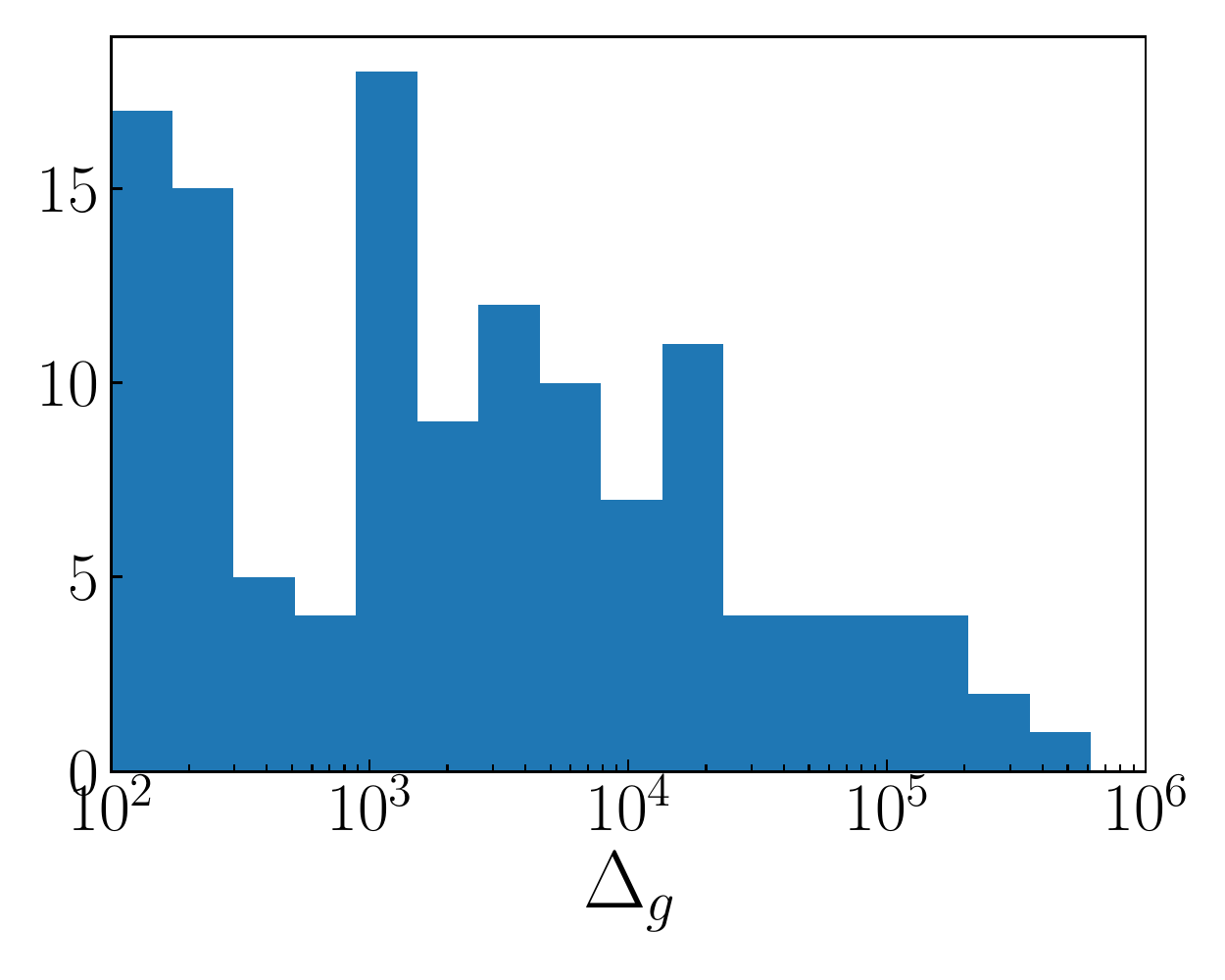}
\includegraphics[width=0.3\textwidth]{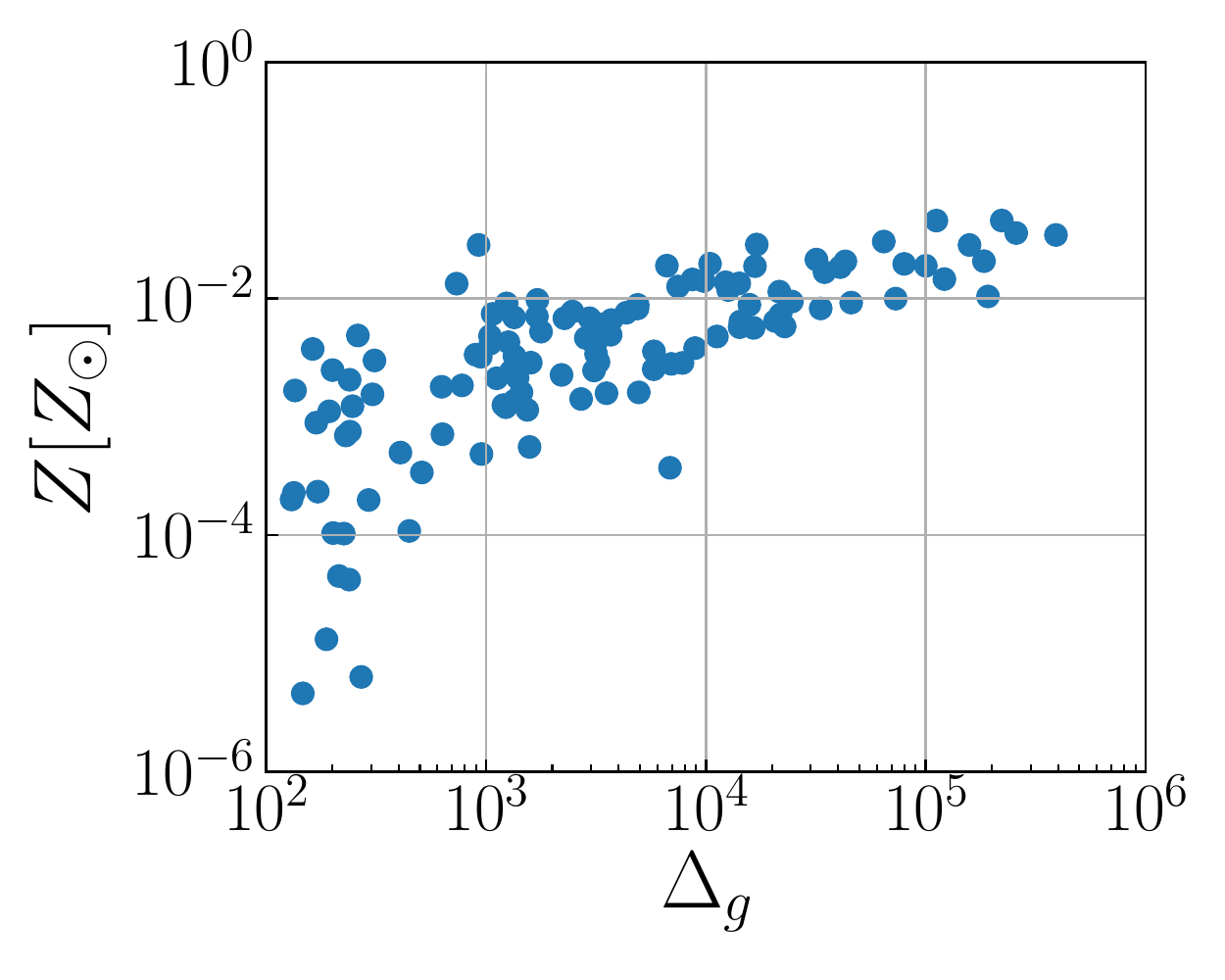}
\caption{Left: { Distribution of the neutral hydrogen column density $N_{\rm HI}$ for the $93$ sightlines with $R_{\rm obs}<1$ pMpc. The column density $N_{\rm HI}$ for each sightline is calculated by summing the $N_{\rm HI}$ for all pixels with $\Delta_g>100$.  Middle:} Distribution of gas densities for all density peaks (red dots in Figure \ref{fig:DLAexample} and \ref{fig:nonDLAexample}) with $\Delta_g>100$ and within $6$ pMpc from the quasar in the $93$ short proximity zone sightlines. Right: Metallicity of these density peaks. }\label{fig:DLAprop}
\end{figure*}

Not every gas clump with $\Delta_g>10^3$ produces the damping wing. Some of them are ionized by the quasar and have $N_{\rm HI}<10^{19} \rm cm^{-2}$. In the sample there are four such systems in total, { corresponding to the four sightlines in the three bins of $N_{\rm HI}\sim 10^{18} \rm cm^{-2}$ in the left panel of Figure \ref{fig:DLAprop}.} One example is shown in the left panel of Figure \ref{fig:nonDLAexample}. The dense gas clump at $0.8$ pMpc from the quasar has $\xHI\approx 10^{-2}$ after $t_Q=30$ Myr, and the column density is $N_{\rm HI}=2.4\times \rm 10^{18} cm^{-2}$.  Note that the LLS cuts the proximity zone extremely short partially because the peculiar velocity difference between the LLS and the quasar host (the blue dot at $0$ pMpc in the first row) projects the LLS closer to the quasar in the velocity space than the pure Hubble flow for that distance (at $z=6.1$, 200 km/s corresponds to $\approx 0.28\rm pMpc$).

The last four extremely small proximity zones are not caused by any dense clump of gas with $\Delta_g>10^3$. Rather, they are caused by a long extent of moderately overdense gas with $10^2 < \Delta_g < 10^3$. These structures are parts of a cosmic filament that happen to be aligned with the quasar sightline. One example is shown in the right panel of Figure \ref{fig:nonDLAexample}. The main feature that terminates $R_{\rm obs}$ is the extended structure at $\sim 1$ pMpc. Over a spatial scale of more than $0.5$ pMpc, this structure has density over the cosmic mean and creates an absorption trough of about $\sim 1$ pMpc in length. But because this gas is not dense enough to block quasar's radiation, the gas behind it is not shielded. Therefore, there are some transmitted spikes outside $R_{\rm obs}$. The percentage of such systems in our sample, $4/6930=0.06\%$, is consistent with that in \citet{eilers2017} simulation (1 in 1100).

In Figure \ref{fig:DLAprop} we show the properties of all the density peaks with $\Delta_g >100$ within $6$ pMpc from the quasar (the red dots in Figure \ref{fig:DLAexample} and Figure \ref{fig:nonDLAexample}).  We can see from the left panel that most of these density peaks have gas overdensity above $10^3$. These $\Delta_g>10^3$ peaks correspond to the DLAs and LLSs in the spectra, while the less dense structures ($10^2\lesssim\Delta_g\lesssim 10^3$) create smaller, localized absorption features.

Sometimes it is hard to identify LLSs because they do not display wide damping wings. In practice, observers usually inspect possible position of these LLSs and try to find any metal lines associated with them. If there are metal lines, then it helps to confirm the existence of an LLS. However, at $z\sim 6$ the properties of LLSs are not well studied and we do no know if all LLSs contain metals. Since in our simulation metal enrichment is modeled along with star formation and stellar feedback, we can investigate the metal content of simulated LLSs. In the right panel of Figure \ref{fig:DLAprop} we plot the relation between the gas overdensity and the metallicity for the dense gas peaks. We find that for very dense gas $\Delta_g>10^5$ the metallicity is above $0.01 Z_\odot$ but for gas with $\Delta_g\sim 10^3$ the metallicity is usually only between $0.001\sim 0.01 Z_\odot$. { These values are lower than what \citet{banados2019} observed of the proximity DLA of a $z=6.4$ quasar.} However, simulated metallicites of intergalactic gas critically depend on the details of the adopted star formation and feedback model. We discuss this topic more in the discussion section below.

\begin{figure}
    \centering
    \includegraphics[width=0.5\textwidth]{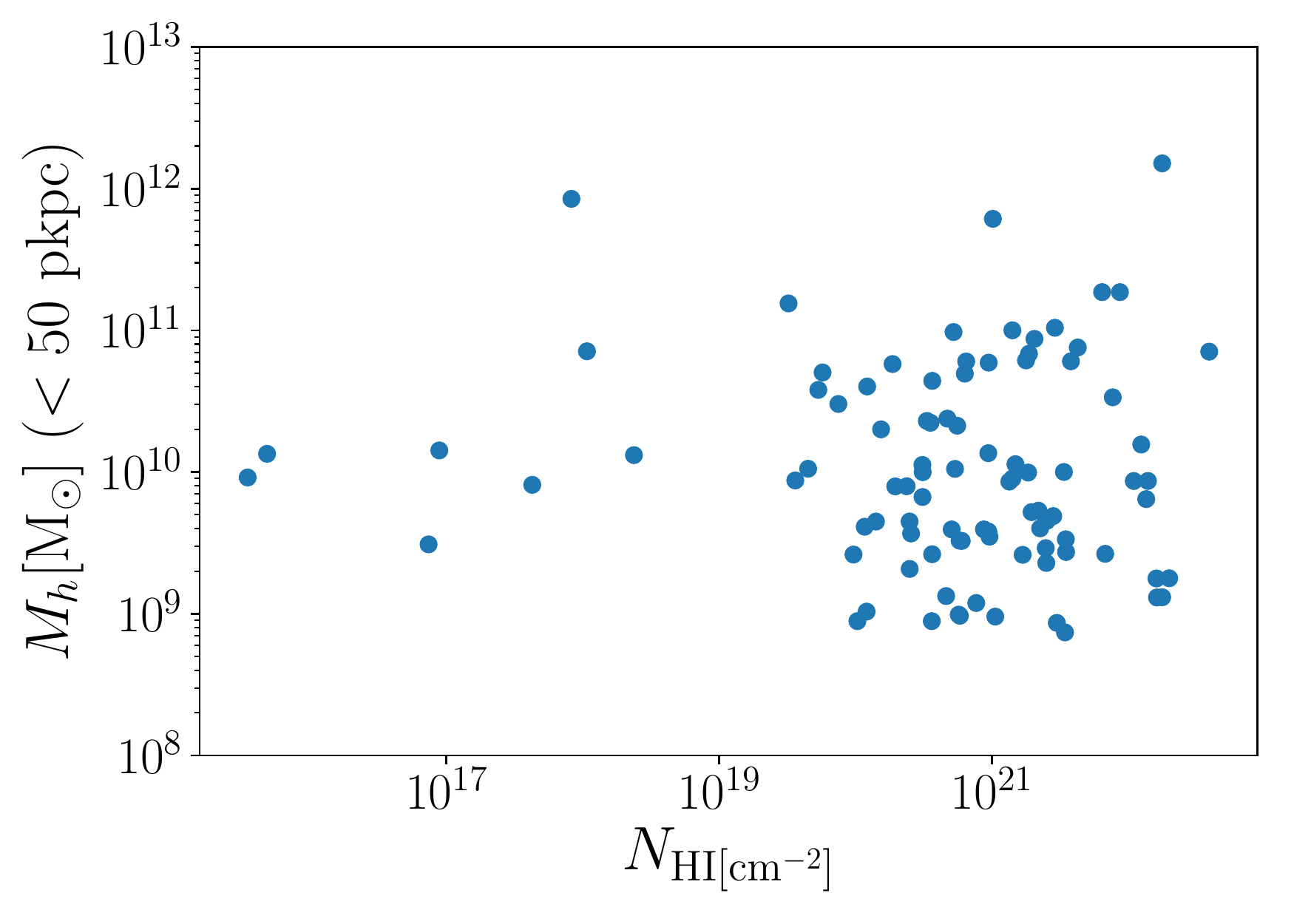}
    \caption{{ The mass of the most massive dark matter halo within $50$ pkpc from the density peaks.}}
    \label{fig:DLAhalos}
\end{figure}

{ One natural question to ask is what halos are associated with the DLAs/LLSs. To answer this question, we search for galaxies around these dense gas structures. In Figure \ref{fig:DLAhalos} we plot the mass of the most massive galactic halos within $50$ pkpc from the density peak for each sightline. We find that most of them are accompanied by massive halos with halo mass of $10^{10} \Msun$. The DLAs themselves may not be bounded to the most massive halo, but is a part of a smaller halo clustering around the massive one. The Pearson's correlation coefficient between $\log_{10} N_{\rm HI}$ of the (sub-)DLAs and $\log_{10} M_h$ of the most massive halo within $50$ pkpc\footnote{The virial radius of the most massive halo in the simulation is $\sim 50$ pkpc.} is $0.02^{+0.12}_{-0.13}$, therefore there is no correlation between these two quantities, although changing the threshold of the associating distance from  $50$ pkpc to $20$ pkpc or $10$ pkpc slightly increases  the Pearson's efficient to $0.17^{+0.12}_{-0.13}$ or $0.22^{+0.11}_{-0.12}$. This non- or weak-correlation is consistent with what has been found at lower redshift \citep[e.g.,][]{theuns2020}.}

\section{Discussion}

\subsection{Differentiating Old and Young Quasars With Small Zones}

\begin{figure}
 \centering \includegraphics[width=0.48\textwidth]{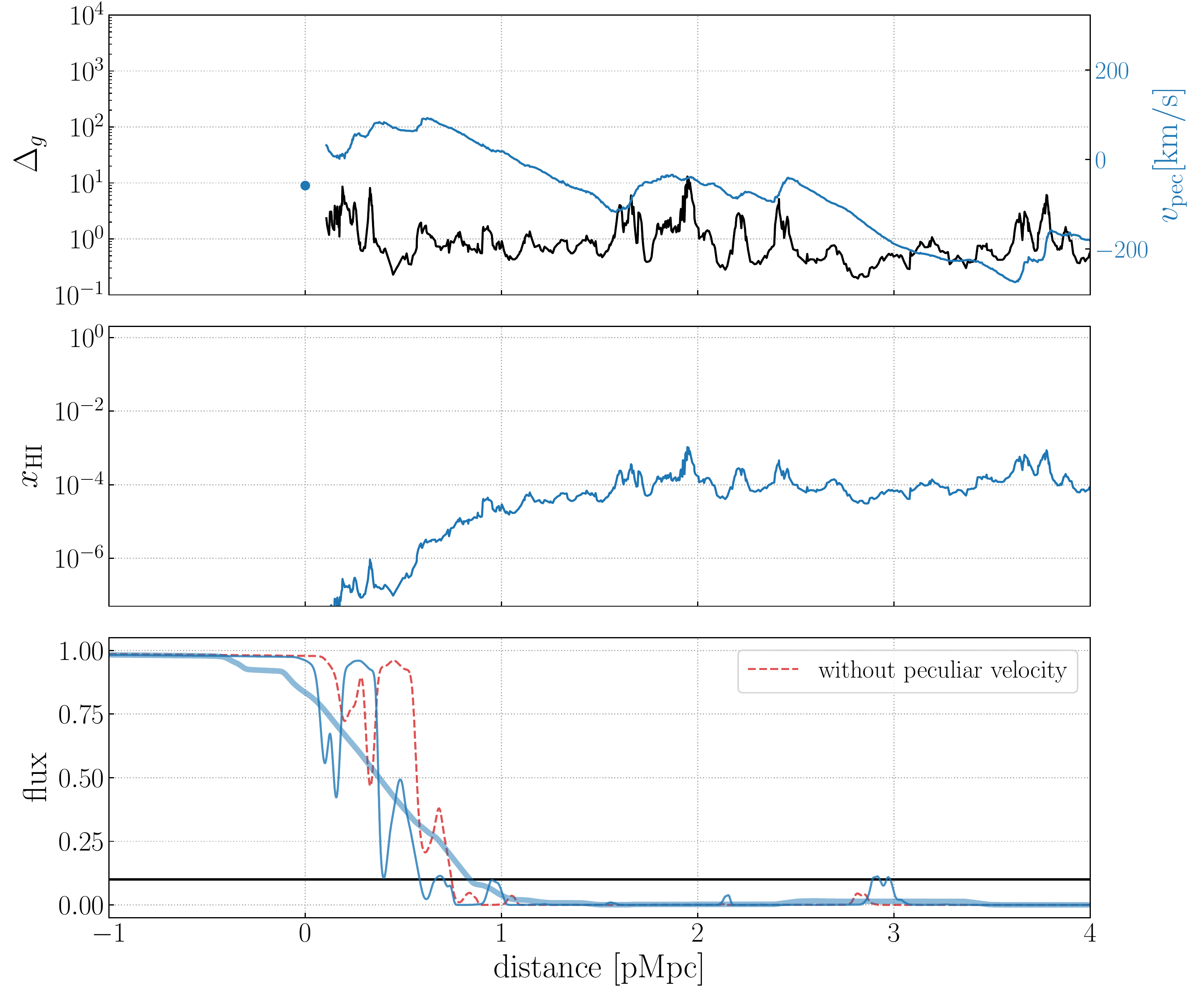}
\caption{Same as Figure \ref{fig:DLAexample} but for a typical young quasar with $t_Q=0.003$ Myr. This sightline does not go through any significant overdense region, but because the quasar is young, gas outside $\sim 1$ pMpc has not reached ionization equilibrium, thus the observational proximity zone size is small, and there are no transmitted spikes outside $R_{\rm obs}$.}\label{fig:youngQexample}
\end{figure}

It is fairly easy for a very young quasar ($t_Q<1\times 10^5$ Myr) to have a small $R_{\rm obs}$. However, the reason why young quasars have extremely small proximity zones is primarily because the ionization equilibrium has not yet been established. This is different from extremely small proximity zones in old quasars, caused by rare overdense regions. In Figure \ref{fig:youngQexample}, we show a typical quasar with $t_Q=3\times 10^3$ yr. In this sightline there are no dense gas clumps but normal density fluctuations around $\Delta_g=1$. Because the quasar only shines for a short time, the neutral fraction at $1$ pMpc is still above $10^{-5}$; such a high neutral fraction causes total absorption. Further away from the quasar, the neutral fraction is even higher, therefore also producing almost complete absorption.

These features in young quasar --- a quick drop in the flux with no more transmission spikes outside $R_{\rm obs}$ --- are distinguished from most old quasars with small proximity zones. However, it is still very hard to distinguish between a typical young quasar with an old quasar whose proximity zone is terminated by a LLS, since both of them have a sharp drop in flux and almost no transmitted spikes outside $R_{\rm obs}$. The best way to differentiate them is to search for metal lines, because in most regions there will be no metals while the LLSs will likely be enriched to $Z>10^{-3} Z_\odot$. 
{ Also, LLSs/DLAs are usually associated with galaxies. Therefore, a detection of a galaxy near the LLS position would favor a LLS explanation for the short proximity zone, although it is not clear how practical it is to observe a galaxy so close to the quasar.}

\begin{figure}
 \centering 
 \includegraphics[width=0.23\textwidth]{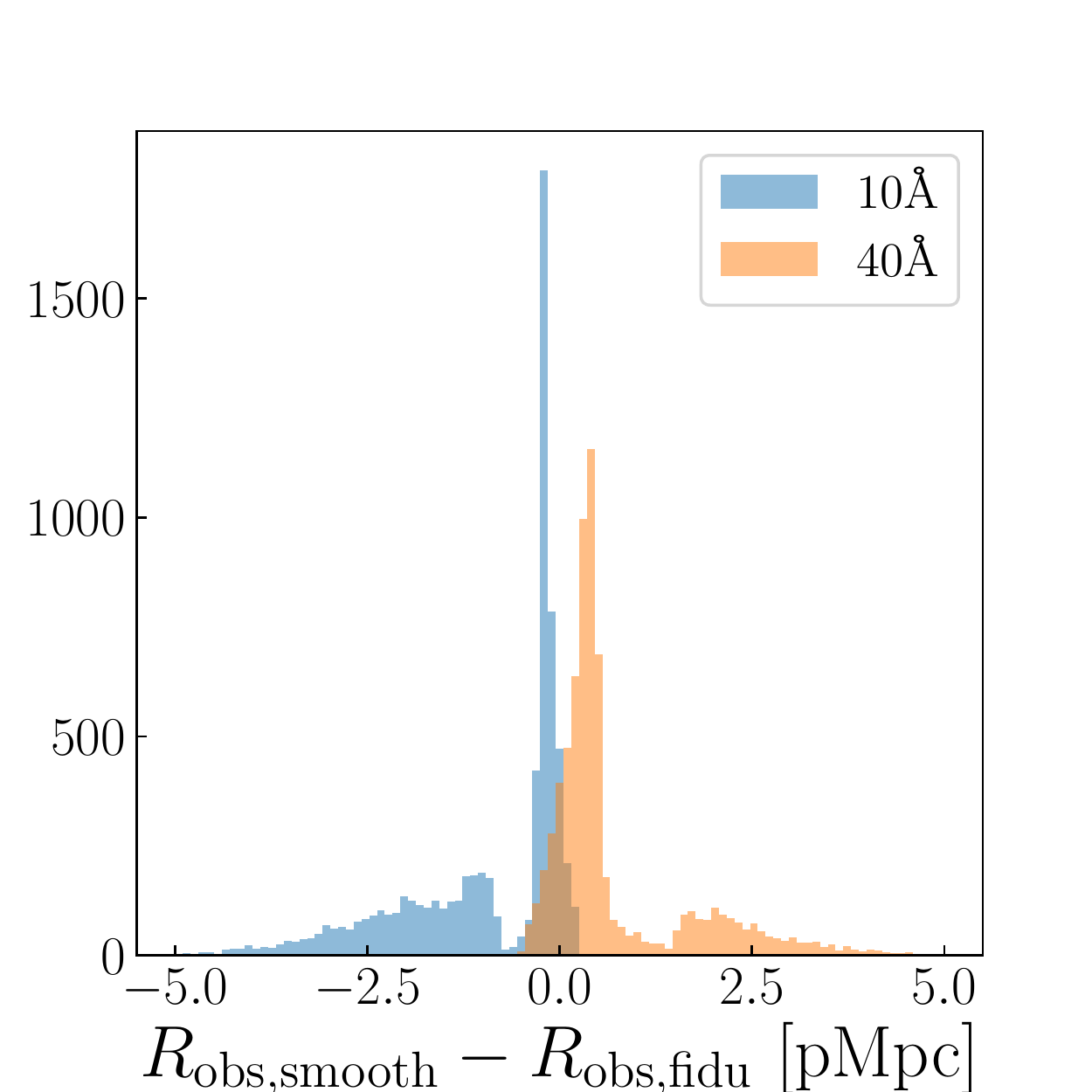}
 \hspace{-0.5cm}
 \includegraphics[width=0.23\textwidth]{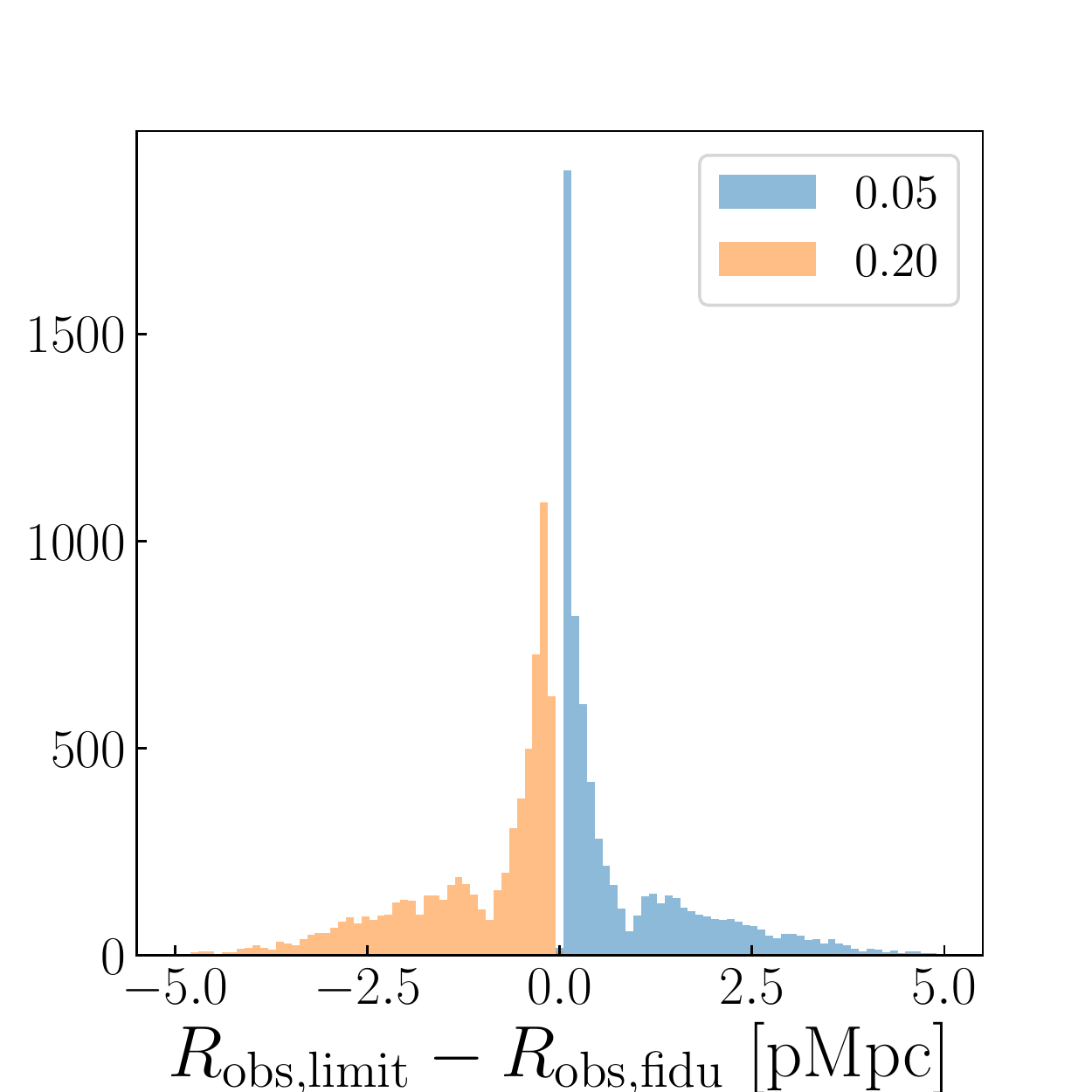}
\caption{Left: histograms of the difference in $R_{\rm obs}$ when using different smoothing kernel sizes. Right: histograms of the difference in $R_{\rm obs}$ when using different limiting thresholds. The fiducial kernel size is $20$ \AA \ and limiting threshold of $0.1$.}\label{fig:diffRobs}
\end{figure}

\subsection{Caveats in Modeling}

\subsubsection{CROC Simulations}

One needs to be aware of the limitations in our modeling. First of all, in this study we only analyzed one relatively small CROC simulation box. This box fully reionizes at $z\approx 6.8$, while in the other five $40 {~\rm cMpc}/h$ boxes, the earliest reionization redshift is $7.1$ and the latest is $6.3$. Therefore, depending on the actual simulation box chosen, the redshifts quoted in this paper may vary by $\Delta z \approx \pm0.5$. Also, these boxes are not large enough to catch the rarest density peaks.

Second, although the properties of IGM can be modeled robustly since the physics is relatively simple, the modeling of galaxies is a subject to a number of uncertainties. Our simulation peak resolution is $100$ pc in proper units; therefore, the detailed structure of the galaxies, such as the vertical structure of the disks, cannot be modeled. In addition to finite spatial resolution, star formation and stellar feedback in galaxies are modeled with sub-grid recipes. These sub-grid recipes are tested against local observations. Although the physics should operate similarly at all redshifts, these recipes have not been  tested against high-$z$ galaxy observations. For example, the stellar masses of massive CROC galaxies are low compared to observations \citep{Zhu2020}. Also, most massive CROC galaxies have metallicities below $0.1 Z_\odot$, while several galaxies with solar metallicity have been observed at $z\sim6$ \citep[e.g.][]{li2020,harikane2020,jiang2006}. This may suggest that the metals in simulated galaxies are also underproduced. It is likely that the stellar feedback model is too strong in our simulation. This may affect the predicted number density of LLSs.

\subsubsection{Post-processing}
\label{sec:pp}

The approach of using post-processing rather than the fully self-consistent 3D RT simulations offers a number of advantages but also has limitations. The full 3D simulation can only model at most a few isolated proximity zones at a time, and such a simulation would take on the order of 300,000 cpu hours to run for 100 Myr \citep{Chen2020} or even higher, as extremely fine temporal sampling would be required to capture the quasar light front. If one wants to study many quasars in different host halos and turning on at different redshifts, hundreds of 3D simulations would be required, which is not practical at present. The post-processing with 1D code is much more efficient and allows to study a wide parameter space.

The primary limitation of post-processing is that it does not account for gas dynamics and evolution of the background radiation field. The effect of gas dynamics obviously depends on the time-scale considered, so in this work we only model reasonably short timescales ($\lesssim 30$ Myr). The rms gas velocity dispersion at z=6 in our simulation is only 56 km/s. At this velocity the gas moves by less than 2 pkpc in 30 Myr - a negligible distance on cosmological scales, although comparable to existing observational constraints on the sizes of LLSs \citep{Fumagalli2016,Zahedy2019}. The latter implies that a specific sightline blocked by a LLS may change on the time-scale of 30 Myr, but statistically our estimate for the fraction of sightlines blocked by LLSs should be reliable.

In order to model timescales of several million years the gas dynamical effects must be accounted for. For example, the time interval from $z=8$ to $z=7.3$ is less than $100$ Myr, and the ionization fraction changes rather significantly in this time (see Figure \ref{fig:snapshots}). Also, the radiative feedback from the quasar can remove gas from small halos after $\sim 100$ Myr, and this may reduce the number of LLSs \citep{Chen2020}.

\subsubsection{Quasar Model}

In this study, we only model one epoch of quasar phase with constant luminosity (the ``lightbulb'' model). Also, here we only model structure outside 0.1 pMpc (several virial radii) from the quasar, so any absorption happening inside 0.1 pMpc is considered to occur during the quasar "obscure" phase.

If the quasar light curve is flickering, the proximity zone evolution can be more complex \citep{davies2020}. 
For a given quasar ``on'' time, flickering light cure may increase the probability of observing a very small proximity zone, especially if the sightline goes though $\Delta_g\sim 1000$ regions. These high density regions have recombination time comparable to $\sim 10$ Myr. Imagine a quasar that has a total ``on'' time of $30$ Myr, but in multiple episodes with gaps $\sim 10$ Myr. The quasar may take the whole episode to ionize a density peak, but during the next ``off'' period the density peak recombines again. Such scenario repeats for every episode, and can keep the proximity zone small every time we observe the quasar. This scenario is particularly interesting, { because it could potentially explain the large fraction of small PZ quasars that also do not show significant LLS features.} By comparing the statistics of proximity zone sizes with observational data, we could not only constrain the quasar lifetime for each episode but also the quasar duty cycle \citep{davies2019}.

\subsection{Observational Definition with Different Thresholds}

The most common definition of the observational proximity zone size is the distance from the quasar to the first point where the flux smoothed by a $20$\AA \  boxcar filter drops below $f_{\rm lim}=10\%$. This definition is motivated in observations because the high-$z$ quasar spectra are usually noisy and need to be smoothed to achieve sufficient signal-to-noise. However, in the physical sense, these thresholds are arbitrary. Here we investigate how the $R_{\rm obs}$ may change if we vary the smoothing size and limiting threshold.

In the left panel of Figure \ref{fig:diffRobs} we show the histogram of the difference in $R_{\rm obs}$ measured with $10$ \AA \  boxcar (blue) and $40$ \AA \  (orange) filters respectively, as compared to the fiducial value of $20$ \AA. We find that when changing the boxcar size to $10$\AA, half of them vary by less than $0.5$ pMpc. However, there is another population on the left that shows significantly reduced $R_{\rm obs}$. These are the sightlines that encounter absorption features with width of $\sim 10$ \AA \ but smaller than $\sim 20$\AA, so the flux drops below $0.1$ when smoothed by a $10$ \AA \ boxcar but not by a $20$ \AA \ boxcar. On the other hand, when apply smoothing with a $40$ \AA \ boxcar, absorption features larger than $20$ \AA \ but smaller than $40$ \AA \ do not terminate $R_{\rm obs}$ like the fiducial one, so $R_{\rm obs}$ defined this way is usually larger. Also, since the $40$ \AA \ kernel size is larger, the ``bump'' of the orange histogram is also further away from zero than that in the $10$ \AA \ kernel case.

The histogram in the right panel shows the difference in $R_{\rm obs}$ when changing the limiting flux only. We can see in this case the two histograms have no overlap by definition, since dropping the threshold always makes the $R_{\rm obs}$ larger. However, like the histograms in the left panel, they also have wide  ``wings", which is related to the strength of small absorption features at the edge of $R_{\rm obs}$.

\section{Summary}

In this study, we have post-processed a CROC simulation and analyzed the proximity zone sizes of quasars with magnitude $M_{1450}=-26.66$. Our simulation models realistic pre-ionized IGM and has high spatial resolution to model LLSs. Our post-processing code uses adaptive time steps and high temporal resolution.
We find that before the global reionization, the median of the observed proximity zone size increases steadily in the first $30$ Myr. After the global reionization, it only grows rapidly in the first $\sim 0.1$ Myr, which is consistent with previous studies \citep{bolton2007a,davies2020}. We find a slow growth of $R_{\rm obs}$ with decreasing turn-on redshift, consistent with the measurements in \citet{eilers2017}.

We also analyzed all the extremely short proximity zones at $z=6.11$ for old quasars ($t_Q=30$ Myr). We find that 93 out of $6930$ sightlines ($1.3\%$) display $R_{\rm obs}<1$ pMpc. The vast majority of them are caused by DLAs or LLSs along the line of sight. These DLAs and LLSs are dense gas with overdensities above $10^3$, and are polluted by metals. The rest of the extremely small proximity zones are caused by absorption from extended regions with overdensity $\gtrsim 100$. There are four such cases, and they all have transmission spikes outside $R_{\rm obs}$.

If the quasar lifetime is long ($>10$ Myr), our simulation shows that the possibility of finding a small proximity zone ($R_{\rm obs}<1$ pMpc) at $z\approx 6$ is $\sim 1\%$. This is smaller than the fraction $\sim 10\%$ reported in observation \citep[e.g.][]{eilers2017}, although currently the number of observed spectra are too limited to draw a firm conclusion. We note that the CROC simulation may have too strong stellar feedback \citep{Zhu2020} that could destroy some LLSs. Also, flickering light curves can increase the probability of observing small proximity zones.

In future work, we will examine more CROC simulations which have slightly different reionization histories. We will also use more complex quasar light curves to study the quasar size distribution with different duty cycles. Also, inside $R_{\rm obs}$ there are many absorption features. These absorption features may contain much more information about quasar age and quasar environments, but have not been exploited yet. The field of quasar proximity zones will open wide up once the thirty-meter-class telescopes go online, since we can obtain high resolution spectra from reionization quasars with much shorter observation time. Investigating the features inside proximity zone is an important part of our project.

\acknowledgements
H.C. and N.G. thank the referee Frederick Davies for very constructive feedback that greatly improved the quality of this paper. The authors also thank James Bolton and Anna-Christina Eilers for valuable comments.
This work was supported by a NASA ATP grant NNX17AK65G and NASA FINESST grant NNH19ZDA005K. This project is carried out on the Midway cluster at the University of Chicago Research Computing Center.

\appendix
\section{Code Tests}

Here we show three tests of our 1D RT code, including the I-front position, the ionization structure of the uniform medium, and a real line of sight running through a LLS/DLA.

\subsection{I-front Position in a Uniform Hydrogen Gas}

Ionization front expanding into a uniform neutral hydrogen background can be described by an analytical solution \citep{shapiro2006,iliev2006}:
$$r_I=r_S[1-\exp(-t/t_{\rm rec})]^{1/3}$$
where $$r_S=[\frac{3\dot{N}_\gamma}{4\pi \alpha^{\rm HI}(T)n_{\rm H}^2}]^{1/3}.$$

\begin{figure*}
    \centering
    \includegraphics[width=0.33\textwidth]{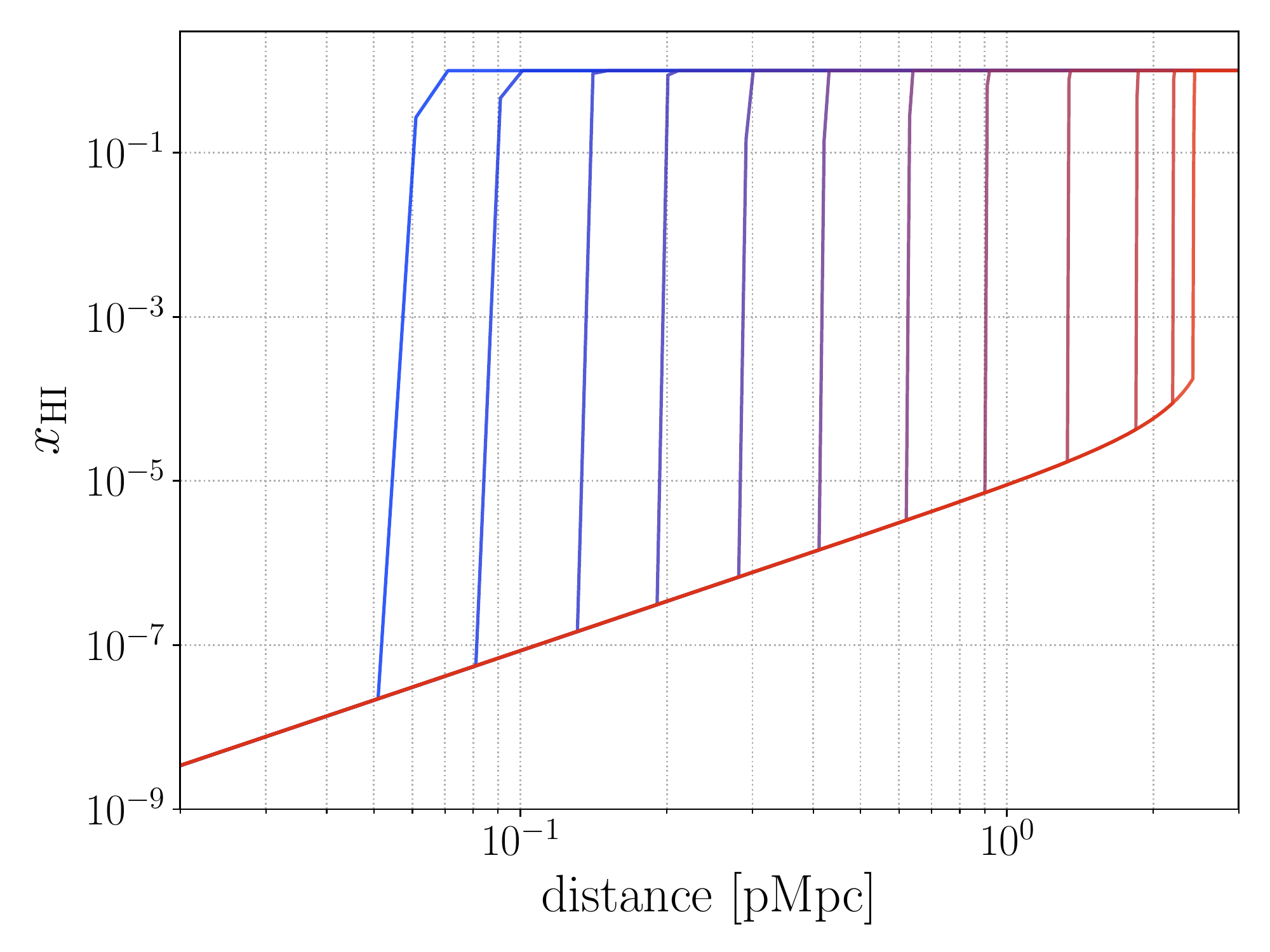}
    \includegraphics[width=0.33\textwidth]{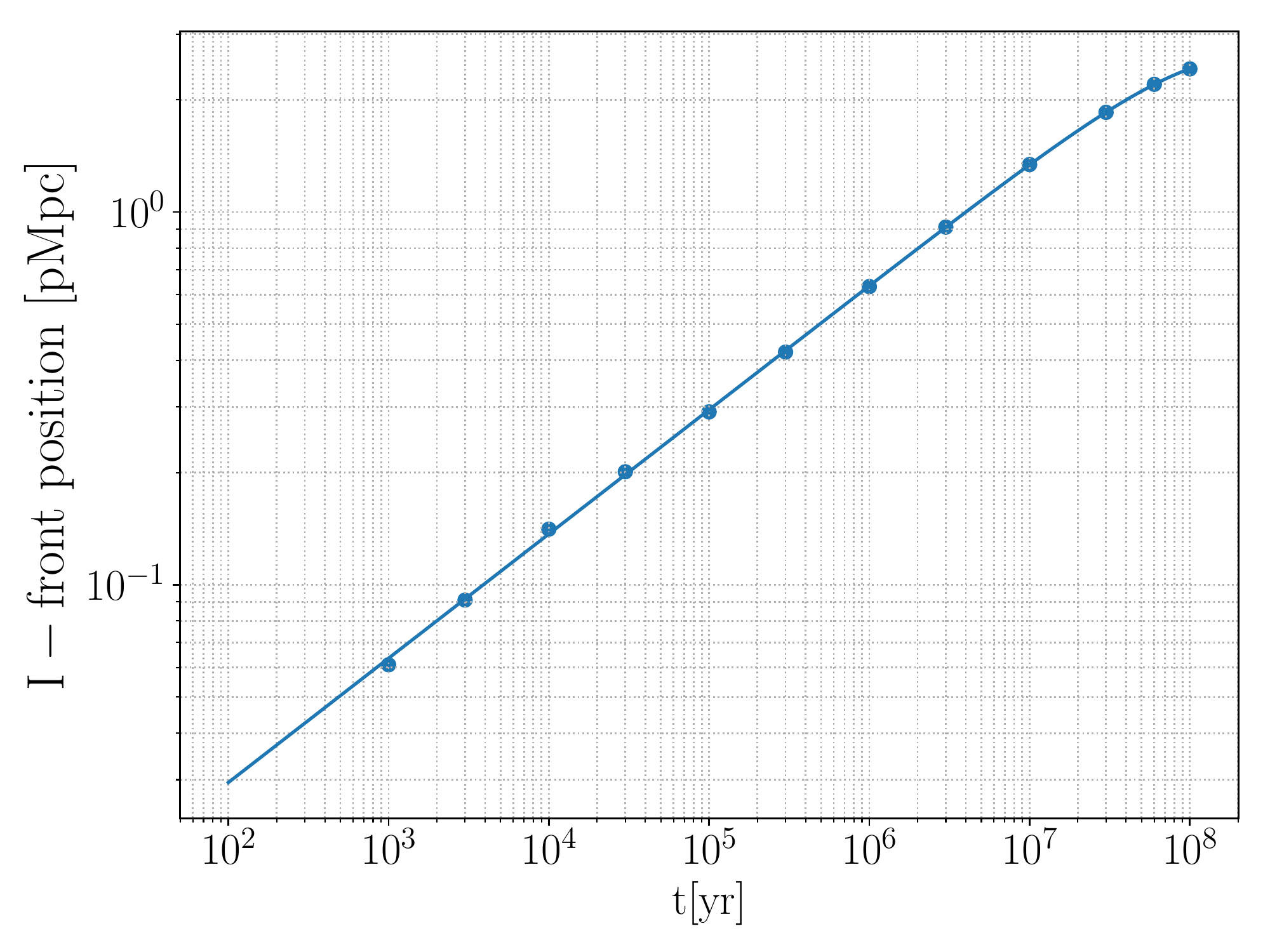}
    \includegraphics[width=0.33\textwidth]{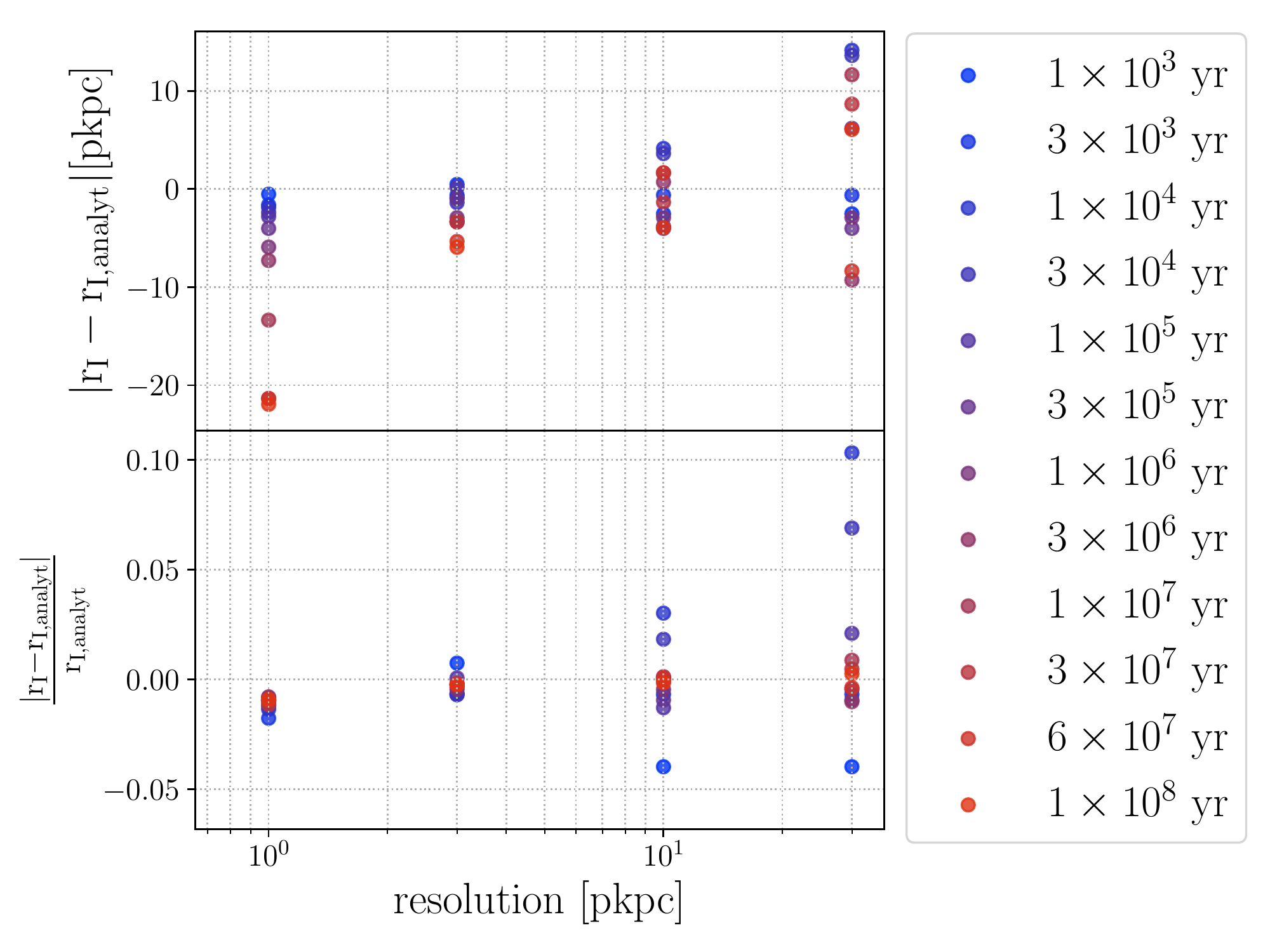}
    \caption{Left: neutral hydrogen fraction for the test problem of the I-front propagating in the uniform static medium, run with the spatial resolution of $10$ pkpc. Colors represent different times listed in the rightmost colormap. Middle: the position of the I-front at different quasar ages. The blue line is the analytical solution. Right: absolute and relative errors on the position of the I-front at different quasar ages, run with different spatial resolution.}
    \label{fig:I-front}
\end{figure*}

We test the scenario with $\dot{N}_\gamma=1\times 10^{57} \rm~ s^{-1}$, $n_{\rm H}=10^{-3} \rm~cm^{-3}$, $T=10^4\rm~K$, and the spatial resolution of $10$ pkpc. The I-front position vs time is shown in Figure \ref{fig:I-front}. The left panel shows the I-front propagation at different times, with the time-color map plotted at the right edge of the figure. In the middle  panel the solid line shows the analytical solution for the I-front position as a function of time. The blue points are the I-front positions obtained from the left panel. They agree very well with the analytical solution. In the right panel we show a resolution test. The upper panel shows the difference between the simulated I-front position and the analytical solution as a function of resolution, color-coded by quasar age. The lower panel shows the relative differences of the I-front positions, most of which are within $1\%$. Note that when the resolution is high, the absolute error in I-front position accumulates as quasar shines longer, but the relative error is still very small. The error can be suppressed if we reduce the tolerance in the $\alpha$-qss scheme. We choose the tolerance in the $\alpha$-qss scheme to be $1\%$ so to achieve both a reasonably small error and fast code speed.

\subsection{Ionization Structure}

\begin{figure*}
    \centering
    \includegraphics[width=\textwidth]{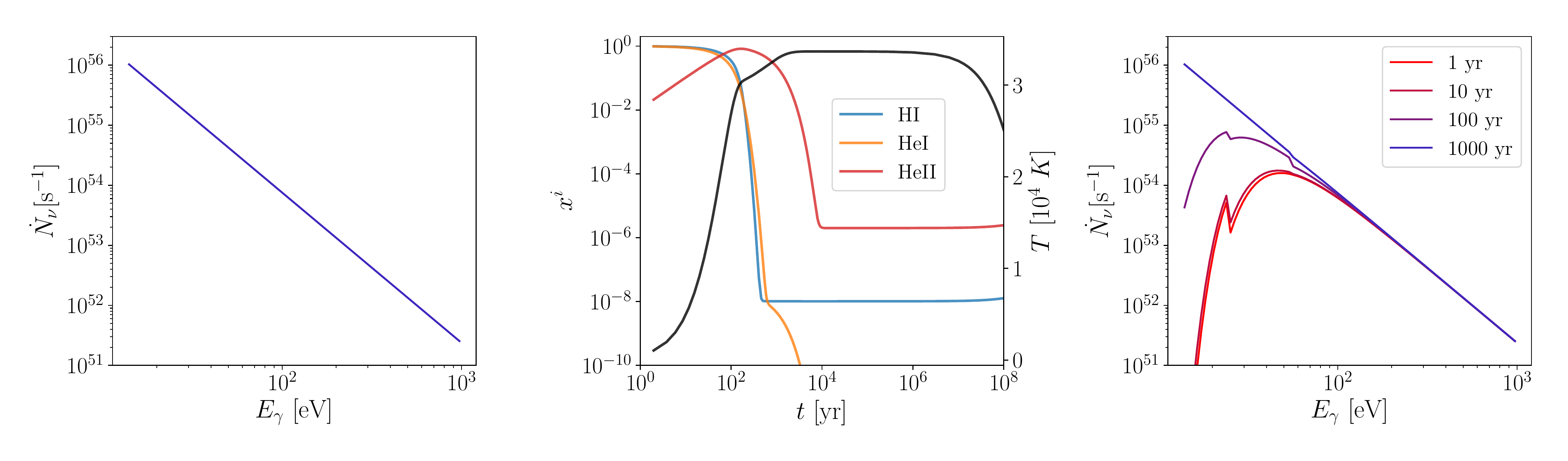}
    \includegraphics[width=\textwidth]{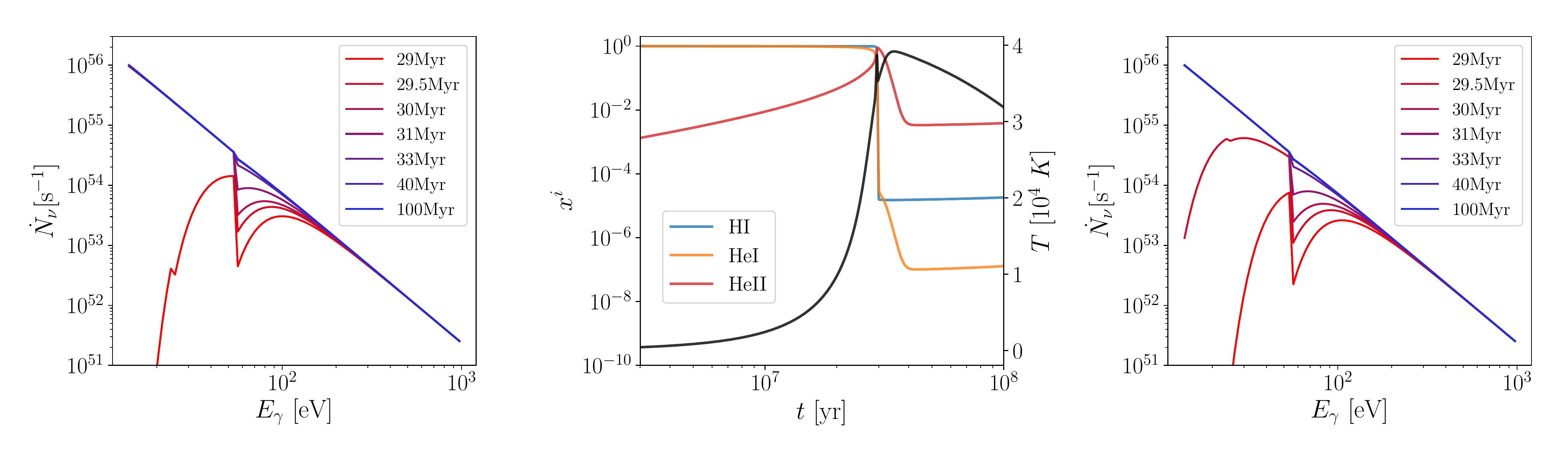}
    \caption{Incidental spectra (left column), the evolution of ionization fractions (middle panel), and the transmitted spectra (right column) of the test problem in A.2. The upper panels show the cell $0.1$ pMpc away from the quasar, while the bottom panels show the cell $4$ pMpc away. The incidental spectra for the first cell at $0.1$ pMpc is the quasar spectrum which does not vary, while the transmitted spectrum changes rapidly when ionization fraction changes the fastest. Note that only some example spectra are plotted.}
    \label{fig:cell}
\end{figure*}

\begin{figure*}
    \centering
    \includegraphics[width=\textwidth]{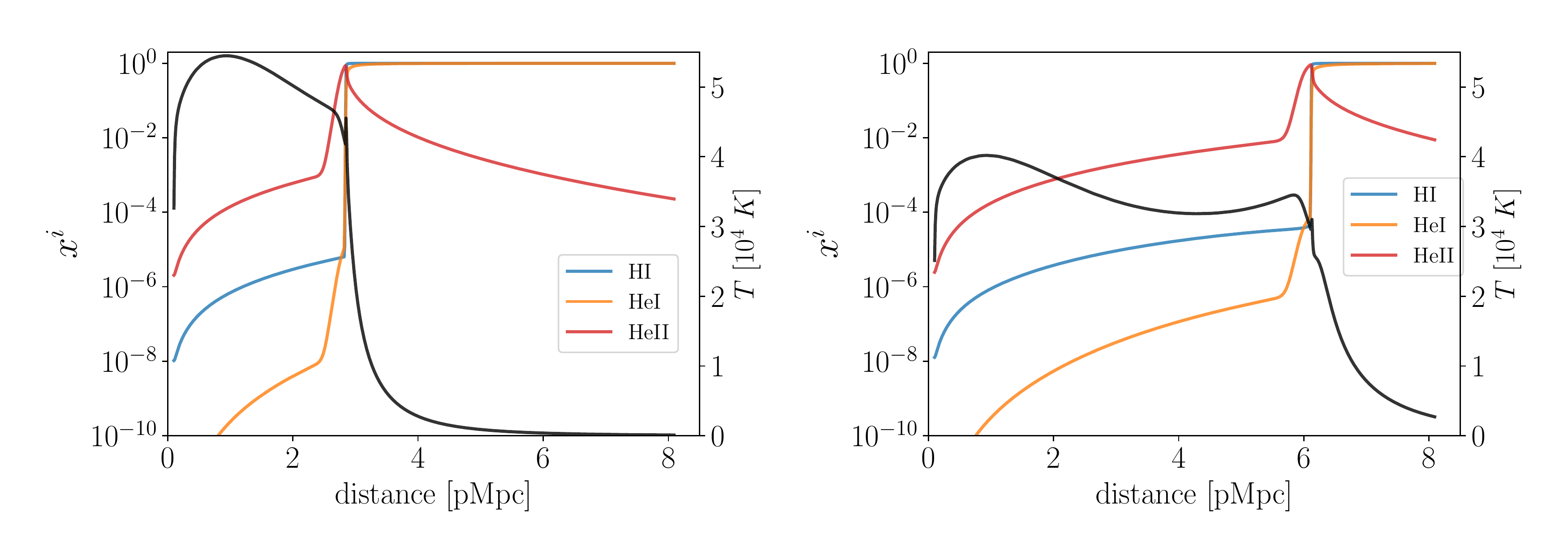}
    \caption{Ionization structure and the temperature structure for the test problem in A.2. The left panel shows the result at $10$ Myr and the right at $100$ Myr.}
    \label{fig:uniformlos}
\end{figure*}

In this section we show a scenario of a quasar ($\dot{N}=1\times10^{57} \rm~ s^{-1}$ , $\alpha=1.5$) turning on in the uniform, static IGM with neutral H and He at the mean cosmic density at $z=7$. Each cell in this line of sight is $10$ pkpc in size. In Figure \ref{fig:cell}, we show the incident spectra (left column), the ionization fraction and temperature evolution (middle column), and the transmitted spectra (right column) for the first cell at $0.1$pMpc (upper row) and a cell at $4$ pMpc (lower row) respectively. The spectra shown are a subset of all spectra stored during the calculation at several times when ionization fractions change the fastest. The quasar intrinsic spectrum (i.e.\ the incident spectrum on the first cell) does not vary, as is shown in the upper left panel. After the first cell, however, the transmitted spectra are hardened at early times because the cell is initially neutral. After thousand years the cell is fully ionized and the transmitted spectrum becomes identical to the incident one. For the cell at $4$ pMpc, the evolution is similar, except that the incident spectra also vary because the optical depth changes between the quasar and the cell and the ionization timescale increases from $1000$ yr to $\sim 1$ Myr. In Figure \ref{fig:uniformlos} we show the ionization and temperature structure of this line of sight at $t_Q=10$ Myr and $100$ Myr respectively. The I-front position and shape of each line are as expected.

\bibliographystyle{apj}
\bibliography{main}

\begin{thebibliography}{58}
\expandafter\ifx\csname natexlab\endcsname\relax\def\natexlab#1{#1}\fi

\bibitem[{{Abel} {et~al.}(1997){Abel}, {Anninos}, {Zhang}, \&
  {Norman}}]{abel1997}
{Abel}, T., {Anninos}, P., {Zhang}, Y., \& {Norman}, M.~L. 1997, \na, 2, 181

\bibitem[{{Ba{\~n}ados} {et~al.}(2019){Ba{\~n}ados}, {Rauch}, {Decarli},
  {Farina}, {Hennawi}, {Mazzucchelli}, {Venemans}, {Walter}, {Simcoe},
  {Prochaska}, {Cooper}, {Davies}, \& {Chen}}]{banados2019}
{Ba{\~n}ados}, E., {Rauch}, M., {Decarli}, R., {Farina}, E.~P., {Hennawi},
  J.~F., {Mazzucchelli}, C., {Venemans}, B.~P., {Walter}, F., {Simcoe}, R.~A.,
  {Prochaska}, J.~X., {Cooper}, T., {Davies}, F.~B., \& {Chen}, S.-F.~S. 2019,
  \apj, 885, 59

\bibitem[{{Becker} {et~al.}(2015){Becker}, {Bolton}, {Madau}, {Pettini},
  {Ryan-Weber}, \& {Venemans}}]{becker2015}
{Becker}, G.~D., {Bolton}, J.~S., {Madau}, P., {Pettini}, M., {Ryan-Weber},
  E.~V., \& {Venemans}, B.~P. 2015, \mnras, 447, 3402

\bibitem[{{Becker} {et~al.}(2001){Becker}, {Fan}, {White}, {Strauss},
  {Narayanan}, {Lupton}, {Gunn}, {Annis}, {Bahcall}, {Brinkmann}, {Connolly},
  {Csabai}, {Czarapata}, {Doi}, {Heckman}, {Hennessy}, {Ivezi{\'c}}, {Knapp},
  {Lamb}, {McKay}, {Munn}, {Nash}, {Nichol}, {Pier}, {Richards}, {Schneider},
  {Stoughton}, {Szalay}, {Thakar}, \& {York}}]{becker2001}
{Becker}, R.~H., {Fan}, X., {White}, R.~L., {Strauss}, M.~A., {Narayanan},
  V.~K., {Lupton}, R.~H., {Gunn}, J.~E., {Annis}, J., {Bahcall}, N.~A.,
  {Brinkmann}, J., {Connolly}, A.~J., {Csabai}, I., {Czarapata}, P.~C., {Doi},
  M., {Heckman}, T.~M., {Hennessy}, G.~S., {Ivezi{\'c}}, {\v{Z}}., {Knapp},
  G.~R., {Lamb}, D.~Q., {McKay}, T.~A., {Munn}, J.~A., {Nash}, T., {Nichol},
  R., {Pier}, J.~R., {Richards}, G.~T., {Schneider}, D.~P., {Stoughton}, C.,
  {Szalay}, A. e.~S., {Thakar}, A.~R., \& {York}, D.~G. 2001, \aj, 122, 2850

\bibitem[{{Behroozi} {et~al.}(2013){Behroozi}, {Wechsler}, \&
  {Wu}}]{Behroozi2013}
{Behroozi}, P.~S., {Wechsler}, R.~H., \& {Wu}, H.-Y. 2013, \apj, 762, 109

\bibitem[{{Bolton} \& {Haehnelt}(2007)}]{bolton2007a}
{Bolton}, J.~S. \& {Haehnelt}, M.~G. 2007, \mnras, 374, 493

\bibitem[{{Bosman} {et~al.}(2018){Bosman}, {Fan}, {Jiang}, {Reed}, {Matsuoka},
  {Becker}, \& {Haehnelt}}]{bosman2018}
{Bosman}, S. E.~I., {Fan}, X., {Jiang}, L., {Reed}, S., {Matsuoka}, Y.,
  {Becker}, G., \& {Haehnelt}, M. 2018, \mnras, 479, 1055

\bibitem[{{Carilli} {et~al.}(2010){Carilli}, {Wang}, {Fan}, {Walter}, {Kurk},
  {Riechers}, {Wagg}, {Hennawi}, {Jiang}, {Menten}, {Bertoldi}, {Strauss}, \&
  {Cox}}]{carilli2010}
{Carilli}, C.~L., {Wang}, R., {Fan}, X., {Walter}, F., {Kurk}, J., {Riechers},
  D., {Wagg}, J., {Hennawi}, J., {Jiang}, L., {Menten}, K.~M., {Bertoldi}, F.,
  {Strauss}, M.~A., \& {Cox}, P. 2010, \apj, 714, 834

\bibitem[{{Cen}(1992)}]{cen1992}
{Cen}, R. 1992, \apjs, 78, 341

\bibitem[{{Cen} \& {Haiman}(2000)}]{cen2000}
{Cen}, R. \& {Haiman}, Z. 2000, \apjl, 542, L75

\bibitem[{{Chen}(2020)}]{Chen2020}
{Chen}, H. 2020, \apj, 893, 165

\bibitem[{{Davies} {et~al.}(2016){Davies}, {Furlanetto}, \&
  {McQuinn}}]{davies2016}
{Davies}, F.~B., {Furlanetto}, S.~R., \& {McQuinn}, M. 2016, \mnras, 457, 3006

\bibitem[{{Davies} {et~al.}(2019){Davies}, {Hennawi}, \& {Eilers}}]{davies2019}
{Davies}, F.~B., {Hennawi}, J.~F., \& {Eilers}, A.-C. 2019, \apjl, 884, L19

\bibitem[{{Davies} {et~al.}(2020){Davies}, {Hennawi}, \& {Eilers}}]{davies2020}
---. 2020, \mnras, 493, 1330

\bibitem[{{Eilers} {et~al.}(2018{\natexlab{a}}){Eilers}, {Davies}, \&
  {Hennawi}}]{eilers2018}
{Eilers}, A.-C., {Davies}, F.~B., \& {Hennawi}, J.~F. 2018{\natexlab{a}}, \apj,
  864, 53

\bibitem[{{Eilers} {et~al.}(2017){Eilers}, {Davies}, {Hennawi}, {Prochaska},
  {Luki{\'c}}, \& {Mazzucchelli}}]{eilers2017}
{Eilers}, A.-C., {Davies}, F.~B., {Hennawi}, J.~F., {Prochaska}, J.~X.,
  {Luki{\'c}}, Z., \& {Mazzucchelli}, C. 2017, \apj, 840, 24

\bibitem[{{Eilers} {et~al.}(2018{\natexlab{b}}){Eilers}, {Hennawi}, \&
  {Davies}}]{eilers2018b}
{Eilers}, A.-C., {Hennawi}, J.~F., \& {Davies}, F.~B. 2018{\natexlab{b}}, \apj,
  867, 30

\bibitem[{{Eilers} {et~al.}(2020){Eilers}, {Hennawi}, {Decarli}, {Davies},
  {Venemans}, {Walter}, {Ba{\~n}ados}, {Fan}, {Farina}, {Mazzucchelli},
  {Novak}, {Schindler}, {Simcoe}, {Wang}, \& {Yang}}]{eilers2020}
{Eilers}, A.-C., {Hennawi}, J.~F., {Decarli}, R., {Davies}, F.~B., {Venemans},
  B., {Walter}, F., {Ba{\~n}ados}, E., {Fan}, X., {Farina}, E.~P.,
  {Mazzucchelli}, C., {Novak}, M., {Schindler}, J.-T., {Simcoe}, R.~A., {Wang},
  F., \& {Yang}, J. 2020, \apj, 900, 37

\bibitem[{{Fan} {et~al.}(2006){Fan}, {Strauss}, {Becker}, {White}, {Gunn},
  {Knapp}, {Richards}, {Schneider}, {Brinkmann}, \& {Fukugita}}]{fan2006b}
{Fan}, X., {Strauss}, M.~A., {Becker}, R.~H., {White}, R.~L., {Gunn}, J.~E.,
  {Knapp}, G.~R., {Richards}, G.~T., {Schneider}, D.~P., {Brinkmann}, J., \&
  {Fukugita}, M. 2006, \aj, 132, 117

\bibitem[{{Fumagalli} {et~al.}(2016){Fumagalli}, {O'Meara}, \&
  {Prochaska}}]{Fumagalli2016}
{Fumagalli}, M., {O'Meara}, J.~M., \& {Prochaska}, J.~X. 2016, \mnras, 455,
  4100

\bibitem[{{Furlanetto} \& {Stoever}(2010)}]{furlanetto2010}
{Furlanetto}, S.~R. \& {Stoever}, S.~J. 2010, \mnras, 404, 1869

\bibitem[{{Gnedin}(2014{\natexlab{a}})}]{ng14}
{Gnedin}, N.~Y. 2014{\natexlab{a}}, \apj, 793, 29

\bibitem[{{Gnedin}(2014{\natexlab{b}})}]{gnedin2014}
---. 2014{\natexlab{b}}, \apj, 793, 29

\bibitem[{{Gnedin} \& {Abel}(2001)}]{gnedin2001}
{Gnedin}, N.~Y. \& {Abel}, T. 2001, \na, 6, 437

\bibitem[{{Haiman} \& {Cen}(2001)}]{haiman2001}
{Haiman}, Z. \& {Cen}, R. 2001, in Astronomical Society of the Pacific
  Conference Series, Vol. 222, The Physics of Galaxy Formation, ed.
  M.~{Umemura} \& H.~{Susa}, 101

\bibitem[{{Harikane} {et~al.}(2020){Harikane}, {Laporte}, {Ellis}, \&
  {Matsuoka}}]{harikane2020}
{Harikane}, Y., {Laporte}, N., {Ellis}, R.~S., \& {Matsuoka}, Y. 2020, \apj,
  902, 117

\bibitem[{{Iliev} {et~al.}(2006){Iliev}, {Ciardi}, {Alvarez}, {Maselli},
  {Ferrara}, {Gnedin}, {Mellema}, {Nakamoto}, {Norman}, {Razoumov},
  {Rijkhorst}, {Ritzerveld}, {Shapiro}, {Susa}, {Umemura}, \&
  {Whalen}}]{iliev2006}
{Iliev}, I.~T., {Ciardi}, B., {Alvarez}, M.~A., {Maselli}, A., {Ferrara}, A.,
  {Gnedin}, N.~Y., {Mellema}, G., {Nakamoto}, T., {Norman}, M.~L., {Razoumov},
  A.~O., {Rijkhorst}, E.-J., {Ritzerveld}, J., {Shapiro}, P.~R., {Susa}, H.,
  {Umemura}, M., \& {Whalen}, D.~J. 2006, \mnras, 371, 1057

\bibitem[{{Ishimoto} {et~al.}(2020){Ishimoto}, {Kashikawa}, {Onoue},
  {Matsuoka}, {Izumi}, {Strauss}, {Fujimoto}, {Imanishi}, {Ito}, {Iwasawa},
  {Kawaguchi}, {Lee}, {Liang}, {Lu}, {Momose}, {Toba}, \&
  {Uchiyama}}]{ishimoto2020}
{Ishimoto}, R., {Kashikawa}, N., {Onoue}, M., {Matsuoka}, Y., {Izumi}, T.,
  {Strauss}, M.~A., {Fujimoto}, S., {Imanishi}, M., {Ito}, K., {Iwasawa}, K.,
  {Kawaguchi}, T., {Lee}, C.-H., {Liang}, Y., {Lu}, T.-Y., {Momose}, R.,
  {Toba}, Y., \& {Uchiyama}, H. 2020, \apj, 903, 60

\bibitem[{{Jiang} {et~al.}(2006){Jiang}, {Fan}, {Hines}, {Shi}, {Vestergaard},
  {Bertoldi}, {Brandt}, {Carilli}, {Cox}, {Le Floc'h}, {Pentericci},
  {Richards}, {Rieke}, {Schneider}, {Strauss}, {Walter}, \&
  {Brinkmann}}]{jiang2006}
{Jiang}, L., {Fan}, X., {Hines}, D.~C., {Shi}, Y., {Vestergaard}, M.,
  {Bertoldi}, F., {Brandt}, W.~N., {Carilli}, C.~L., {Cox}, P., {Le Floc'h},
  E., {Pentericci}, L., {Richards}, G.~T., {Rieke}, G.~H., {Schneider}, D.~P.,
  {Strauss}, M.~A., {Walter}, F., \& {Brinkmann}, J. 2006, \aj, 132, 2127

\bibitem[{{Kakiichi} {et~al.}(2018){Kakiichi}, {Ellis}, {Laporte}, {Zitrin},
  {Eilers}, {Ryan-Weber}, {Meyer}, {Robertson}, {Stark}, \&
  {Bosman}}]{kakiichi2018}
{Kakiichi}, K., {Ellis}, R.~S., {Laporte}, N., {Zitrin}, A., {Eilers}, A.-C.,
  {Ryan-Weber}, E., {Meyer}, R.~A., {Robertson}, B., {Stark}, D.~P., \&
  {Bosman}, S. E.~I. 2018, \mnras, 479, 43

\bibitem[{{Keating} {et~al.}(2015){Keating}, {Haehnelt}, {Cantalupo}, \&
  {Puchwein}}]{keating2015}
{Keating}, L.~C., {Haehnelt}, M.~G., {Cantalupo}, S., \& {Puchwein}, E. 2015,
  \mnras, 454, 681

\bibitem[{{Khrykin} {et~al.}(2019){Khrykin}, {Hennawi}, \&
  {Worseck}}]{khrykin2019}
{Khrykin}, I.~S., {Hennawi}, J.~F., \& {Worseck}, G. 2019, \mnras, 484, 3897

\bibitem[{{Kravtsov}(1999)}]{kravtsov1999}
{Kravtsov}, A.~V. 1999, PhD thesis, NEW MEXICO STATE UNIVERSITY

\bibitem[{{Kravtsov} {et~al.}(2002){Kravtsov}, {Klypin}, \&
  {Hoffman}}]{kravtsov2002}
{Kravtsov}, A.~V., {Klypin}, A., \& {Hoffman}, Y. 2002, \apj, 571, 563

\bibitem[{{Li} {et~al.}(2020){Li}, {Wang}, {Cox}, {Gao}, {Walter}, {Wagg},
  {Menten}, {Bertoldi}, {Shao}, {Venemans}, {Decarli}, {Riechers}, {Neri},
  {Fan}, {Omont}, \& {Narayanan}}]{li2020}
{Li}, J., {Wang}, R., {Cox}, P., {Gao}, Y., {Walter}, F., {Wagg}, J., {Menten},
  K.~M., {Bertoldi}, F., {Shao}, Y., {Venemans}, B.~P., {Decarli}, R.,
  {Riechers}, D., {Neri}, R., {Fan}, X., {Omont}, A., \& {Narayanan}, D. 2020,
  \apj, 900, 131

\bibitem[{{Lidz} {et~al.}(2007){Lidz}, {McQuinn}, {Zaldarriaga}, {Hernquist},
  \& {Dutta}}]{lidz2007}
{Lidz}, A., {McQuinn}, M., {Zaldarriaga}, M., {Hernquist}, L., \& {Dutta}, S.
  2007, \apj, 670, 39

\bibitem[{{Lu} {et~al.}(2020){Lu}, {Goto}, {Tang}, {Hashimoto}, {Wong},
  {Chiang}, {Wu}, {Kim}, {Ho}, {Wang}, {On}, \& {D. Santos}}]{lu2020}
{Lu}, T.-Y., {Goto}, T., {Tang}, J.-J., {Hashimoto}, T., {Wong}, Y.-H.~V.,
  {Chiang}, C.-Y., {Wu}, Y.-H., {Kim}, S.~J., {Ho}, S. C.~C., {Wang}, T.-W.,
  {On}, A. Y.~L., \& {D. Santos}, D.~J. 2020, \apj, 893, 69

\bibitem[{{Lusso} {et~al.}(2015){Lusso}, {Worseck}, {Hennawi}, {Prochaska},
  {Vignali}, {Stern}, \& {O'Meara}}]{Lusso2015}
{Lusso}, E., {Worseck}, G., {Hennawi}, J.~F., {Prochaska}, J.~X., {Vignali},
  C., {Stern}, J., \& {O'Meara}, J.~M. 2015, \mnras, 449, 4204

\bibitem[{{Madau} \& {Rees}(2000)}]{madau2000}
{Madau}, P. \& {Rees}, M.~J. 2000, \apjl, 542, L69

\bibitem[{{Martini}(2004)}]{martini2004}
{Martini}, P. 2004, in Coevolution of Black Holes and Galaxies, ed. L.~C. {Ho},
  169

\bibitem[{{Mazzucchelli} {et~al.}(2017){Mazzucchelli}, {Ba{\~n}ados},
  {Venemans}, {Decarli}, {Farina}, {Walter}, {Eilers}, {Rix}, {Simcoe},
  {Stern}, {Fan}, {Schlafly}, {De Rosa}, {Hennawi}, {Chambers}, {Greiner},
  {Burgett}, {Draper}, {Kaiser}, {Kudritzki}, {Magnier}, {Metcalfe}, {Waters},
  \& {Wainscoat}}]{mazzucchelli2017}
{Mazzucchelli}, C., {Ba{\~n}ados}, E., {Venemans}, B.~P., {Decarli}, R.,
  {Farina}, E.~P., {Walter}, F., {Eilers}, A.~C., {Rix}, H.~W., {Simcoe}, R.,
  {Stern}, D., {Fan}, X., {Schlafly}, E., {De Rosa}, G., {Hennawi}, J.,
  {Chambers}, K.~C., {Greiner}, J., {Burgett}, W., {Draper}, P.~W., {Kaiser},
  N., {Kudritzki}, R.~P., {Magnier}, E., {Metcalfe}, N., {Waters}, C., \&
  {Wainscoat}, R.~J. 2017, \apj, 849, 91

\bibitem[{{McGreer} {et~al.}(2015){McGreer}, {Mesinger}, \&
  {D'Odorico}}]{mcgreer2015}
{McGreer}, I.~D., {Mesinger}, A., \& {D'Odorico}, V. 2015, \mnras, 447, 499

\bibitem[{Mott \& Oran(2001)}]{mott01}
Mott, D.~R. \& Oran, E.~S. 2001, CHEMEQ2: A solver for the stiff ordinary
  differential equations of chemical kinetics, Tech. rep., Naval Research Lab
  Washington DC

\bibitem[{{Osterbrock}(1989)}]{osterbrock1989}
{Osterbrock}, D.~E. 1989, {Astrophysics of gaseous nebulae and active galactic
  nuclei}

\bibitem[{{Peebles}(1971)}]{peebles1971}
{Peebles}, P.~J.~E. 1971, {Physical cosmology}

\bibitem[{{Rudd} {et~al.}(2008){Rudd}, {Zentner}, \& {Kravtsov}}]{rudd2008}
{Rudd}, D.~H., {Zentner}, A.~R., \& {Kravtsov}, A.~V. 2008, \apj, 672, 19

\bibitem[{{Shapiro} {et~al.}(2006){Shapiro}, {Iliev}, {Alvarez}, \&
  {Scannapieco}}]{shapiro2006}
{Shapiro}, P.~R., {Iliev}, I.~T., {Alvarez}, M.~A., \& {Scannapieco}, E. 2006,
  \apj, 648, 922

\bibitem[{{Shull} \& {van Steenberg}(1985)}]{shull1985}
{Shull}, J.~M. \& {van Steenberg}, M.~E. 1985, \apj, 298, 268

\bibitem[{{Smith} \& {Bromm}(2019)}]{smith2019}
{Smith}, A. \& {Bromm}, V. 2019, Contemporary Physics, 60, 111

\bibitem[{{Tepper-Garc{\'\i}a}(2006)}]{tepper-garcia2006}
{Tepper-Garc{\'\i}a}, T. 2006, \mnras, 369, 2025

\bibitem[{{Theuns}(2020)}]{theuns2020}
{Theuns}, T. 2020, \mnras, 500, 2741

\bibitem[{{Theuns} {et~al.}(1998){Theuns}, {Leonard}, {Efstathiou}, {Pearce},
  \& {Thomas}}]{theuns1998}
{Theuns}, T., {Leonard}, A., {Efstathiou}, G., {Pearce}, F.~R., \& {Thomas},
  P.~A. 1998, \mnras, 301, 478

\bibitem[{{Turk} {et~al.}(2011){Turk}, {Smith}, {Oishi}, {Skory}, {Skillman},
  {Abel}, \& {Norman}}]{turk2011}
{Turk}, M.~J., {Smith}, B.~D., {Oishi}, J.~S., {Skory}, S., {Skillman}, S.~W.,
  {Abel}, T., \& {Norman}, M.~L. 2011, \apjs, 192, 9

\bibitem[{{Venemans} {et~al.}(2015){Venemans}, {Ba{\~n}ados}, {Decarli},
  {Farina}, {Walter}, {Chambers}, {Fan}, {Rix}, {Schlafly}, {McMahon},
  {Simcoe}, {Stern}, {Burgett}, {Draper}, {Flewelling}, {Hodapp}, {Kaiser},
  {Magnier}, {Metcalfe}, {Morgan}, {Price}, {Tonry}, {Waters}, {AlSayyad},
  {Banerji}, {Chen}, {Gonz{\'a}lez-Solares}, {Greiner}, {Mazzucchelli},
  {McGreer}, {Miller}, {Reed}, \& {Sullivan}}]{venemans2015}
{Venemans}, B.~P., {Ba{\~n}ados}, E., {Decarli}, R., {Farina}, E.~P., {Walter},
  F., {Chambers}, K.~C., {Fan}, X., {Rix}, H.~W., {Schlafly}, E., {McMahon},
  R.~G., {Simcoe}, R., {Stern}, D., {Burgett}, W.~S., {Draper}, P.~W.,
  {Flewelling}, H., {Hodapp}, K.~W., {Kaiser}, N., {Magnier}, E.~A.,
  {Metcalfe}, N., {Morgan}, J.~S., {Price}, P.~A., {Tonry}, J.~L., {Waters},
  C., {AlSayyad}, Y., {Banerji}, M., {Chen}, S.~S., {Gonz{\'a}lez-Solares},
  E.~A., {Greiner}, J., {Mazzucchelli}, C., {McGreer}, I., {Miller}, D.~R.,
  {Reed}, S., \& {Sullivan}, P.~W. 2015, \apjl, 801, L11

\bibitem[{{Wolfe} {et~al.}(2005){Wolfe}, {Gawiser}, \& {Prochaska}}]{wolfe2005}
{Wolfe}, A.~M., {Gawiser}, E., \& {Prochaska}, J.~X. 2005, \araa, 43, 861

\bibitem[{{Yang} {et~al.}(2020){Yang}, {Wang}, {Fan}, {Hennawi}, {Davies},
  {Yue}, {Eilers}, {Farina}, {Wu}, {Bian}, {Pacucci}, \& {Lee}}]{yang2020}
{Yang}, J., {Wang}, F., {Fan}, X., {Hennawi}, J.~F., {Davies}, F.~B., {Yue},
  M., {Eilers}, A.-C., {Farina}, E.~P., {Wu}, X.-B., {Bian}, F., {Pacucci}, F.,
  \& {Lee}, K.-G. 2020, \apj, 904, 26

\bibitem[{{Zahedy} {et~al.}(2019){Zahedy}, {Chen}, {Johnson}, {Pierce},
  {Rauch}, {Huang}, {Weiner}, \& {Gauthier}}]{Zahedy2019}
{Zahedy}, F.~S., {Chen}, H.-W., {Johnson}, S.~D., {Pierce}, R.~M., {Rauch}, M.,
  {Huang}, Y.-H., {Weiner}, B.~J., \& {Gauthier}, J.-R. 2019, \mnras, 484, 2257

\bibitem[{{Zhu} {et~al.}(2020){Zhu}, {Avestruz}, \& {Gnedin}}]{Zhu2020}
{Zhu}, H., {Avestruz}, C., \& {Gnedin}, N.~Y. 2020, \apj, 899, 137

\end{thebibliography}

\end{document}